# Spectroscopic Studies of Quasiparticle Low-Energy Excitations in Cuprate and Iron-Based High-Temperature Superconductors

Nai-Chang Yeh[1]

*Department of Physics and Kavli Nanoscience Institute, California Institute of Technology, Pasadena, California 91125, USA*

**Abstract:** Recent development in the physics of high-temperature superconductivity is reviewed, with special emphasis on the studies of the low-energy excitations of cuprate and iron-based superconductors. For cuprate superconductors, a phenomenology based on coexisting competing orders with superconductivity in the ground state of these doped Mott insulators is shown to provide a consistent account for a wide range of experimental findings. In the case of iron-based superconductors, studies of the low-energy excitations reveal interesting similarities and differences when compared with cuprate superconductors. In contrast to the single-band cuprate superconductivity with an insulating parent state, the ferrous superconductors are multi-band materials with a semi-metallic parent state and exhibit two-gap superconductivity when doped. On the other hand, both systems exhibit strong antiferromagnetic correlation and Fermi-surface distortion, leading to unconventional pairing symmetries with sign-changing order parameters on different parts of the Fermi surface. These findings suggest that the pairing potentials in both the cuprate and the ferrous superconductors are generally repulsive, thus favor a pairing mechanism that is electronically driven and a pairing strength that is closely related to the electronic correlation. The physical implications of the unified phenomenology based on antiferromagnetic correlations and remaining open issues associated with the cuprate and ferrous superconductivity are discussed.

## 1. Introduction

Since the discovery of superconductivity about 100 years ago, the subject of superconductivity has remained one of the most intellectually challenging topics in condensed matter physics. Although much progress has been made in the physics and applications of superconductivity, to date it is still impossible to predict from first principle whether a specific type of material or compound would become superconducting below a given temperature. Moreover, the discovery of high-temperature superconducting (high-$T_c$) cuprates in 1986 [1] and the subsequent discovery of the iron-based superconductors in 2008 [2] have completely defied the conventional wisdom to avoid oxides and magnetic materials in search of high-$T_c$ superconductors. Despite intense research efforts worldwide, the pairing mechanism for high-$T_c$ superconductivity remains elusive to date. On the other hand, the discovery of iron-based high-$T_c$ superconductors provides interesting comparisons with the cuprate superconductors, which help shed new light on the mystery of high-$T_c$ pairing mechanism.

The objective of this review is to survey the up-to-date status of experimental manifestations of unconventional low-energy excitations in the cuprates, explore the feasibility of a unified phenomenology for all cuprates, compare the findings in the cuprates with those of the iron-based superconductors, and then discuss the implications of these studies on the microscopic pairing mechanism of high-$T_c$ superconductivity.

This article is structured as follows. In Section 2 an overview is given for the representative empirical findings of unconventional low-energy excitations in the cuprate superconductors, followed by discussions of the underlying physics associated with these phenomena in various states, including the zero-field pairing state below the superconducting transition, the zero-field normal state above the superconducting transition, and the vortex state in the presence of finite magnetic fields. In Section 3 we review the basic properties of iron-based superconductors and the characteristics of low-energy quasiparticle and spin excitations, and then discuss the theoretical implications associated with these empirical findings. In Section 4 we compare the low-energy excitations of the cuprate and ferrous superconductors, and discuss the physical implications on the mechanism for high-temperature superconductivity. Additionally, important open issues and possible clues in the quest for high-temperature superconductivity are summarized. Finally, Section 5 concludes the status of our current understanding of cuprate and ferrous superconductivity and the challenges required to unravel the mystery of the high-temperature superconducting mechanism.

[1] E-mail: ncyeh@caltech.edu

## 2. Unconventional Low-Energy Excitations in the Cuprate Superconductors

### 2.1. Overview

Cuprate superconductors are doped antiferromagnetic (AFM) Mott insulators with strong electronic correlation [3-7]. Mott insulators differ from conventional "band insulators" in that the latter are dictated by the Pauli exclusion principle when the highest occupied band contains two electrons per unit cell, whereas the former are influenced by the strong on-site Coulomb repulsion such that double occupancy of electrons per unit cell is energetically unfavorable and the electronic system behaves like an insulator rather than a good conductor at half filling. An important signature of doped Mott insulators is the strong electronic correlation among the carriers due to poor screening and the sensitivity of their ground state to the doping level. In the cuprates, the ground state of the undoped perovskite oxide is an antiferromagnetic (AFM) Mott insulator, with nearest-neighbor $Cu^{2+}$-$Cu^{2+}$ AFM exchange interaction in the $CuO_2$ planes [8]. Depending on doping with either electrons or holes into the $CuO_2$ planes [8,9], the Néel temperature ($T_N$) for the AFM-to-paramagnetic transition decreases with increasing doping level. Upon further doping of carriers, long-range AFM vanishes, spin fluctuations become important, and various competing orders (COs) begin to appear in the ground state, followed by the occurrence of superconductivity (SC). As schematically illustrated in the phase diagrams for the hole- and electron type cuprates in Fig. 1(a), the superconducting transition temperature ($T_c$) first increases with increasing doping level ($\delta$), reaching a maximum $T_c$ at an optimal doping level, then decreases and finally vanishes with further increase of doping.

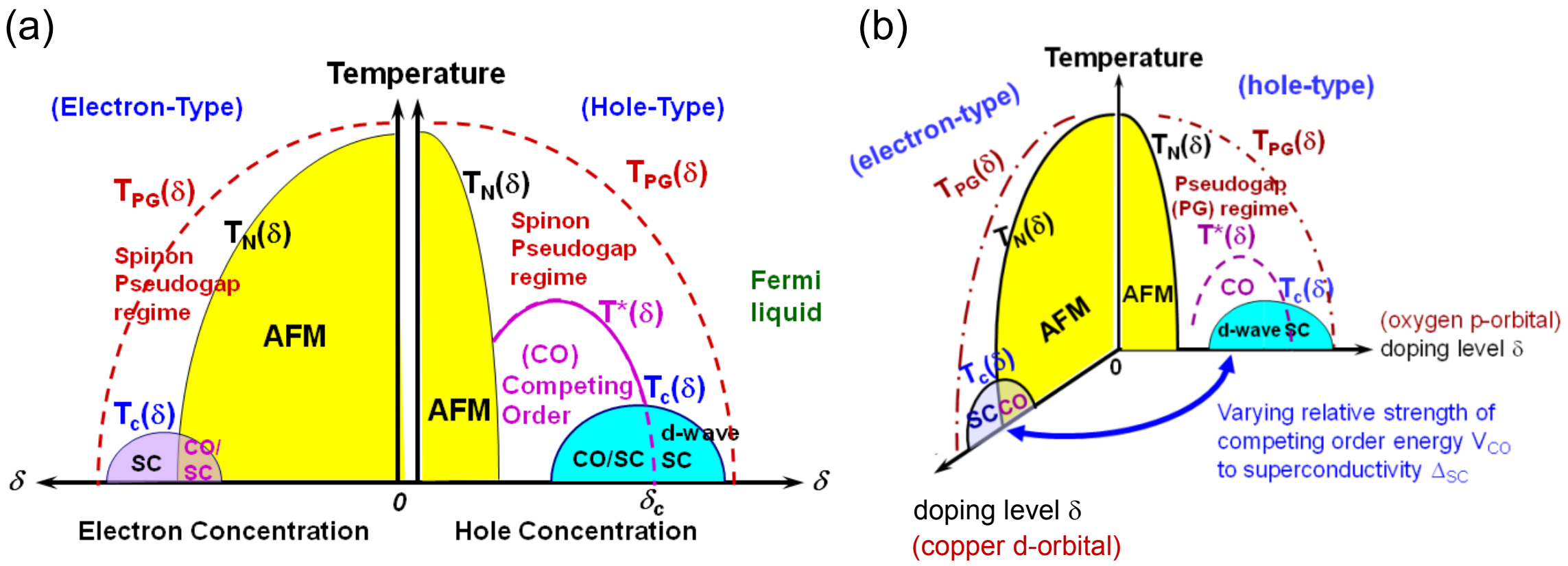

**Fig. 1: (a)** A schematic zero-field temperature ($T$) versus doping level ($\delta$) generic phase diagram for electron- and hole-type cuprates [7]. (AFM: antiferromagnetism, CO: competing order, SC: superconductivity, $\delta$: doping level, $T_N$: Néel temperature, $T_c$: superconducting transition temperature, $T^*$: low-energy pseudogap (PG) temperature, $T_{PG}$: high-energy pseudogap temperature). **(b)** A feasible explanation for the asymmetric hole- and electron-type phase diagrams may be attributed to the differences in the ratio of the SC energy gap ($\Delta_{SC}$) relative to the competing order energy gap ($V_{CO}$): For hole-type cuprates revealing low-energy PG phenomena, empirical evidences suggest $V_{CO} > \Delta_{SC}$ [7,10-16]. In contrast, tunneling experiments indicate a "hidden CO gap" only revealed in the vortex state of the electron-type cuprate superconductors with $V_{CO} < \Delta_{SC}$ [7,10,13,17].

Although much similarity exists between the phase diagrams for the hole- and electron-type cuprates, closer inspection indicates asymmetric characteristics: For hole-type cuprates in the under- and optimally doped regime, the physical properties above $T_c$ but below a crossover temperature $T^*$, known as the low-energy pseudogap (PG) temperature, are significantly different from those of Fermi liquids, including slightly suppressed electronic density of states (DOS) referred to as the low-energy PG phenomenon [18-20] and incompletely recovered Fermi surfaces in the momentum space known as the Fermi arc phenomenon [16,19,21]. Here the PG phenomenon refers to the observation of a soft gap without coherence peaks in the quasiparticle excitation spectra above $T_c$ in hole-type cuprates and below a PG temperature $T^*$, as exemplified in Fig. 2. Evidence for the PG in hole-type cuprates has been reported in tunneling measurements [7,10-15,20,22-25], $^{63}$Cu spin-lattice relaxation rate and $^{63}$Cu Knight shift in nuclear magnetic resonance (NMR) experiments [18,26-31], optical conductivity experiments [18,32], Raman scattering experiments [33,34] and angle resolved photoemission spectroscopy (ARPES) measurements [19,21]. Intimately related to the PG is the observation of Fermi arcs in the hole-type cuprates, which refers to an incomplete recovery of the full Fermi surface for temperature in the range $T_c < T < T^*$ [16,19,21], and will be discussed in more details in Section 2.3. On the other hand, neither low-energy PG [7,17,35] nor Fermi arc phenomena [36] can be found in the electron-type cuprate superconductors in the absence of magnetic fields. Interestingly, however, break-junction tunneling spectra of a one-layer electron-type cuprate revealed PG phenomena at $T < T_c$ in the vortex state [37,38]. Similarly, spatially resolved scanning tunneling spectroscopic studies of the vortex state of the infinite-layer electron-type cuprate $Sr_{0.9}La_{0.1}CuO_2$ revealed PG features with a characteristic energy $\Delta_{PG} < \Delta_{SC}$ inside the vortex core [7,10,17]. Moreover, electronic Raman scattering experiments on $Nd_{2-x}Ce_xCuO_4$ also revealed contributions of an additional small energy gap in the SC state [39]. Thus, many of the seemingly puzzling asymmetric properties between the hole- and electron-type cuprates may be explained by the differences in the ratio of the SC energy gap ($\Delta_{SC}$) relative to a competing order (CO) energy gap ($V_{CO}$) and by attributing the origin of the low-energy PG phenomena to the presence of a CO energy gap so that $V_{CO} \sim \Delta_{PG}$ [7,10-17]. Thus, the presence (absence) of the zero-field low-energy PG phenomena in the hole-type (electron-type) cuprate superconductors may be considered as the result of $V_{CO} > \Delta_{SC}$ ($V_{CO} < \Delta_{SC}$), as schematically illustrated in Fig. 1(b). A feasible physical cause for such differences will be discussed later.

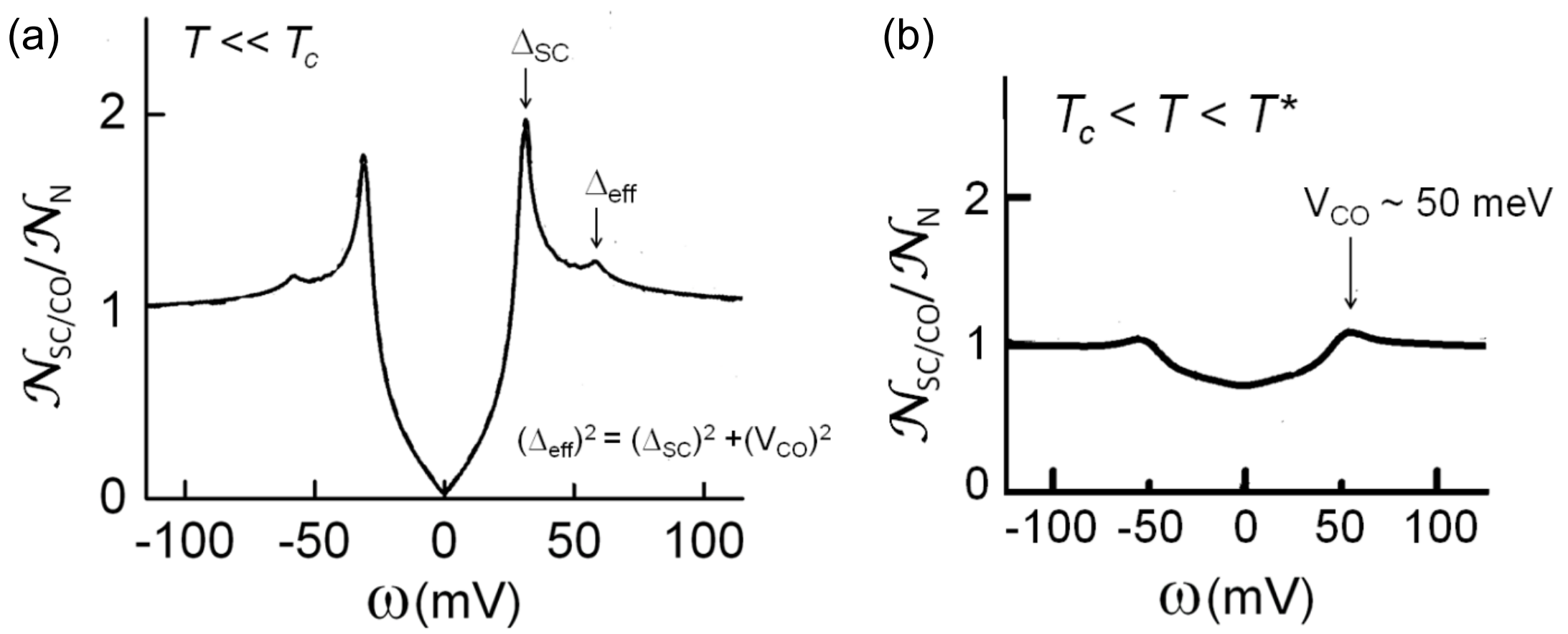

**Fig. 2:** Manifestations of the pseudogap (PG) phenomena in quasiparticle tunneling spectra based on the assumptions of coexisting superconductivity (SC) with a competing order (CO) in the ground state of a cuprate superconductor [7,10-13] and $V_{CO} > \Delta_{SC}$, where $V_{CO}$ ($\Delta_{SC}$) denotes the CO (SC) energy gap. **(a)** The quasiparticle density of states ($\mathcal{N}_{SC/CO}$) at $T << T_c$, as calculated from the CO scenario elaborated in Section 2.2.2, is normalized relative to the normal-state density of states ($\mathcal{N}_N$) as a function of the quasiparticle energy $\omega$. Here it is assumed $V_{CO} = 50$ meV and $\Delta_{SC} = 33$ meV, which correspond to the spectral parameters obtained from fitting the data of optimally doped Bi-2212 system with $T_c \sim 92$ K [12,13]. The calculated spectrum exhibits sharp coherence peaks at energy $\omega = \pm\Delta_{SC}$ and small "humps" at $\omega = \pm\Delta_{eff}$, where the effective gap energy $\Delta_{eff}$ is defined by $(\Delta_{eff})^2 \equiv (\Delta_{SC})^2 + (V_{CO})^2$. This finding is in general agreement with the tunneling spectra obtained from under- and optimally hole-doped cuprate superconductors [20]. **(b)** For temperature in the range $T_c < T << T^*$, a spectral PG feature remains at $\omega <\sim \pm V_{CO}$, which is consistent with the fact that $\Delta_{SC}$ vanishes above $T_c$ [12,13].

In addition to the contrasting low-energy charge excitations among the hole- and electron-type cuprates, the low-energy spin excitations also exhibit electron-hole asymmetry: Neutron scattering experiments on the hole-type $La_{2-x}Sr_xCuO_{4-y}$ reveal incommensurate spin correlations in the superconducting state, with a temperature independent spin gap observed both below and above $T_c$ [40–42]. One the other hand, the one-layer electron-type cuprate $Nd_{2-x}Ce_xCuO_{4-y}$ (NCCO) [43] displayed commensurate spin correlations and a temperature dependent spin gap in neutron scattering experiments [43]. Furthermore, the spin gap of NCCO was observed to reach a maximum as $T \rightarrow 0$ and disappeared as $T \rightarrow T_c$ [43]. Therefore, the spin gap in both hole- and electron-type cuprates may be related to the charge PG, as the spin gap and PG are both absent above $T_c$ in the electron-doped cuprates and present above $T_c$ in hole-doped cuprates. Moreover, enhancement of a static commensurate magnetic order by application of a magnetic field up to 9 T in electron-doped $Pr_{0.89}LaCe_{0.11}CuO_4$ (PLCCO) is observed [44], and a commensurate quasi-2D spin density wave (SDW) enhanced by the application of a magnetic field equal to 5 T in underdoped $Pr_{0.88}LaCe_{0.12}CuO_4$ is also reported [45].

In addition to the low-energy PG phenomenon that is correlated with the Fermi arcs and only found in hole-type cuprates slightly above $T_c$, there is a high-energy PG (denoted by $T_{PG}$ in Fig. 1(a)) that is present in both electron- and hole-type cuprates according to optical [46,47] and neutron scattering [48] experiments. As further elaborated in this review, the low-energy PG may be associated with the onset of competing orders (COs). In contrast, the high-energy PG appears to be related to the short-range magnetic exchange coupling in the cuprates [3].

Physically, the existence of various competing orders (COs) besides superconductivity (SC) in the ground state of the cuprates may be attributed to the complexity of the cuprates and the strong electronic correlation, which is in stark contrasts to conventional superconductors where SC is the sole ground state. The presence of COs has been manifested by a wide range of experiments, including x-ray and neutron scattering [40-45,48-55], muon spin resonance ($\mu$SR) [56], nuclear magnetic resonance (NMR) [57,58], optical conductivity [46,47] and Raman scattering [33,34,39], angle resolved photoemission spectroscopy (ARPES) [21,59,60] and scanning tunneling microscopy/spectroscopy (STM/STS) [7,10-17,24,61-64]. Moreover, theoretical evidences for COs have been provided by analytical modeling and numerical simulations [3-6,65-75]. The occurrence of a specific type of CO such as the spin density wave (SDW) [5,6,68-70], pair density wave (PDW) [71,72], $d$-density wave (DDW) [73,74] or charge density wave (CDW) [4,75] depends on the microscopic properties of a given cuprate, such as electron or hole-doping, the doping level ($\delta$), and the number of $CuO_2$ layers per unit cell ($n$) [7,10-17,76-78].

Although the relevance of competing orders to cuprate superconductivity remains unclear to date, the existence of COs has a number of important physical consequences. In addition to the aforementioned non-universal phenomena among different cuprates, quantum criticality naturally emerges naturally as the result of competing phases in the ground state [5-7,66,67]. Moreover, strong quantum fluctuations are expected due to the proximity to quantum criticality [7,14,15,66,77,78], and the low-energy excitations from the ground state become unconventional due to the redistributions of the spectral weight between SC and COs in the ground state [7,10-17]. Macroscopically, the presence of COs and strong quantum fluctuations naturally lead to weakened superconducting stiffness upon increasing $T$ and magnetic field $H$ [7,14,77-80], which contributes to the extreme type-II nature and the novel vortex dynamics of cuprate superconductors [81-97]. Additionally, a novel vortex liquid may exist in the presence high magnetic fields at low temperatures [77,78], giving rise to quantum oscillations in the "strange metallic state" of the cuprates when the applied magnetic fields ($H$) are still much smaller than the upper critical field $H_{c2}$ [98-102]. Indeed, theoretical analysis of the experimental data taken on underdoped hole-type cuprates $YB_2Cu_3O_{6+x}$ (Y-123) reveals that the commonly held assumption that the oscillation period is given by the underlying Fermi-surface area via the Onsager relation becomes invalid [103], prompting conjectures for reconstructed Fermi surfaces due to incommensurate SDW [102,104]. However, the observation of negative Hall effects [105] in the low-temperature high-field limit where quantum oscillations appear is similar to the commonly observed anomalous sign-reversal Hall conductivity of both hole- and electron-type cuprates in the flux flow limit [106-111]. These findings suggest that quantum fluctuations and COs may be both relevant to the appearance of quantum oscillations.

Given that the manifestation of unconventional low-energy excitations is one of the natural consequences of COs in the ground state of cuprates, investigation of the low-energy excitations can help reveal the characteristics of relevant COs. Here we summarize the best known unconventional phenomena to be discussed in this section: the appearance of satellite features [7,10-15,20] and energy-independent local density of states (LDOS) modulations in the quasiparticle spectra below $T_c$ [7,10,11,24,61-64]; the existence of a low-energy PG [23-25] and the appearance of the Fermi arcs for $T_c < T < T^*$ in hole-type cuprates [16,19,21]; "dichotomy" in the momentum dependence of quasiparticle coherence in hole-type cuprates [12,19,112,113]; PG-like vortex-core states in both electron- and hole-type cuprate superconductors [10,11,17,114]; strong quantum fluctuations found in all cuprates for $H^* < H << H_{c2}$ at $T \to 0$, where the crossover field $H^*$ is dependent on the doping level, the electronic anisotropy, and the number of $CuO_2$ layers per unit cell [14,15,77,78]; the occurrence of quantum oscillations in the vortex liquid phase of hole-type cuprate superconductors for magnetic fields applied perpendicular to the $CuO_2$ planes [98-102]; the anomalous sign-reversal of the Hall conductivity in the vortex liquid state [106-111]; and the anomalous Nernst effect appearing in the normal state of hole-type cuprate superconductors [115,116].

Various theoretical attempts have been made to explain the aforementioned anomalous behaviors amongst the cuprates, which may be largely categorized into two types of scenarios known as the "one-gap" [117] and "two-gap" models [7]. The former is associated with the "pre-formed pair" conjecture that asserts strong phase fluctuations in the cuprates so that formation of Cooper pairs may occur at a temperature well above the superconducting transition [117]. The latter considers coexistence of COs and SC with different energy scales in the ground state of the cuprates [7], as described in this overview. While the unconventional and asymmetric low-energy excitations amongst the hole- and electron-type cuprates have puzzled researchers and derailed the successful development of a microscopic theory that consistently accounts for all the differences found in the cuprates of varying doping levels, there appear to be a converging consensus recently based on the two-gap model to consistently account for most experimental phenomena [7,10,21,33,34,57,58,60,64]. Moreover, the occurrence of COs does not exclude the possibility of pre-formed Cooper pairs [25] above the superconducting transition, because there is no apparent reason for a CO that coexists with coherent Cooper pairs below $T_c$ to be incompatible with incoherent Cooper pairs above $T_c$ and below the PG temperature $T^*$. Nonetheless, several open issues remain. Further, whether the presence of COs is relevant or even devastating to the occurrence of cuprate superconductivity is still inconclusive.

In this section we survey the up-to-date status of experimental manifestations of unconventional low-energy excitations in the cuprates, explore the feasibility of a unified phenomenology for all cuprates based on the CO scenario, and then discuss the implications of the phenomenology in the context of microscopic pairing mechanism of cuprate superconductivity. The survey is divided into three subsections for the pairing state and normal state in zero fields, and for the vortex state in finite fields.

## 2.2. The Zero-Field Pairing State

As mentioned in the overview, the proximity of the cuprates to Mott insulators implies strong electronic correlation in the parent compound and in the small doping limit. To understand the formation of holes in the strongly correlated cuprates, an effective one-band *t-J* model was formulated by Zhang and Rice [118], in which the eigen-state of a single $CuO_4$ cluster is considered. That is, when a hole is introduced into the $CuO_4$ cluster, an orbital with $d_{x^2-y^2}$ symmetry among the four oxygen atoms can form, which will hybridize with the $3d^9$-state of the central copper site, whereas the $3d^8$-state of the copper site (which corresponds to a double occupancy of holes at the same site) is forbidden due to the large (up to ~ 8 eV) onsite Coulomb repulsion energy. More specifically, the lowest-energy hybrid configuration with Coulomb interaction included would involve a singlet combination of the two states: $[(1p^2 2p^6)_3(1p^2 2p_\sigma^5)(3d_{-\sigma}^9)]$ and $[(1p^2 2p^6)_4(3d^{10})]$, which is known as the "Zhang-Rice singlet". Here $\sigma$ denotes that spin state of an electron in the *p* or *d*-orbital. By considering the tunneling among these singlets in a half-filled $CuO_2$ plane, Zhang and Rice were able to derive an effective one-band *t-J* model [118]. Such singlet pairing for delocalized holes could reduce the strong onsite Coulomb repulsion, and would favor the unconventional $d_{x^2-y^2}$–wave pairing symmetry.

### *2.2.1. Unconventional pairing symmetry*

Historically, one of the most heated debates in the first decade of cuprate superconductivity research was the pairing symmetry [119-122]. From the symmetry point of view, the pairing channels of a singlet superconductor with the square-lattice symmetry must be consistent with even orbital quantum numbers (such as *s*, *d*, *g* ... for $\ell$ = 0, 2, 4 ... or their linear combinations). Given the quasi-two dimensional nature of most cuprates and the strong on-site Coulomb repulsion, it is feasible that the pairing symmetry is predominantly associated with the *d*-channel rather than the *s*-channel in order to minimize the Coulomb repulsion and to accommodate the quasi-two dimensional nature of the cuprates at the price of a higher kinetic energy.

The issue of the pairing symmetry in cuprate superconductors was eventually settled by means of phase sensitive measurements [119-122]. It was also realized later on that directional quasiparticle tunneling spectra by means of STS studies of the cuprate single crystals along different crystalline planes were also good experimental verifications of the pairing symmetry [123-127], because different pairing symmetries would result in distinctly different characteristics in the directional quasiparticle tunneling spectra due to the phase-changing pairing potential with varying quasiparticle momentum $\mathbf{k}$ [128-130], as schematically exemplified in Fig. 3 for the *s*-wave and $d_{x^2-y^2}$-wave pairing potentials. Indeed, overwhelming experimental evidences [119-127] have revealed signatures for predominantly $d_{x^2-y^2}$ pairing symmetry in all hole-type cuprate superconductors that are under-doped or optimally doped.

The principle for verifying the pairing symmetry of a superconductor with a pairing potential $\Delta_\mathbf{k}$ is based the generalized Blonder-Tinkham-Klapwijk (BTK) model [131] first derived by Hu, Tanaka and Kashiwaya [128-130]. Specifically, for an N-I-S junction (N: normal metal, I: insulator, S: superconductor) such as in the case of an STS experiment with a metallic tip, the tunneling conductance ($dI_{NS}/dV$) under a biased voltage $V$ and at $T$ = 0 is given by [128-130]:

$$\frac{dI_{NS}}{dV}(V) \propto \int d\theta \left[1 + |a(E,\theta)|^2 - |b(E,\theta)|^2\right] e^{-(\theta^2/\beta^2)}, \tag{1}$$

$$\sim \frac{1 + \sigma_N |\Gamma_+|^2 + (\sigma_N - 1)|\Gamma_+ \Gamma_-|^2}{\left|1 + (\sigma_N - 1)\Gamma_+ \Gamma_- e^{i(\phi_- - \phi_+)}\right|^2}, \tag{2}$$

where $|a(E,\theta)|^2$ and $|b(E,\theta)|^2$ in Eq. (1) refer to the Andreev reflection and normal reflection coefficients, respectively, $E = eV$ is the quasiparticle energy, $\theta$ is the incident angle of quasiparticles relative to the N-I-S junction, $\beta$ denotes the tunneling cone, $\sigma_N$ is the normal state conductance, and $\Gamma_{+,-}$ and $\phi_{+,-}$ in Eq. (2) are related to the electron-like (+) and hole-like (−) quasiparticle pairing potentials $\Delta_{+,-}$ as follows [128-130]:

$$\Gamma_{+,-} = \frac{E - \sqrt{E^2 - |\Delta_{+,-}|^2}}{|\Delta_{+,-}|}, \qquad \left(\Delta_{+,-} \equiv |\Delta_{+,-}| e^{i\phi_{+,-}}\right). \tag{3}$$

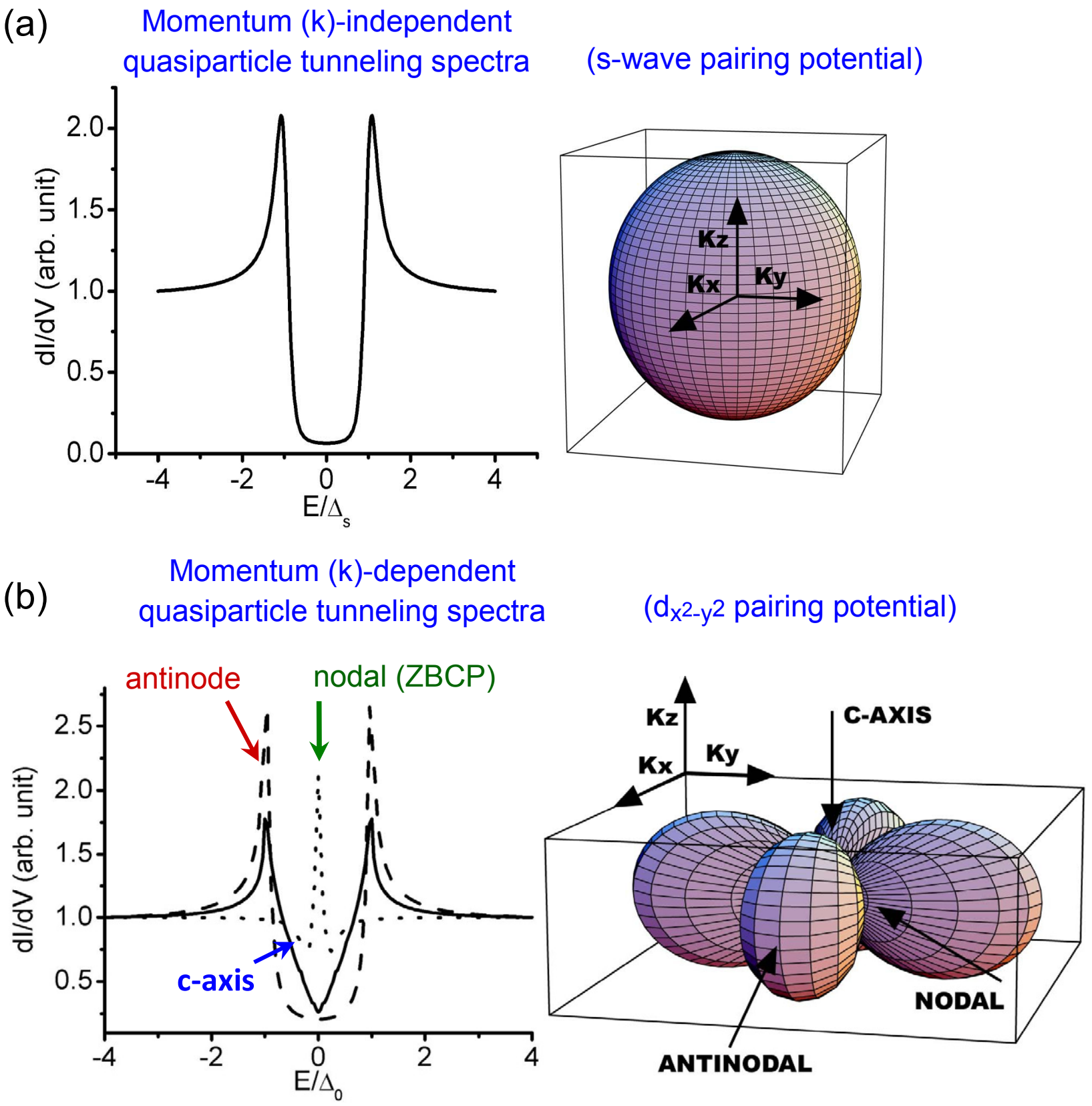

**Fig. 3:** Comparison of the quansiparticle tunneling spectra obtained from the *s*-wave and $d_{x^2-y^2}$-wave pairing symmetries [128-130,132]: **(a)** Momentum (**k**) independent quasiparticle tunneling conductance (*dI*/*dV*) vs. energy (*E*) normalized to the isotropic *s*-wave superconducting gap ($\Delta_s$). The right panel depicts the isotropic *s*-wave superconducting gap in the **k**-space. **(b)** Momentum (**k**)-dependent quasiparticle tunneling conductance spectra for a $d_{x^2-y^2}$-wave pairing superconductor, where the $d_{x^2-y^2}$-wave pairing potential is approximated by the relation $\Delta_{\mathbf{k}} = \Delta_0 \cos(2\theta_k)$, and $\theta_k$ denotes the incident angle of the quasiparticle momentum **k** relative to the anti-nodal direction of the $d_{x^2-y^2}$-wave pairing potential. The right panel depicts the anisotropic $d_{x^2-y^2}$-wave superconducting gap in the **k**-space. We note strong **k**-dependent spectral characteristics. In particular, for **k** along the nodal direction, a sharp peak known as the zero-bias conductance peak (ZBCP) [128-130] appears due to the presence of Andreev bound states. For more details, see Refs. [123-130].

Thus, given a pairing potential of a superconductor, the incident angle of the quasiparticles relative to the N-I-S junction and the tunneling cone for the incident quasiparticles, the resulting tunneling spectra can be derived from Eqs. (1) – (3), provided that SC is the only ground state and that the sample is relatively clean so that spectral broadening due to the finite quasiparticle lifetime may be neglected.

While the pairing symmetry of most cuprates is predominantly $d_{x^2-y^2}$, mixed pairing symmetries (such as $d_{x^2-y^2}+s$) have been widely reported in a number of cuprates, including in the tunneling junction studies [124,133,134], phase sensitive measurements [135,136], microwave spectra [137], optical spectra [138], and $\mu$SR penetration depth measurements [139,140]. In particular, the subdominant $s$-wave component appears to increase with increasing hole doping, as demonstrated by both STS and Raman spectroscopic studies [124,138]. As exemplified in Fig. 4(a) for the theoretical c-axis tunneling spectra associated with different pairing symmetries, and in Fig. 4(b) and Fig. 4(c) for representative experimental tunneling spectra of an optimally doped hole-type cuprate superconductor $YBa_2Cu_3O_{7-\delta}$ (with $T_c$ = 93 K) and an overdoped cuprate $(Y_{0.7}Ca_{0.3})Ba_2Cu_3O_{7-\delta}$ (with $T_c$ ~ 78 K), it is apparent that a subdominant $s$-wave component becomes non-negligible with increasing hole doping. This finding is consistent with the notion that the $d_{x^2-y^2}$-wave pairing is more favorable when onsite Coulomb repulsion is significant near the Mott insulator limit, whereas the $s$-wave pairing component may become energetically preferred in the overdoped limit when cuprate superconductors become more like conventional superconductors. We further note that the ($d_{x^2-y^2}+s$)-pairing leads to a reduction in the rotation symmetry from four- to two-fold symmetry while maintaining nodes as well as changing signs in the pairing potential as a function of the quasiparticle momentum.

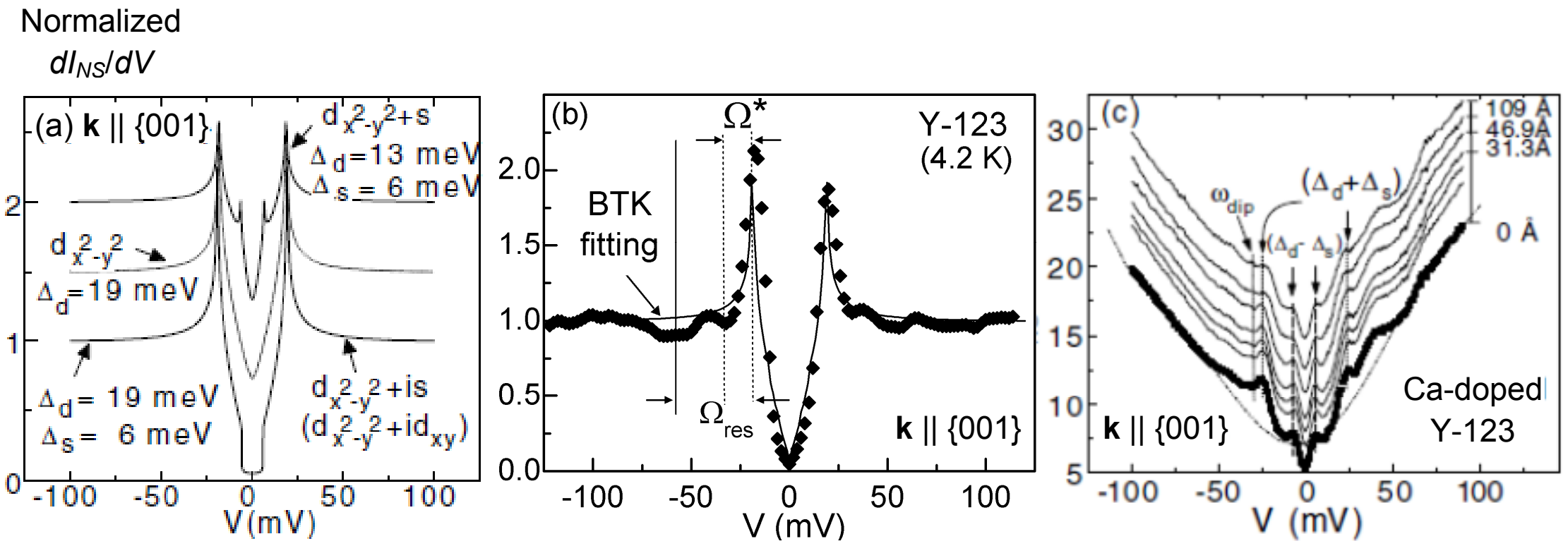

**Fig. 4:** Evidence for increasing subdominant $s$-wave component with increasing doping in hole-type cuprate superconductors with ($d_{x^2-y^2}+s$)-wave pairing potentials [124,125]: **(a)** Theoretically tunneling spectra for different pairing potentials of pure $d_{x^2-y^2}$, ($d_{x^2-y^2}+s$) and ($d_{x^2-y^2}+is$). **(b)** A representative c-axis (**k** || {001}) tunneling spectrum of an optimally doped cuprate superconductor $YBa_2Cu_3O_{7-\delta}$ (Y-123, with $T_c$ = 93 K), showing predominantly a $d_{x^2-y^2}$-wave pairing potential. **(c)** A series of c-axis (**k** || {001}) tunneling spectra of an overdoped cuprate $(Y_{0.7}Ca_{0.3})Ba_2Cu_3O_{7-\delta}$ (with $T_c$ ~ 78 K), showing spatially homogeneous tunneling spectra with tunneling characteristics consistent with the theoretical curve in (a) for a ($d_{x^2-y^2}+s$)-wave pairing potential. For more details, see Refs. [124,125].

An alternative approach to revealing the underlying pairing symmetry of cuprate superconductors is to investigate the spatially evolution of the quasiparticle low-energy excitations near quantum impurities. It is well known that magnetic quantum impurities can suppress conventional superconductivity effectively [141-146], whereas non-magnetic impurities in the dilute limit appear to inflict negligible effects on conventional superconductivity, as explained by the Anderson theory for dirty superconductors [147]. However, the findings of strong effects of spinless quantum impurities on the hole-type cuprate superconductors [124-126,148-159] and related theoretical studies suggest that the effects of quantum impurities depend on the pairing symmetry and the existence of magnetic correlation in cuprate superconductors [160-168]. For instance, fermionic nodal quasiparticles in the cuprates with either $d_{x^2-y^2}$ or ($d_{x^2-y^2}+s$) pairing symmetry can interact strongly with the quantum impurities in the $CuO_2$ planes and incur significant suppression of superconductivity regardless of the spin configuration of the impurity [160-164]. Moreover, the spatial evolution of the quasiparticle spectra near quantum impurities would differ significantly if a small component of complex order parameter existed in the cuprate. For instance, should the pairing symmetry contain a complex component such as ($d_{x^2-y^2}+id_{xy}$) that broke the time-reversal ($\mathcal{T}$) symmetry, the quasiparticle spectrum at a non-magnetic impurity site would have revealed two resonant scattering peaks at energies of equal magnitude but opposite signs in the electron-like and hole-like quasiparticle branches [24]. In contrast, for either $d_{x^2-y^2}$ or ($d_{x^2-y^2}+s$) pairing symmetry, only one resonant scattering peak at the impurity site is expected [160,162-164]. All empirical data to date [124-126,148-159] are consistent with the latter scenario.

In addition, the existence of nearest-neighbor AFM $Cu^{2+}$-$Cu^{2+}$ correlation in the superconducting state of the cuprates can result in an unusual Kondo-like behavior near a spinless impurity [165-167] due to an induced spin-1/2 ($S$ = 1/2) moment when one of the $Cu^{2+}$ ions is substituted with a spinless ion such as $Zn^{2+}$, $Mg^{2+}$, $Al^{3+}$ and $Li^{+}$ [124-126,148-159]. Indeed, the Kondo-like behavior associated with isolated spinless impurities in hole-type cuprates has been confirmed from the nuclear magnetic resonance (NMR) [148,149,157] and the inelastic neutron scattering (INS) experiments [154,155], and the spinless impurities are found to have more significant effects on broadening the NMR linewidth, damping the collective magnetic excitations and reducing the superfluid density than the magnetic impurities such as $Ni^{2+}$ with $S$ = 1 [124-126,148-159]. On the other hand, both types of impurities exhibit similar effects on suppressing superconducting transition temperature ($T_c$), increasing the microwave surface resistance in the superconducting state and increasing the normal state resistivity [124-126,148-159].

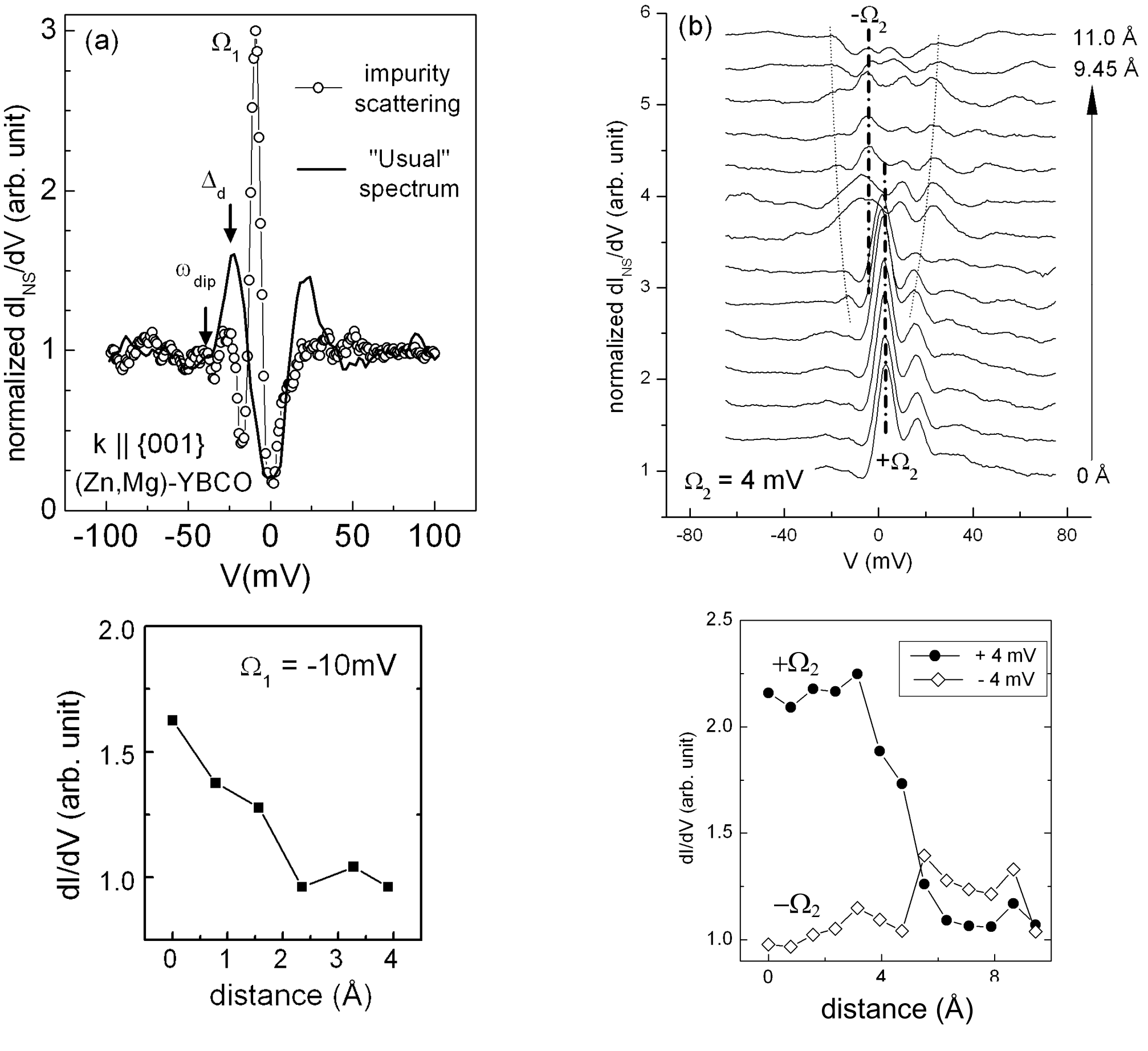

**Fig.5:** Normalized c-axis differential conductance (*dI/dV*) versus bias voltage (*V*) quasiparticle tunneling spectra of the (Zn,Mg)-Y123 single crystal near impurity sites at 4.2 K [124-126]. **(a)** Upper panel: A representative impurity scattering spectrum with a resonant peak at $\Omega_1 \sim -10$ meV and a typical spectrum away from impurities. Lower panel: Spatial variation of the impurity-induced resonant peak intensity. **(b)** Upper panel: Representative spectra revealing spatial variations in the quasiparticle spectra along the Cu-O bonding direction from an impurity with a maximum scattering at $\Omega_2 \sim +4$ meV. We note the spatially alternating resonant peak energies between + 4 meV and – 4 meV and the particle-hole asymmetry in the degrees of suppression of the SC coherence peaks. Lower panel: Spatial variation of the impurity-induced resonant peak intensity, showing alternating peak intensities at energies + 4 meV and – 4 meV with distance from an impurity.

The overall stronger suppression of superconductivity due to non-magnetic impurities in *d*-wave cuprates has been attributed to the slower spatial relaxation of spin polarization near the spinless impurities than that near the $S = 1$ impurities, the latter being partially screened by the surrounding antiferromagnetically coupled $Cu^{2+}$ spins [167,168]. The detailed spatial evolution of the quasiparticle tunneling spectra near these quantum impurities in the cuprates can further provide useful insights into the pairing state of the cuprates, and has been investigated in impurity-substituted $YBa_2Cu_3O_{7-\delta}$ and $Bi_2Sr_2CaCu_2O_{8+\delta}$ systems using the low-temperature scanning tunneling microscopy (STM) techniques [124-126,158,159]. As exemplified in Fig. 5 for an optimally doped $YBa_2Cu_3O_{7-\delta}$ (Y123) single crystal with 0.26% Zn and 0.4% Mg substituted into the Cu sites in the $CuO_2$ planes, hereafter denoted as (Zn,Mg)-Y123. The superconducting transition temperature of the sample is $T_c = 82.0$ K, which is substantially lower than that of the pure optimally doped Y123 with $T_c = 93.0$ K. For STM tip significantly far away from any impurities, the tunneling spectra were similar to the typical c-axis quasiparticle tunneling spectra in pure Y123, as shown in the upper panel of Fig. 5 (a). However, the global superconducting energy gap $\Delta_d$ was suppressed to $(25 \pm 2)$ meV from the value $\Delta_d = (29 \pm 1)$ meV in pure Y123 [124-126]. Moreover, the energy $\omega_{dip}$ associated with the "dip-hump" satellite features had also shifted substantially relative to that in pure Y123. We note that the dip-hump feature has been attributed to the effects of quasiparticle damping by the background many-body excitations such as spin fluctuations [169,170] or phonons [171], and the resonant energy of the many-body excitation may be empirically given by $|\Omega_{res}| = |\omega_{dip} - \Delta_d|$. The finding that $|\Omega_{res}|$ in the (Zn,Mg)-Y123 sample decreased significantly to $(7 \pm 1)$ meV from the value $|\Omega_{res}| = (17 \pm 1)$ meV in the pure Y123 with a very small impurity concentration in our (Zn,Mg)-Y123 clearly rules out phonons as the relevant many-body excitations associated with the satellite features. Therefore, one may associate the $\omega_{dip}$ satellite features with the inelastic scattering of quasiparticles by spin fluctuations. Further, the strong suppression of $\omega_{dip}$ by a small amount of non-magnetic impurities implies that the $Cu^{2+}$ spin fluctuations in the optimally doped cuprates are susceptible to the type of quantum impurities that interrupt the coherence of pair formation [124-126].

Interestingly, detailed studies on the surface of (Zn,Mg)-Y123 revealed apparent impurity scattering spectra that could be associated with two types of impurities, with maximum scattering intensity occurring at either $\Omega_1 \sim -10$ meV or $\Omega_2 \sim 4$ meV [124-126]. Further, regardless the impurity species, the intensity of each resonant peak decreased rapidly within approximately one Fermi wavelength along the Cu-O bonding direction, as exemplified in the lower panels of Fig. 5 (a) and 5 (b), while the coherence peaks associated with the SC gap were significantly suppressed near the quantum impurities, and the degree of suppression was asymmetric between the electron-like and hole-like branches. Moreover, as the STM tip was moved away from the impurity site, the resonant scattering peak appeared to alternate between energies of the same magnitude and opposite signs, as exemplified in the upper panel of Fig. 5 (b). Such spatial variations are expected for both Kondo-like and charge-like impurities [160-168]. Finally, the impurity effects on the variations in the quasiparticle spectra appeared to have completely diminished at approximately two coherence lengths (~ 3 nm) away from the impurity, as shown in lower panel of Fig. 5 (b).

Theoretically, the consideration for the effect of quantum impurities has been limited to a perturbative and one-band approximation without self-consistently solving for the spatially varying pairing potential in the presence of impurities [160-167]. Further, the existence of COs and the interaction among impurities have been neglected, which may be justifiable in the limit of dilute impurities and for overdoped cuprates. In this simplified approximation, the Hamiltonian $\mathcal{H}$ is approximated by $\mathcal{H} = \mathcal{H}_{BCS} + \mathcal{H}_{imp}$, where $\mathcal{H}_{BCS}$ is the $d_{x^2-y^2}$-wave BCS Hamiltonian that contains the normal (diagonal) one-band single-particle eigen-energy and anomalous (off-diagonal) $d_{x^2-y^2}$-wave pairing potential $\Delta_k$ ($= \Delta_d \cos 2\theta_k$, $\theta_k$ being the angle relative to the anti-node of the order parameter in the momentum space) of the unperturbed host, and

$$\mathcal{H}_{imp} = \mathcal{H}_{pot} + \mathcal{H}_{mag} = U \sum_\sigma c_{0\sigma}^\dagger c_{0\sigma} + \sum_{\mathbf{R}} J_{\mathbf{R}} \boldsymbol{S} \bullet \boldsymbol{\sigma}_{\mathbf{R}} \quad (4)$$

denotes the impurity perturbation due to both the localized potential scattering term $\mathcal{H}_{pot}$ ($= U \sum_\sigma c_{0\sigma}^\dagger c_{0\sigma}$; $U$: the on-site Coulomb scattering potential) and the Kondo-like magnetic exchange interaction term $\mathcal{H}_{mag}$ ($= \sum_{\mathbf{R}} J_{\mathbf{R}} \boldsymbol{S} \bullet \boldsymbol{\sigma}_{\mathbf{R}}$) between the spins of the conduction carriers on the **R** sites ($\boldsymbol{\sigma}_{\mathbf{R}}$) and those of the localized magnetic moments ($S$), with $J_{\mathbf{R}}$ being the exchange coupling constant. Assuming the aforementioned model Hamiltonian $\mathcal{H}$, one can obtain the quasiparticle spectra due to impurities from the Green function derived from $\mathcal{H}$. If the effects of the tunneling matrix are further neglected for simplicity, one obtains in the pure potential scattering limit (where $\mathcal{H}_{imp} = \mathcal{H}_{pot}$) a resonant energy at $\Omega$ on the impurity site that satisfies the following relation [160,161]:

$$|\Omega/\Delta_d| \approx \{(\pi/2)\cot\delta_0 \,/\, \ln[8/(\pi \cot\delta_0)]\}, \quad (5)$$

where $\delta_0$ is the impurity-induced phase shift in the quasiparticle wavefunction at a long distance. Generally $\delta_0 \to (\pi/2)$ is true in the strong potential scattering (unitary) limit. On the other hand, in the case of magnetic impurities with both contributions from $\mathcal{H}_{pot}$ and $\mathcal{H}_{mag}$, one expects two spin-polarized impurity states at energies $\Omega_{\pm}$, which are given by [163]:

$$|\Omega_{\pm}/\Delta_d| = 1/[2\mathcal{N}_F(U \pm W)\ \ln|8\mathcal{N}_F(U \pm W)|], \tag{6}$$

where $\mathcal{N}_F$ is the density of states at the Fermi level, $W \equiv J(\mathbf{S}\bullet\boldsymbol{\sigma})$ implies that magnetic impurities are isolated and equivalent at all sites, and the two energies $\Omega_+$ and $\Omega_-$ are associated with the upper and lower signs in Eq. (6), respectively.

Comparing the STS studies of (Zn,Mg)-Y123 with those of a 0.6% Zn-substituted $Bi_2Sr_2CaCu_2O_{8-x}$ (Bi-2212) [158], it is found that the primary features such as the appearance of single resonant scattering peak and strong suppression of the superconducting coherence peaks at the impurity site, as well as the rapidly decreasing intensity of the resonant peak with the displacement from the impurity site, are generally comparable in both systems. These findings are consistent with the simplified theoretical model outlined above in the unitary limit, and further imply the preservation of time-reversal ($\mathcal{T}$) symmetry in both systems, suggesting the absence of any discernible complex order parameter in the pairing symmetry. The agreement of experimental findings with the simplified model Hamiltonian may be understood in terms of the presence of nodal quasiparticles in hole-type cuprates. As elaborated later in Section 2.2.2, the relevant competing orders in the hole-type cuprates are primarily associated with the CDW or PDW orders with a wave-vector parallel to the Cu-O bonding directions, and therefore the effective energy gap of the hole-type cuprates in the pairing state always vanishes at $\mathbf{k} = (\pm\pi,\pm\pi)$. Thus, nodal quasiparticles most responsible for the quantum impurity-induced low-energy excitations are always presence in the hole-type cuprates, leading to experimental observation qualitatively consistent with the model Hamiltonian. In fact, the model Hamiltonian may be generalized to including the existence of competing orders by replacing the superconducting $d$-wave gap $\Delta_d$ in Eqs. (2) and (3) by the effective gap $\Delta_{eff} = [(\Delta_d)^2+(V_{CO})^2]^{1/2}$, where $V_{CO}$ denotes the energy gap associated with a given competing order.

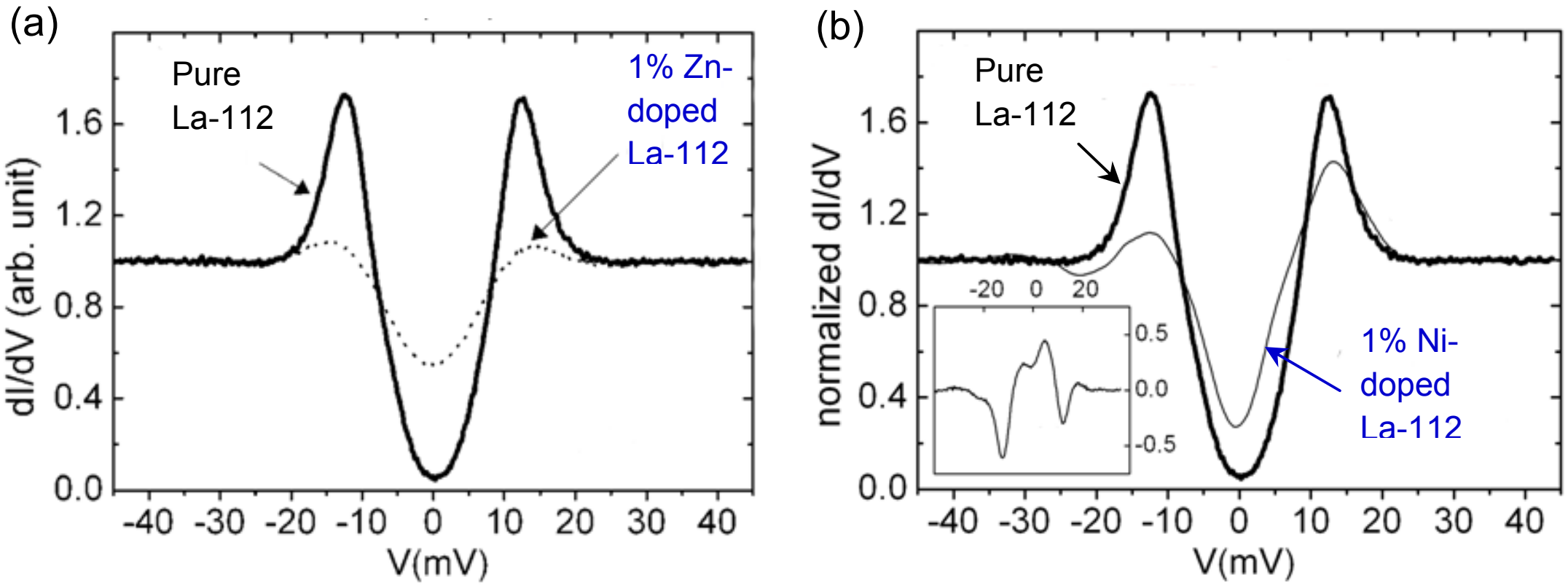

**Fig. 6:** Comparison of the quasiparticle tunneling spectrum (thick solid line) of the pure electron-type optimally doped cuprate $Sr_{0.9}La_{0.1}CuO_2$ (La-112) with those of the quantum impurity-substituted La-112 at $T$ = 4.2 K [35]: **(a)** A representative tunneling spectrum of 1% Zn-substituted La-112, $Sr_{0.9}La_{0.1}(Cu_{0.99}Zn_{0.01})O_2$ shown by the dashed line reveals symmetrically suppressed superconducting coherence peaks and a substantially increased sub-gap spectral weight, suggesting strong enhancement of low-energy quasiparticle excitations induced by the Zn impurities. **(b)** A representative tunneling spectrum of 1% Ni-substituted La-112, $Sr_{0.9}La_{0.1}(Cu_{0.99}Ni_{0.01})O_2$ (shown in the main panel by the thin line), reveals asymmetrically suppressed superconducting coherence peaks and moderately increased sub-gap spectral weight. This finding suggests weaker effects induced by the Ni impurities. The inset shows the spectral difference due to Ni-impurities. The spectral difference for impurity bound states is found to extend over a long range [35], similar to the Shiba impurity bands [142].

On the other hand, several spectral differences are noteworthy between the (Zn,Mg)-Y123 and the Zn-substituted Bi-2212. First, the strength of impurity scattering appears weaker and longer-ranged in Y123. Second, various phenomena that are more consistent with the Kondo effect, such as alternating resonant peak energies between $+\Omega$ and $-\Omega$ with the distance from a non-magnetic impurity and temporal variations of the resonant peak, have only been observed in Y123. Third, global suppression of the SC energy gap and of the collective magnetic excitation energy has only been revealed in Y123. Such a difference may be attributed to the fact that pure Y123 generally exhibits long-range spectral homogeneity [124,125,172], whereas nano-scale spectral variations commonly observed in nominally pure Bi-2212 samples [173,174] (probably due to the inherent non-stoichiometric nature of Bi-2212) yield difficulties in identifying the global effect of impurity substitutions.

While the hole-type cuprates are shown to be strongly affected by both magnetic and non-magnetic quantum impurities in the $CuO_2$ planes, in the case of an electron-type cuprate superconductor, the infinite-layer system $Sr_{0.9}La_{0.1}CuO_2$ (La-112), it is found to be insensitive to a small concentration of non-magnetic impurities so that the bulk magnetization studies [175] revealed no suppression in the bulk $T_c$ up to 3% substitutions of Zn into the Cu sites, beyond which the compound became inhomogeneous and phase segregated. On the other hand, the substitution of magnetic Ni-impurities into the Cu sites of La-112 yielded significant $T_c$ suppression: With 1% Ni, $T_c$ already decreased from 43 K to 32 K; 2% Ni dropped $T_c$ to below 4 K, and 3% Ni completely suppressed the bulk superconductivity although the sample was still stoichiometrically homogeneous from X-ray diffraction. The quasiparticle spectra near quantum impurities in La-112 [35] also differ significantly from those of hole-type cuprates [124-126,148-159]. Specifically, the substitutions of either Zn or Ni do not result in strong resonant peaks near the impurities [35]. Rather, the quasiparticle spectra only exhibit modifications to the height of the coherence peaks and an increase in the spectral weight within the superconducting gap, with symmetric suppression of both coherence peaks in the case of Zn impurities (see Fig. 6(a)) [35] and asymmetric suppression of the particle-like and hole-like coherence peaks in the case of Ni impurities (see Fig. 6(b)) [35]. Moreover, substantially increased sub-gap spectral weight is found in the case of Zn impurities, suggesting strong enhancement of low-energy quasiparticle excitations induced by the Zn substitution. In contrast, the spectral difference due to Ni impurities reveals long-range impurity bound states [35] that are similar to the Shiba impurity bands for magnetic impurities in fully gapped conventional superconductors [142]. The assumption of impurity bands may be justified by noting that the average Ni-Ni separation for 1% Ni substitutions (~ 1.8 nm) is shorter than the in-plane superconducting coherence lengths of the pure La-112 sample (~ 4.8 nm) [176] so that substantial overlap of the impurity wave functions may be expected.

The puzzling spectral response of the electron-type La-112 to Zn and Ni impurities was initially considered as a supporting evidence for *s*-wave pairing symmetry in the La-112 system [35]. However, further investigations of the electron-type cuprates [7,10,16,17] (see Section 2.2.2 for more details) suggest that the coexistence of a SDW competing order with a $(\pi,\pi)$ wave-vector and an energy gap small than the SC gap $(\Delta_{SDW} < \Delta_{SC})$ could qualitatively account for the experimental observation. That is, the effective gap of the electron-type cuprate superconductors is fully gapped and anisotropic for all **k**-values even though the $d_{x^2-y^2}$-wave SC gap $\Delta_{SC}$ vanishes at $(\pm\pi/4,\pm\pi/4)$ [7,10,16], which is consistent with the ARPES finding [177]. Hence, there are effectively no nodal quasiparticles interacting with quantum impurities so that no sharp spectral resonances could be found at quantum impurities in the electron-type cuprate superconductors. In this context, the characteristics of the low-energy excitations in electron-type cuprates due to either non-magnetic or magnetic impurities are analogous to those of fully gapped, anisotropic conventional superconductors. However, to date there has not been theoretical investigation of the spectral effects of quantum impurities in electron-type cuprates so that no quantitative comparison can be made with experimental observation.

### *2.2.2. Unconventional low-energy excitations in zero fields*

As mentioned in the introduction, one of the most widely debated issues in cuprate superconductivity is the possibility of preformed Cooper pairs and the origin of the pseudogap (PG) phenomenon. In this section the zero-field spectral characteristics of various types of hole- and electron-type cuprates are examined, with special emphasis on the analysis of the quasiparticle local density of state (LDOS) spectra taken by means of scanning tunneling microscopy/spectroscopy (STM/STS), and the momentum dependent quasparticle spectra taken by means of the angle-resolved photoemission spectroscopy (ARPES). We shall show below that many spectral details as a function of the energy, momentum and temperature cannot be explained by considering Bogoliubov quasiparticles as the sole low-energy excitations in the cuprate superconductors. In contrast, the incorporation of competing orders (COs) coexisting with superconductivity (SC) can provide consistent descriptions for all data.

Our theoretical analysis begins with a mean-field Hamiltonian $\mathcal{H}_{MF} = \mathcal{H}_{SC} + \mathcal{H}_{CO}$ that consists of coexisting SC and a CO at $T = 0$ [10-17]. We further assume that the SC gap $\Delta_{SC}$ vanishes at $T_c$ and the CO order parameter vanishes at $T^*$, and that both $T_c$ and $T^*$ are second-order phase transitions. The SC Hamiltonian is given by:

$$\mathcal{H}_{\rm SC} = \sum_{\mathbf{k},\alpha} \xi_{\mathbf{k}} c^{\dagger}_{\mathbf{k},\alpha} c_{\mathbf{k},\alpha} - \sum_{\mathbf{k}} \Delta_{\rm SC}(\mathbf{k}) \left( c^{\dagger}_{\mathbf{k},\uparrow} c^{\dagger}_{-\mathbf{k},\downarrow} + c_{-\mathbf{k},\downarrow} c_{\mathbf{k},\uparrow} \right), \tag{7}$$

where $\Delta_{\rm SC}(\mathbf{k}) = \Delta_{\rm SC}(\cos k_x - \cos k_y)/2$ for $d_{x^2-y^2}$-wave pairing, **k** denotes the quasiparticle momentum, $\xi_{\mathbf{k}}$ is the normal-state eigen-energy relative to the Fermi energy, $c^{\dagger}$ and $c$ are the creation and annihilation operators, and $\alpha = \uparrow, \downarrow$ refers to the spin states. The CO Hamiltonian is specified by the energy $V_{\rm CO}$, a wave-vector **Q**, and a momentum distribution δ**Q** that depends on a form factor, the correlation length of the CO, and also on the degree of disorder [10-17]. We have previously considered the effect of various types of COs on the quasiparticle spectral density function $A(\mathbf{k},\omega)$ and the density of states $\mathcal{N}(\omega)$. For instance, in the case that charge density waves (CDW) is the relevant CO, we have a wave-vector $\mathbf{Q}_1$ parallel to the $CuO_2$ bonding direction (π,0) or (0,π) in the CO Hamiltonian [10-17]:

$$\mathcal{H}_{\rm CDW} = -\sum_{\mathbf{k},\alpha} V_{\rm CDW}(\mathbf{k}) \left( c^{\dagger}_{\mathbf{k},\alpha} c_{\mathbf{k}+\mathbf{Q}_1,\alpha} + c^{\dagger}_{\mathbf{k}+\mathbf{Q}_1,\alpha} c_{\mathbf{k},\alpha} \right). \tag{8}$$

On the other hand, for commensurate SDW being the relevant CO, the SDW wave-vector becomes $\mathbf{Q}_2 = (\pi,\pi)$, and the corresponding CO Hamiltonian is [178]:

$$\mathcal{H}_{\rm SDW} = -\sum_{\mathbf{k},\alpha,\beta} V_{\rm SDW}(\mathbf{k}) \left( c^{\dagger}_{\mathbf{k}+\mathbf{Q}_2,\alpha} \sigma^3_{\alpha\beta} c_{\mathbf{k},\beta} + c^{\dagger}_{\mathbf{k},\alpha} \sigma^3_{\alpha\beta} c_{\mathbf{k}+\mathbf{Q}_2,\beta} \right), \tag{9}$$

where $\sigma^3_{\alpha\beta}$ denotes the matrix element $\alpha\beta$ of the Paul matrix $\sigma^3$. Similarly, other types of COs such as the disorder-pinned SDW [179] and the *d*-density wave (DDW) [73] may be considered by using the following CO Hamiltonians [7,10]:

$$\mathcal{H}^{\rm pinned}_{\rm SDW} = -g^2 \sum_{\mathbf{k},\alpha} V_{\rm SDW}(\mathbf{k}) \left( c^{\dagger}_{\mathbf{k},\alpha} c_{\mathbf{k}+\mathbf{Q}_3,\alpha} + c^{\dagger}_{\mathbf{k}+\mathbf{Q}_3,\alpha} c_{\mathbf{k},\alpha} \right). \tag{10}$$

$$\mathcal{H}_{\rm DDW} = -\sum_{\mathbf{k},\alpha} \frac{1}{2} V_{\rm DDW} \left( \cos k_x - \cos k_y \right) \left( i c^{\dagger}_{\mathbf{k}+\mathbf{Q}_2,\alpha} c_{\mathbf{k},\alpha} - i c^{\dagger}_{\mathbf{k},\alpha} c_{\mathbf{k}+\mathbf{Q}_2,\alpha} \right), \tag{11}$$

where $\mathbf{Q}_3 = \mathbf{Q}_1/2$ for disorder pinned SDW [178], and $g$ denotes the coupling strength between disorder and SDW. Thus, by incorporating realistic bandstructures and Fermi energies for different families of cuprates with given doping and by specifying the SC pairing symmetry and the form factor for the CO, $\mathcal{H}_{\rm MF}$ can be diagonalized to obtain the bare Green function $G_0(\mathbf{k},\omega)$ for momentum **k** and energy $\omega$. Further, quantum phase fluctuations between the CO and SC may be included by solving the Dyson's equation self-consistently for the full Green function $G(\mathbf{k},\omega)$ [10-17], which gives the quasiparticle spectral density function $A(\mathbf{k},\omega) = -\,{\rm Im}\,[G(\mathbf{k},\omega)]/\pi$ for comparison with ARPES [16] and the quasiparticle density of states $\mathcal{N}(\omega) = \sum_{\mathbf{k}} A(\mathbf{k},\omega)$ for comparison with STM spectroscopy [10-17].

Based on the Green function analysis outlined above for coexisting $d_{x^2-y^2}$-wave SC and a specific CO, the zero-field quasiparticle spectra $\mathcal{N}(\omega)$ and $A(\mathbf{k},\omega)$ at $T = 0$ can be fully determined by the parameters $\Delta_{\rm SC}$, $V_{\rm CO}$, **Q**, δ**Q**, $\Gamma_{\mathbf{k}}$ (the quasiparticle linewidth), and $\eta$ (the magnitude of quantum phase fluctuations), which is proportional to the mean-value of the velocity-velocity correlation function [12,13]. For finite temperatures, the temperature Green function is employed to account for the thermal distributions of quasiparticles.

Using the aforementioned theoretical analysis we have been able to consistently account for the $T$-dependent quasiparticle tunneling spectra in both hole- and electron-type cuprates if we assume Fermi-surface nested CDW as the CO in the hole-type cuprates such as Y-123 and Bi-2212, and commensurate SDW as the CO in the electron-type La-112 and PCCO [10-17], which are consistent with findings from neutron scattering experiments [180,181]. On the other hand, it is found that quasiparticle spectra obtained from considering the DDW scenario generally do not agree with experimental observation.

Specifically, for hole-type cuprates such as in the spectra of Y-123 and Bi-2212, the sharp peaks and satellite "hump" features at $T \ll T_c$ in Fig. 7(a) and also in Fig. 2(a) can be associated with $\omega = \pm\Delta_{\rm SC}$ and $\omega = \pm\Delta_{\rm eff}$, respectively, where $\Delta_{\rm eff} \equiv [(\Delta_{\rm SC})^2 + (V_{\rm CO})^2]^{1/2}$ is an effective excitation gap. Hence, the condition $V_{\rm CO} > \Delta_{\rm SC}$ in hole-type cuprates is responsible for the appearance of the satellite features at $T \ll T_c$ and the PG phenomena at $T^* > T > T_c$ [10-16]. In contrast, the condition $V_{CO} < \Delta_{SC}$ in electron-type cuprates, as exemplified in Fig. 7(b), is responsible for only one set of characteristic features at $\pm\Delta_{eff}$ and the absence of PG above $T_c$ [17].

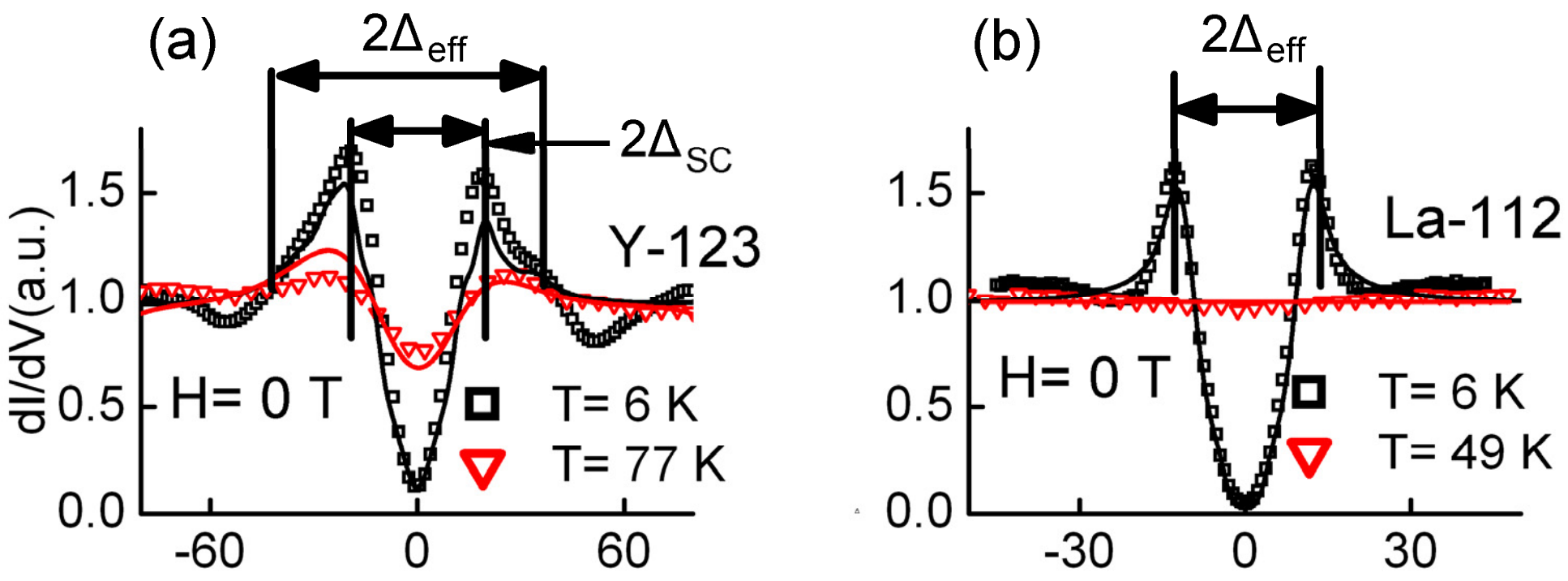

**Fig. 7:** Implication of CO from zero- and finite-field STS in Y-123 and La-112 [10-17]: **(a)** Normalized zero-field tunneling spectra of Y-123 taken at $T = 6$ K (black) and 77 K (red). The solid lines represent fittings to the $T = 6$ and 77 K spectra by assuming coexisting SC and CDW, with fitting parameters of $\Delta_{SC} = 20$ meV, $V_{CDW} = 32$ meV and $Q_{CDW} = (0.25\pi \pm 0.05\pi, 0) / (0, 0.25\pi \pm 0.05\pi)$, following Refs. [10-16]. **(b)** Normalized zero-field tunneling spectra of La-112 taken at $T = 6$ K (black) and 49 K (red). The solid lines represent fittings to the $T = 6$ and 49 K spectra by assuming coexisting SC and SDW, with fitting parameters $\Delta_{SC} = 12$ meV, $V_{SDW} = 8$ meV, and $Q_{SDW} = (\pm\pi, \pm\pi)$, following Refs. [10-17]. We further note that the CO energies are consistent with those obtained from neutron scattering experiments [180,181].

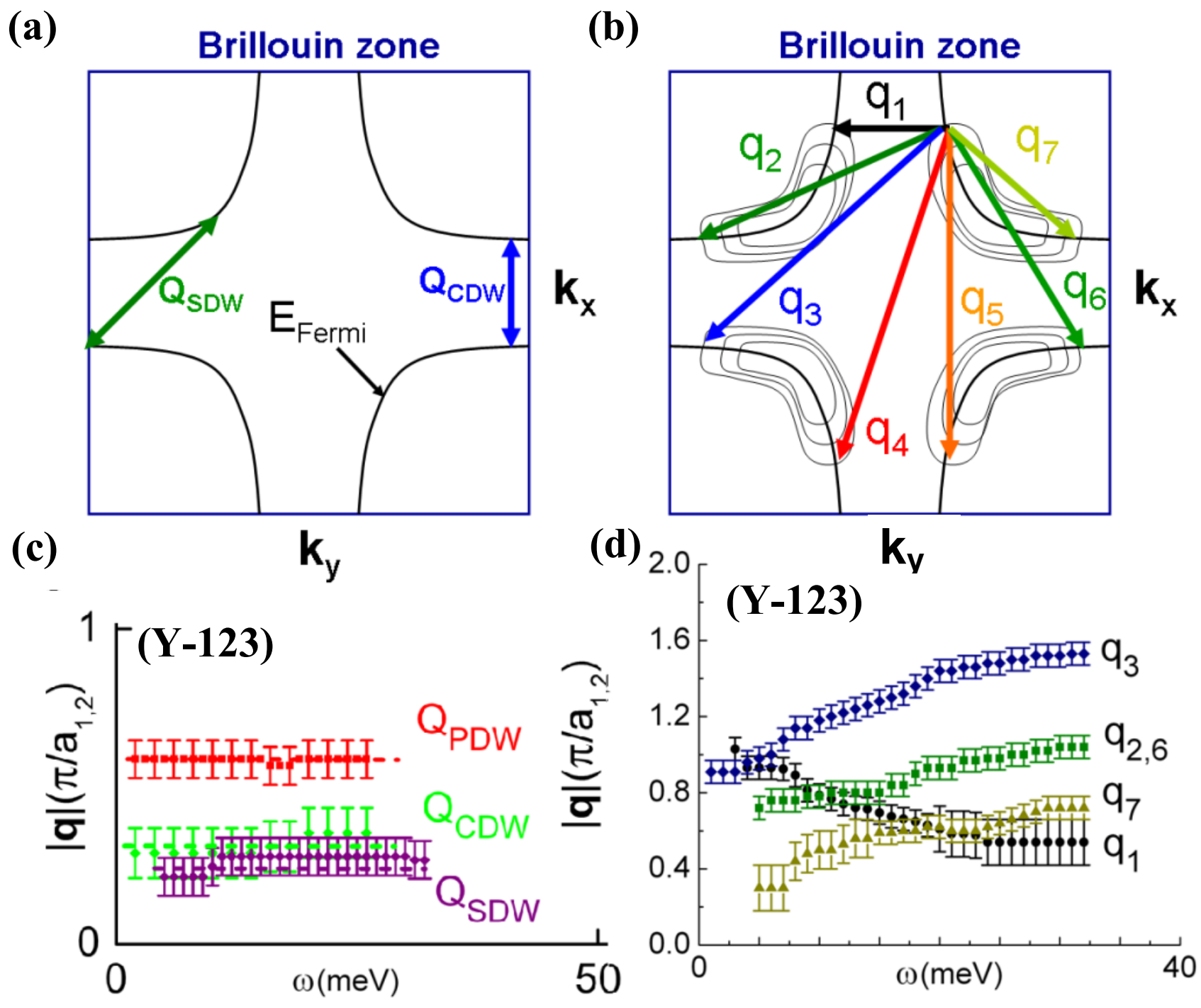

**Fig. 8:** Scattering wave-vectors obtained from the FT-LDOS of quasiparticle tunneling spectra [10,11]: **(a)** Illustration of the wave-vectors associated with SDW and CDW excitations. **(b)** Illustration of the wave-vectors associated with elastic quasiparticle interferences (QPI) between pairs of points on equal energy contours with maximum joint density of states. **(c)** The nearly energy-independent collective modes $|\mathbf{Q}_{PDW}|$, $|\mathbf{Q}_{CDW}|$ and $|\mathbf{Q}_{SDW}|$ obtained from the FT-LDOS data of Y-123 [10,11]. **(d)** The QPI momentum ($|\mathbf{q}_i|$) versus energy ($\omega$) dispersion relations derived from the FT-LDOS data of Y-123 [10,11], which are found to be in excellent agreement with the QPI results obtained from the optimally doped Bi-2212 [112].

By extending the above analysis to the tunneling spectra of different doping levels ($\delta$) associated with the hole-type cuprates, it is found that $\Delta_{SC}(\delta)$ generally follows the same non-monotonic dependence of $T_c(\delta)$ [7,13]. In contrast, $V_{CO}(\delta)$ increases with decreasing $\delta$, which is consistent with the general trend of the zero-field PG temperature in hole-type cuprates [7,13]. Moreover, analysis of the FT-LDOS of hole-type cuprates of Y-123 [10,11] and $Bi_2Sr_2CuO_{6+x}$ (Bi-2201) [64] reveals energy-*independent* scattering wave-vectors that are in stark contrast to the strongly energy-dependent wave-vectors due to quasiparticle interferences (QPI) from elastic impurity scattering of Bogoliubov quasiparticles, as exemplified in Fig. 8(b) for the optimally doped Y-123 single crystal [10,11]. Additionally, the energy-independent scattering wave-vector found in Bi-2201 reveals a strong doping dependence [64], which is consistent with a CDW nesting wave-vector on the Fermi surface so that the wave-vector decreases with increasing hole doping, as schematically sketched in Fig. 8(a). Finally, in the optimally doped Bi-2212 [61,62] and under-doped $Ca_{2-x}Na_xCuO_2Cl_2$ [63], the observed strong FT-LDOS intensity associated with an energy-independent wave-vector along the $(\pm\pi,0)/(0,\pm\pi)$ directions at $T << T_c$ and the remaining finite intensity of these four spots for $T_c < T < T^*$ [23] are all in contradiction to the "one-gap" scenario while naturally accounted for if the energy-independent wave-vector is attributed to the CDW nesting wave-vector that coexist with SC at $T << T_c$ [61-63] and still remains for $T_c < T < T^*$ [23]. These viewpoints have been discussed in details by quantitative analysis of the experimental data [7,10-16].

### *2.2.3. Dichotomy of quasiparticle coherence*

Another empirical finding of unconventional low-energy excitations associated with the cuprates is the dichotomy of quasiparticle coherence revealed by spectroscopic studies of the pairing state of hole-type cuprates [19,112,113,182]. This finding can be naturally explained by considering the broadening of the spectral density function $A(\mathbf{k},\omega)$ by the increasing magnitude of quantum phase fluctuations $\eta$ [12,13], and both the quasiparticle coherence, as manifested by the inverse linewidth ($\Gamma^{-1}$) of $A(\mathbf{k},\omega)$, and the renormalized effective gap $\Delta_{eff}(\mathbf{k})$ exhibit "dichotomy" in the momentum space [12,13], showing different evolution in the Cu-O bonding direction $(0,\pi)/(\pi,0)$ from that in the $(\pi,\pi)$ nodal direction, as confirmed by recent ARPES results [113]. Specifically, dichotomy in the quasiparticle coherence can be manifested by comparing the linewidth of fluctuation-renormalized $A(\mathbf{k},\omega \sim \Delta_{eff})$ for $\mathbf{k}$ along $(\pi,\pi)$ with that for $\mathbf{k}$ along $(0,\pi)/(\pi,0)$, as exemplified in Figs. 9(a) and 9(b) for coexisting $d_{x^2-y^2}$-wave SC and disorder-pinned SDW or CDW [12]. The degree of dichotomy in the quasiparticle coherence decreases with increasing quantum fluctuations, as shown in the inverse linewidth $\Gamma^{-1}$-vs.-$\eta$ plots in Figs. 9(a) and 9(b). In particular, we note that quasiparticles of $d_{x^2-y^2}$-wave SC with disorder-pinned SDW or CDW exhibit better coherence along $(\pi,\pi)$ than along $(0,\pi)/(\pi,0)$ for small $\eta$, which is consistent with the ARPES and STS data [112,113,182]. The CO-induced dichotomy in the momentum-dependent effective gap $\Delta_{eff}(\mathbf{k})$ is illustrated in the first quadrant of the first Brillouin zone (BZ) in Fig. 9(c) [12,13].

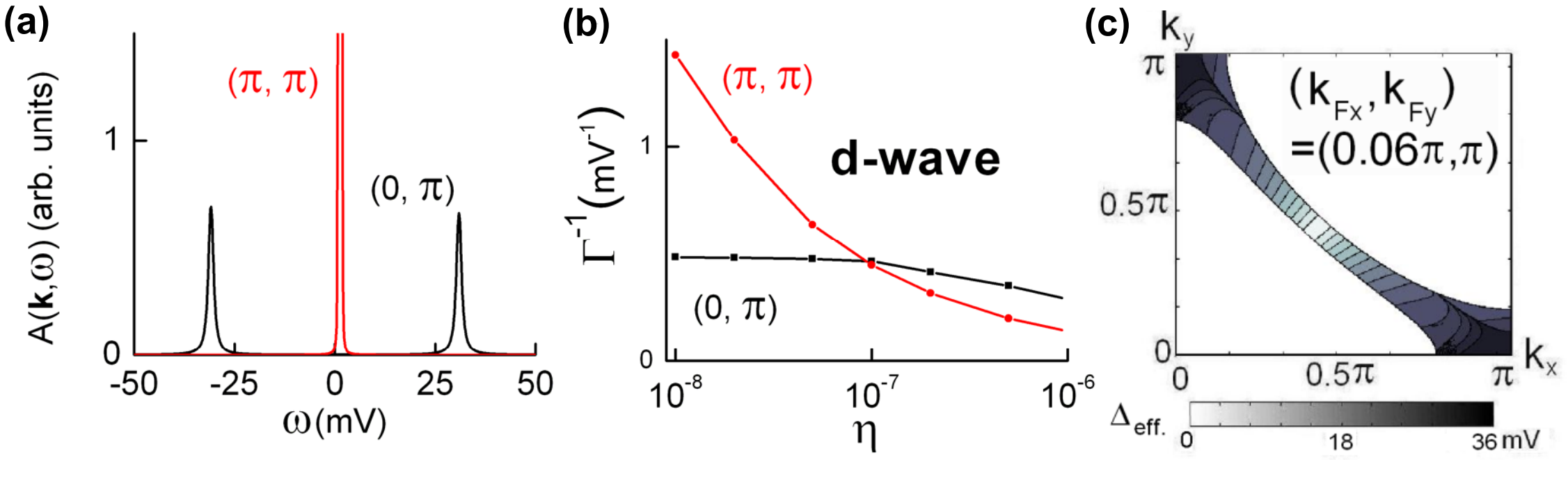

**Fig. 9:** Dichotomy in the spectral density function and excitation gap due to coexisting SC/CO and quantum phase fluctuations [12,13]: **(a)** Contrasts in the fluctuation renormalized $A(\mathbf{k},\omega)$ at the Fermi level for $\mathbf{k}$ along $(\pi,0)/(0,\pi)$ [darker (black) lines] and along $(\pi,\pi)$ [lighter (red) lines] for coexisting $d_{x^2-y^2}$-wave SC and disorder-pinned SDW with $\eta = 3 \times 10^{-8}$, showing a narrower linewidth in $A(\mathbf{k},\omega)$ along $(\pi,\pi)$. **(b)** Dichotomy in the quasiparticle lifetime ($\propto \Gamma^{-1}$), showing better quasiparticle coherence along $(\pi,\pi)$ than along $(\pi,0)/(0,\pi)$ if the quantum fluctuations are sufficiently small, as manifested by the $\Gamma^{-1}$ vs. $\eta$ plot. The latter finding is consistent with the empirical observation of more coherent nodal quasiparticles in hole-type cuprate superconductors [112,113,182]. **(c)** Competing order-induced dichotomy in the momentum-dependent effective gap $\Delta_{eff}(\mathbf{k})$ is illustrated in the first quadrant of the first Brillouin zone (BZ).

### 2.3. Normal State

#### *2.3.1. Pseudogap phenomena*

The low-energy PG phenomena described in the overview are most notably observed in $Bi_2Sr_2CaCu_2O_x$ (Bi-2212) and $Bi_2Sr_2CuO_x$ (Bi-2201) as a function of hole doping [18,19,21-25]. The persistence of gapped quasiparticle spectral density functions near the (π,0) and (0,π) portions of the Brillouin zone above $T_c$ in hole-type cuprates are the source of the incomplete recovery of the Fermi surface [16,19,21,60]. In contrast, electron-type cuprates exhibit neither the low-energy PG nor the Fermi arc above $T_c$ [35-37], although "hidden pseudogap" features in the quasiparticle excitation spectra have been observed under the superconducting dome in doping dependent grain-boundary tunneling experiments on $Pr_{2-x}Ce_xCuO_{4-y}$ (PCCO) and $La_{2-x}Ce_xCuO_{4-y}$ (LCCO) when a magnetic field $H > H_{c2}$ is applied to suppress superconductivity [37,38]. Further, a gap enhancement near the "hot spots" (π,π) of the Fermi surface in the SC state of the electron-type cuprate $Nd_{2-x}Ce_xCuO_{4-y}$ (NCCO) [39] is also consistent with the notion of a hidden PG with $\Delta_{PG} < \Delta_{SC}$ at $T < T_c$.

In addition to the PG phenomena manifested in the energy dependence of the LDOS spectra as exemplified in Fig. 2(b), STS studies further suggest that the PG phase in the hole-type cuprates in fact stems from lattice translational symmetry breaking, as represented by the energy-independent wave vector obtained from the FT-LDOS for $T_c < T < T^*$ [23] and exemplified in Fig. 10 for theoretically calculated FT-LDOS as a function of temperature in the presence (absence) of a CDW-like competing order [183]. Moreover, the low-energy excitation spectra for $T_c < T < T^*$ reveal particle–hole asymmetric spectral characteristics that differ fundamentally from the particle-hole symmetric Bogoliubov quasiparticle spectra for superconductivity [60]. Further, Raman scattering spectroscopic studies of the doping dependence of $YBa_2Cu_3O_{6+x}$ and $Bi_2Sr_2(Ca_xY_{1-x})Cu_2O_8$ and ARPES studies of $Bi_2Sr_2CuO_{6+x}$ (Bi-2201) also confirmed particle–hole symmetry breaking and pronounced spectral broadening, indicative of spatial symmetry breaking without long-range order at the opening of the PG, in agreement with the STS findings that the PG state is a broken-symmetry state that is distinct from homogeneous superconductivity [60].

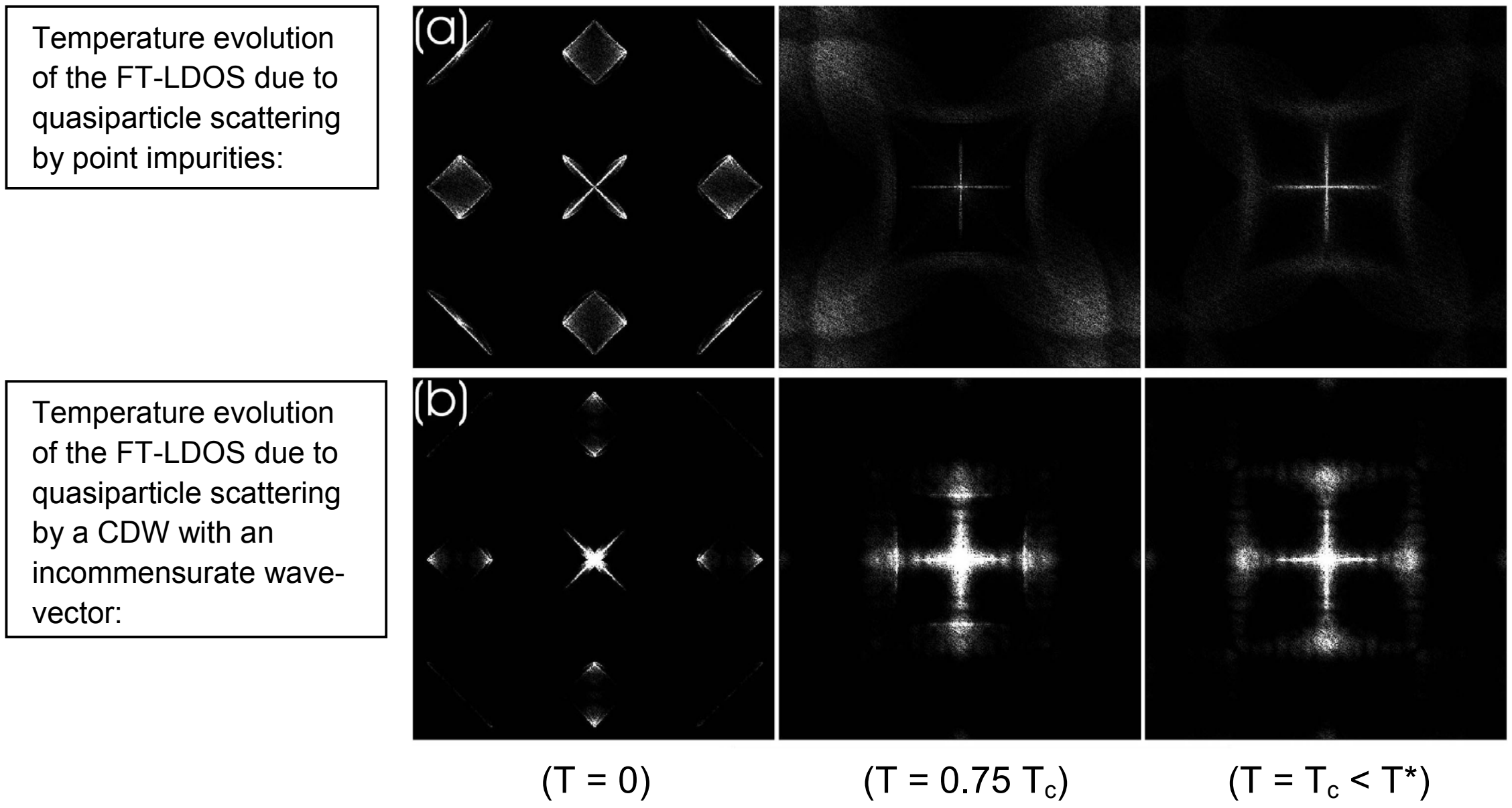

**Fig. 10**: The temperature evolution of the FT-LDOS maps in the first Brillouin zone at $T = 0$, $0.75T_c$, $T_c$ (from left to right) for quasiparticle scattering in a $d_{x^2-y^2}$–wave superconductor [183] by **(a)** point impurities and **(b)** a competing order such as a charge-density-wave (CDW) or a disorder-pinned spin-density-wave (SDW) with an incommensurate wave-vector parallel to the (π,0)/(0,π) direction. Here we have assumed that the CO has an energy gap $V_{CO}$ larger than the SC gap and that $V_{CO}$ does not vanish until the PG temperature $T^* > T_c$ [183]. The non-vanishing FT-LDOS intensities for a constant |**k**| value above $T_c$ are consistent with the experimental observation [23].

### *2.3.2. Fermi arcs*

Intimately related to the PG phenomena found in the quasiparticle tunneling spectra is the Fermi arc observed from ARPES studies of hole-type cuprate superconductors [18,21], where the Fermi arc refers to the truncated Fermi surface not fully recovered at $T_c < T < T^*$ [18,21,182,184]. Specifically, the appearance of the low-energy PG in hole-type cuprates may be correlated with the appearance of the Fermi arc above $T_c$ and below the PG temperature $T^*$ [18,21,182,184] within the CO scenario [7,10,16].

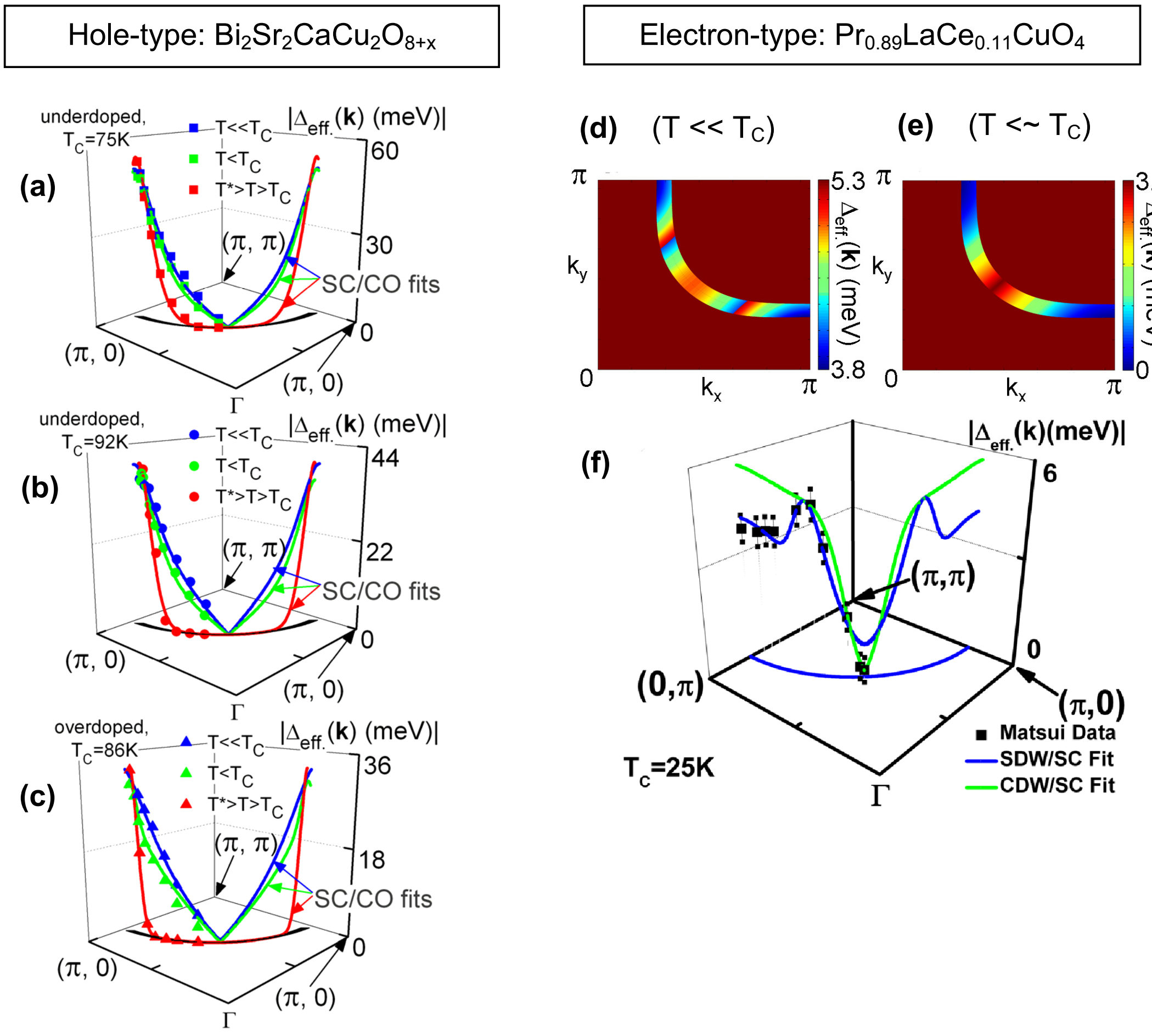

**Fig. 11**: **(a) – (c)** Theoretical fittings (solid color curves) [16] to the momentum (**k**) dependent effective excitation gap $\Delta_{eff}$ (color symbols) determined from ARPES spectra on Bi-2212 for three different doping levels and at three temperatures [21]. By considering the scenario of coexisting $d_{x^2-y^2}$-SC and CDW [16], the calculated ARPES results agree well with experimental findings [21]: Below $T_c$, $\Delta_{eff}(\mathbf{k})$ only vanishes at the nodal point, whereas above $T_c$ and below $T^*$, $\Delta_{eff}(\mathbf{k})$ vanishes over an "arc"-shaped of wave-vectors. In contrast, the **k**- and $T$-dependence of $\Delta_{eff}(\mathbf{k})$ in electron-type cuprates $Pr_{0.89}LaCe_{0.11}CuO_4$ differs from that in the hole-type cuprate, as exemplified in **(d) – (f)** [35]. The CO scenario can account for both the absence of Fermi arcs above $T_c$ and the $\Delta_{eff}(\mathbf{k})$-vs.-**k** behavior below $T_c$ by assuming a SDW with a $(\pi,\pi)$ wave-vector as the relevant CO and $V_{SDW} < \Delta_{SC}$ [16,39,181]. On the other hand, fitting with a CDW does not agree with the k-dependence of $\Delta_{eff}(\mathbf{k})$.

As exemplified in Fig. 11 and detailed elsewhere [16], we find that the Fermi arc as a function of the quasiparticle momentum **k**, temperature ($T$) and doping level ($\delta$) in Bi-2212 [21] may be explained consistently by assuming a CDW (or disorder-pinned incommensurate SDW) as the relevant CO, which occurs at the PG temperature $T^* > T_c$ with an energy gap $V_{CDW} > \Delta_{SC}$ and an *incommensurate* wave-vector $Q_{CDW} \parallel (\pi,0)/(0,\pi)$ [7,10-16]. Thus, the PG gap is associated with the CO energy gap $V_{CO}$ for $T_c < T < T^*$, whereas the effective gap $\Delta_{eff}(\mathbf{k})$ for $T < T_c$ is given by $\Delta_{eff}(\mathbf{k}) \equiv \{[\Delta_{SC}(\mathbf{k})]^2 + [V_{CO}(\mathbf{k})]^2\}^{1/2}$. In contrast, the **k**- and *T*-dependence of the effective gap $\Delta_{eff}(\mathbf{k})$ and the absence of Fermi arcs in electron-type cuprates (*e.g.* $Pr_{0.89}LaCe_{0.11}CuO_4$) [35] can also be explained by incorporating a SDW with $V_{SDW} < \Delta_{SC}$ [39,181] and a *commensurate* wave-vector $Q_{SDW} = (\pi,\pi)$ (see Eq. (9)) into spectral characteristics [16]. Hence, the CO scenario is shown to provide adequate phenomenology for a wide variety of experimental findings from electron-type to hole-type cuprate superconductors and as a function of temperature, doping level and quasiparticle momentum **k**.

#### *2.3.3. Phase fluctuations above $T_c$*

One of the major experimental signatures that strongly favor the "two-gap" scenario over the "one-gap" model is the finding of vanishing SC coherence at temperatures well below the PG temperature $T^*$ [80]. Specifically, the one-gap conjecture [185-188] suggests that in underdoped cuprate superconductors, Cooper pairs could form at $T^*$ while significantly SC phase fluctuations prevent Cooper pairs from condensing into a true SC state until the temperature is lowered below $T_c$. Hence, one would expect remnants of SC phase coherence for $T_c < T < T^*$, which may be captured by means of high-frequency optical conductivity measurements if the relaxation time of the "preformed" Cooper pairs becomes sufficiently short as $T \to T^*$ from below. However, optical conductivity measurements for underdoped hole-type cuprates Bi-2212 over a range of doping levels [80] reveal that the complex paraconductivity associated with the SC coherence generally follows the Kosterlitz-Thouless-Berezinskii (KTB) theory for thermally generated vortices and only survives over a small temperature window above $T_c$ and much below $T^*$ [80]. The finding of rapidly vanishing superconducting phase coherence above $T_c$ together with the breaking of particle-hole symmetry at $T_c < T < T^*$, and the quantitative studies outlined above for both the pairing state and the normal state properties of a variety of cuprate superconductors strongly suggest that the physical origin of the PG differs from the SC gap, and hence favors the two-gap scenario. As further discussed in the following section for the vortex-state properties, it is found that the PG phenomena observed above $T_c$ in zero magnetic fields may be revealed by suppressing SC using an external magnetic field at $T << T_c$, again confirming the scenario of the PG physical origin being associated with COs.

### 2.4. Vortex State

High-temperature superconducting cuprates are extreme type-II superconductors that exhibit strong thermal, disorder, and quantum fluctuations in their vortex states [77,78,81-97]. While much research has focused on the *macroscopic* vortex dynamics of cuprate superconductors with phenomenological descriptions [81-97], little effort has been made to address the *microscopic* physical origin of their extreme type-II nature until recently when spatially resolved vortex-state quasiparticle tunneling spectra became available [7,10,11,17]. As discussed in previous sections, competing orders (COs) can coexist with superconductivity (SC) in the ground state of cuprate superconductors [7,10], which lead to the occurrence of quantum criticality [5,6,67,68,189]. The proximity to quantum criticality and the existence of COs can significantly affect the low-energy excitations of the cuprates due to strong quantum fluctuations [77,78] and the redistribution of quasiparticle spectral weight among SC and COs [7,12-15]. Moreover, external variables such as temperature ($T$) and applied magnetic field ($H$) can vary the interplay of SC and COs, such as inducing or enhancing [41,52,68] the COs at the price of more rapid suppression of SC, thereby leading to weakened SC stiffness and strong thermal and field-induced fluctuations [7,77,78]. On the other hand, the quasi-two-dimensional nature of the cuprates can also result in quantum criticality in the limit of decoupling of $CuO_2$ planes [190]. In this section we review experimental studies of the unconventional low-energy excitations of the cuprates in the vortex state from both microscopic and macroscopic viewpoints.

#### *2.4.1. Intra-vortex pseudogap and energy-independent wave-vectors in the quasiparticle tunneling spectra*

In conventional type-II superconductors, superconductivity is suppressed inside periodic Abrikosov vortices [191], leading to continuous quasiparticle bound states and a peak of local density of states (LDOS) at zero energy [192–194]. In contrast, the effect of magnetic field on high-$T_c$ superconductors is much more complicated than that on conventional type-II superconductors. Microscopically, neutron scattering experiments on the hole-doped cuprate $La_{1.84}Sr_{0.16}CuO_4$ reported an effective radius of vortices substantially larger than the superconducting coherence length $\xi_{SC}$ [41,52]. Scanning tunneling spectroscopic (STS) studies of optimally doped Bi-2212 found PG-like features rather than zero-bias conductance peaks inside vortices [20,195]. Further detailed spatially resolved STS studies of Bi-2212 in one magnetic field $H$ = 5 T revealed a field-induced ($4a_0 \times 4a_0$) conductance modulation inside each vortex, where $a_0$ = 0.385 nm is the planar lattice constant of Bi-2212 [61].

The latter finding has been attributed to the presence of a coexisting competing order (CO) such as pair-density waves (PDW) [71,72], pinned spin-density waves (SDW) [5,6,68-70], or charge-density waves (CDW) [4,75] upon suppression of SC inside the vortices.

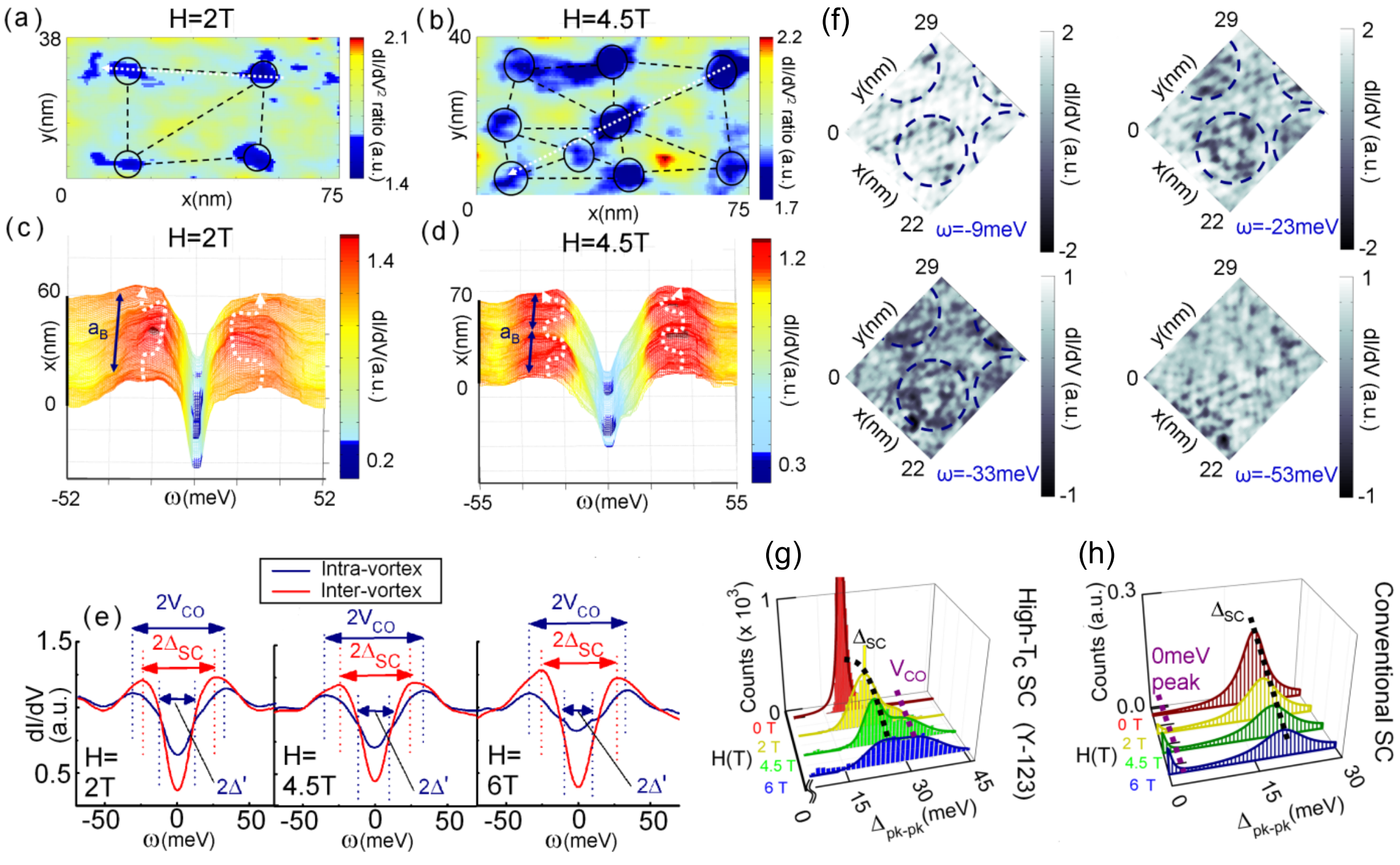

**Fig. 12:** Spatially resolved STS studies of the vortex-state of Y-123 at $T = 6$ K [10,11]: **(a)** Tunneling conductance power ratio $r_G$ map over a (75×38) nm$^2$ area for $H = 2$ T, showing disordered vortices with an average vortex-vortex separation $a_B = (33.2 \pm 9.0)$ nm. Here the conductance power ratio at each pixel is defined by the ratio of $(dI/dV)^2$ at $V = (\Delta_{SC}/e)$ to that at $V = 0$. **(b)** The $r_G$ map over a (75×40) nm$^2$ area for $H = 4.5$ T, showing $a_B = (23.5 \pm 8.0)$ nm. **(c)** Conductance spectra along the white line in (a), showing SC peaks at $\omega = \pm\Delta_{SC}$ outside vortices and PG features at $\omega = \pm V_{CO}$ inside vortices. **(d)** Conductance spectra along the dashed line indicated in (b). **(e)** Spatially averaged intra- and intervortex spectra for $H = 2.0$ T, 4.5 T and 6 T from left to right. **(f)** The LDOS modulations of Y-123 at $H = 5$ T over a (22×29) nm$^2$ area, showing patterns associated with density-wave modulations and vortices (circled objects) for $\omega = -9$ meV ~ $-\Delta'$, $\omega = -23$ meV ~ $-\Delta_{SC}$, $\omega = -33$ meV ~ $-V_{CO}$ and $\omega = -53$ meV, which is comparable to the longitudinal optical phonon frequency [196]. The vortex contrasts are the most apparent at $|\omega| \sim \Delta_{SC}$ and become nearly invisible for $|\omega| \sim V_{CO}$. **(g)** Energy histograms for the field-dependent spectral weight derived from the STS data for $H = 0$, 2, 4.5, and 6T, showing a spectral shift from $\Delta_{SC}$ to $V_{CO}$ and $\Delta'$ with increasing $H$. **(h)** Schematic of the histograms for a conventional type-II superconductor in the limit of $T \ll T_c$ and $H \ll H_{c2}$.

More recently, spatially resolved STS studies of the optimally doped hole-type cuprate superconductor Y-123 and the optimally doped electron-type cuprate superconductor La-112 in the vortex state have been carried out as a function of applied magnetic fields [10,11,17], which reveal rich information and interesting contrasts between the hole and electron-type cuprate superconductors. In the case of optimally doped Y-123, [10,11] while the zero-field LDOS revealed highly homogeneous spectral characteristics, the vortex-state STS studies suggest strongly disordered vortices as well as a "vortex halo" radius $\xi_{halo}$ much larger than $\xi_{SC}$, as exemplified in Figs. 12(a) and 12(b) for $H = 2.0$ T and 4.5 T, respectively. This finding is consistent with the report from neutron scattering experiments [41,52]. Moreover, the spatial evolution of the vortex-state spectra reveals modulating gap-like features everywhere without any zero-energy peaks, as shown in Figs. 12(c) and 12(d), and in 12(e) for representative spectra taken inside and outside of vortices at $H = 2.0$ T, 4.5 T and 6.0 T. For each constant field, the inter-vortex spectrum reveals a sharper set of peaks at $\omega = \pm\Delta_{SC} \sim \pm 23$ meV, whereas the intra-vortex spectrum exhibits PG features at $\omega = \pm V_{CO} \sim \pm 32$ meV and $V_{CO} > \Delta_{SC}$. Interesting, the PG energy $V_{CO}$ revealed inside the vortex core is in excellent agreement with the CO energy obtained from theoretical fitting to the zero-field spectra using the relation $V_{CO} = [(\Delta_{eff})^2 - (\Delta_{SC})^2]^{1/2}$, as described in Section 2.2.2. Additionally, subgap features at $\omega = \pm \Delta' = \pm (7 \sim 10)$ meV are found inside vortices, which become more pronounced with increasing $H$. The physical origin of $\Delta'$ is still unknown, although it may be associated with the energy of PDW, while $V_{CO}$ may be associated with the CDW or disorder-pinned SDW. Further, apparent LDOS modulations are visible at constant quasiarticle energies (Fig. 12(f)), showing patterns associated with density-wave modulations and vortices (circled objects) for $\omega = -9$ meV $\sim -\Delta'$, $\omega = -23$ meV $\sim -\Delta_{SC}$, $\omega = -33$ meV $\sim -V_{CO}$ and $\omega = -53$ meV. The vortex contrasts are the most apparent at $|\omega| \sim \Delta_{SC}$ and become nearly invisible for $|\omega| \sim V_{CO}$. The vanishing contrast at high energies may be due to the onset of Cu-O optical phonons (~50 meV for the cuprates [196]) so that both the collective modes and quasiparticles become scattered inelastically. Finally, energy histograms of the gap features exhibit strong spectral shifts from $\Delta_{SC}$ to $V_{CO}$ and $\Delta'$ with increasing magnetic field, as shown in Fig. 12(g), which is in sharp contrast to the vortex-state spectral shifts in conventional type-II superconductors, as schematically illustrated in Fig. 12(h). These vortex-state spectral findings are all consistent with the coexistence of COs with SC in Y-123 [7].

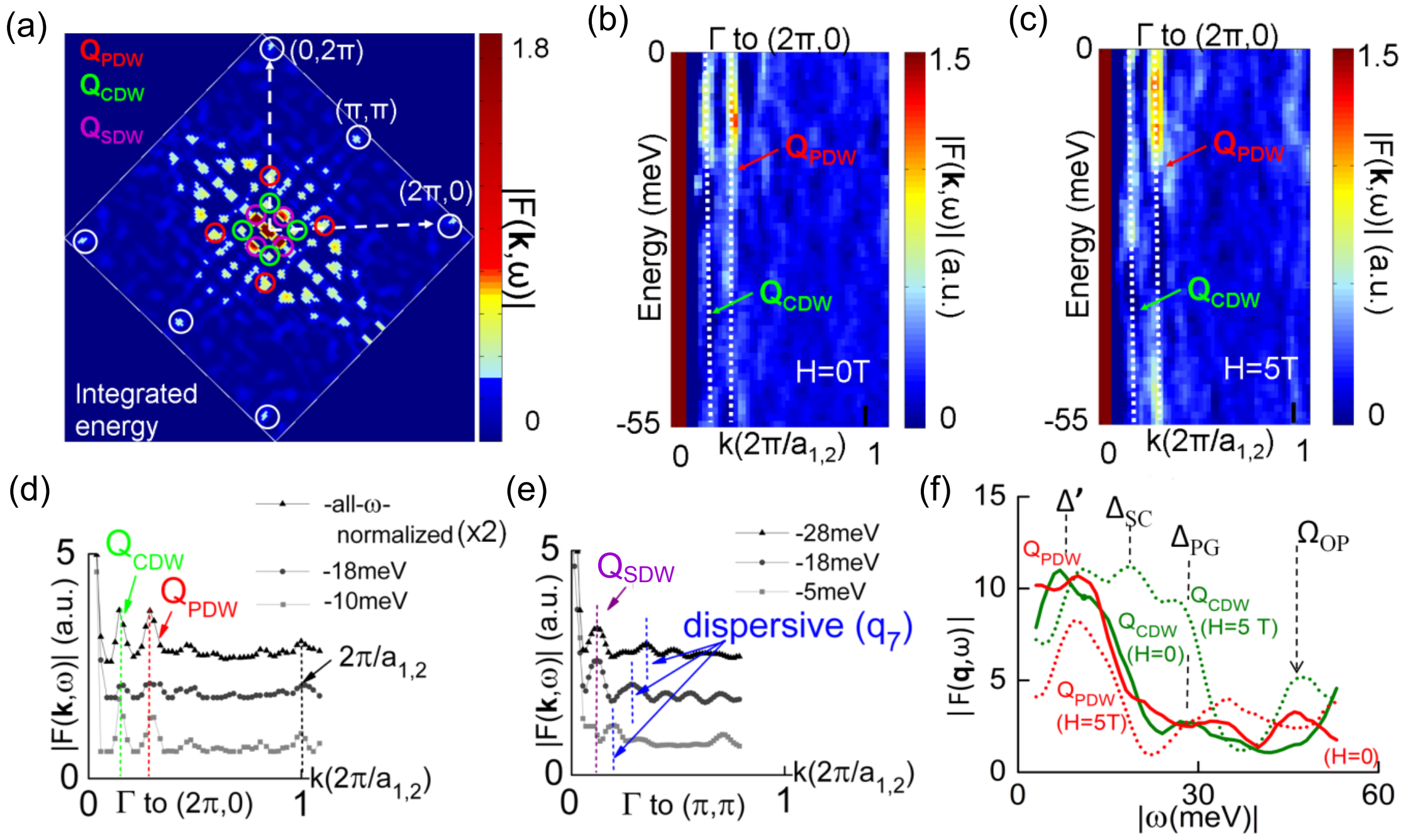

**Fig. 13:** Studies of the vortex-state FT-LDOS maps of Y-123 in the two-dimensional reciprocal space [10,11,197]: **(a)** Normalized FT-LDOS at $H = 5$ T obtained by integrating $|F(\mathbf{k},\omega)|$ from $\omega = -1$ meV to $-30$ meV. There are three sets of $\omega$-independent spots in addition to the reciprocal lattice constants and the $(\pi,\pi)$ resonance, which are circled for clarity. These characteristic wave-vectors include $\mathbf{Q}_{PDW}$ and $\mathbf{Q}_{CDW}$ along the $(\pi,0)/(0,\pi)$ directions and $\mathbf{Q}_{SDW}$ along $(\pi,\pi)$. **(b)** The $\omega$-dependence of $|F(\mathbf{k},\omega)|$ at $H = 0$ is plotted in the $\omega$-*vs.*-$\mathbf{k}$ plot against $\mathbf{k}(\pi,0)$, showing $\omega$-independent modes (bright vertical lines) at $\mathbf{Q}_{PDW}$ and $\mathbf{Q}_{CDW}$. **(c)** The $\omega$-dependence of $|F(\mathbf{k},\omega)|$ at $H = 5$ T is plotted in the $\omega$-*vs.*-$\mathbf{k}$ plot against $\mathbf{k} \parallel (\pi,0)$, showing field enhanced spectral intensities at $\mathbf{Q}_{PDW}$ and $\mathbf{Q}_{CDW}$. **(d)** $|F(\mathbf{k},\omega)|$ for different energies are plotted against $\mathbf{k} \parallel (\pi,0)$, showing peaks at $\omega$-independent $Q_{PDW}$, $Q_{CDW}$ and the reciprocal lattice constants at $(2\pi/a_1)$ along $(\pi,0)$. **(e)** $|F(\mathbf{k},\omega)|$ for different energies are plotted against $\mathbf{k} \parallel (\pi,\pi)$, showing peaks at energy-independent $Q_{SDW}$ along $(\pi,\pi)$. Additionally, dispersive wave vectors due to quasiparticle scattering interferences (QPI) are found, as exemplified by the dispersive QPI momentum $\mathbf{q}_7$ specified in Figs. 8(b) and 8(d). **(f)** The FT-LDOS intensities $|F(\mathbf{q},\omega)|$ of Y-123 for $\mathbf{q} = \mathbf{Q}_{PDW}$ (red) and $\mathbf{Q}_{CDW}$ (green) are shown as a function of $\omega$ and for $H = 0$ (solid lines) and $H = 5$ T (dashed lines) [197]. Here $\Delta_{PG} = V_{CO}$ and $\Omega_{OP}$ denotes the longitudinal optical phonon mode along the Cu-O bond.

In addition to the revelation of COs at characteristic energies $V_{\text{CO}}$ and $\Delta'$, evidences for collective modes at characteristic wave vectors may be identified from studies of the Fourier transformation (FT) of the LDOS. As exemplified in Figs. 13(a)-(f), the spectral intensity $|F(\mathbf{k},\omega)|$ of the FT-LDOS reveal abundant information about the dependence of the cuprate low-energy excitations on momentum ($\mathbf{k}$), energy ($\omega$) and magnetic field ($H$). Similar to the findings in zero fields, the FT-LDOS spectra contains two types of high intensity spots. One type is associated with the strongly $\omega$-dependent Bogoliubov quasiparticle interferences (QPI) due to elastic scattering by impurities, as exemplified in Fig. 13(e) and previously shown in Figs. 8(b) and 8(d). The other type contains three sets of $\omega$-independent spots in addition to the reciprocal lattice constants and the $(\pi,\pi)$ resonance, including $\mathbf{Q}_{\text{PDW}}$ and $\mathbf{Q}_{\text{CDW}}$ along the $(\pi,0)/(0,\pi)$ directions and $\mathbf{Q}_{\text{SDW}}$ along $(\pi,\pi)$, as exemplified in Figs. 13(b)-(e). Further investigation of the FT-LDOS reveals interesting magnetic field dependence, showing field-enhanced spectral intensities $|F(\mathbf{k},\omega)|$ for $\mathbf{k} = \mathbf{Q}_{\text{PDW}}$ and $\mathbf{Q}_{\text{CDW}}$, which may be compared with the significant shifts in spectral weights from $\Delta_{\text{SC}}$ to $V_{\text{CO}}$ and $\Delta'$ with increasing $H$. Additionally, $|F(\mathbf{k},\omega)|$-*vs.*-$\omega$ data for both $\mathbf{k} = \mathbf{Q}_{\text{PDW}}$ and $\mathbf{Q}_{\text{CDW}}$ consistently reveal a spectral peak around $\Delta'$, as shown in Fig. 13(f) [197], whereas significant enhancements in $|F(\mathbf{Q}_{\text{CDW}},\omega)|$ for $\omega > \Delta'$ only occur for $H > 0$, suggesting that the application of a finite magnetic field increases the CDW excitations that break the particle-hole symmetry. It is clear that none of these spectral dependences on $\mathbf{k}$, $\omega$ and $H$ can be simply explained in terms of a pure $d_{x^2-y^2}$-wave SC ground state in the cuprates, nor can they be attributed to simple bandstructure effects because of the sensitive $H$-dependence.

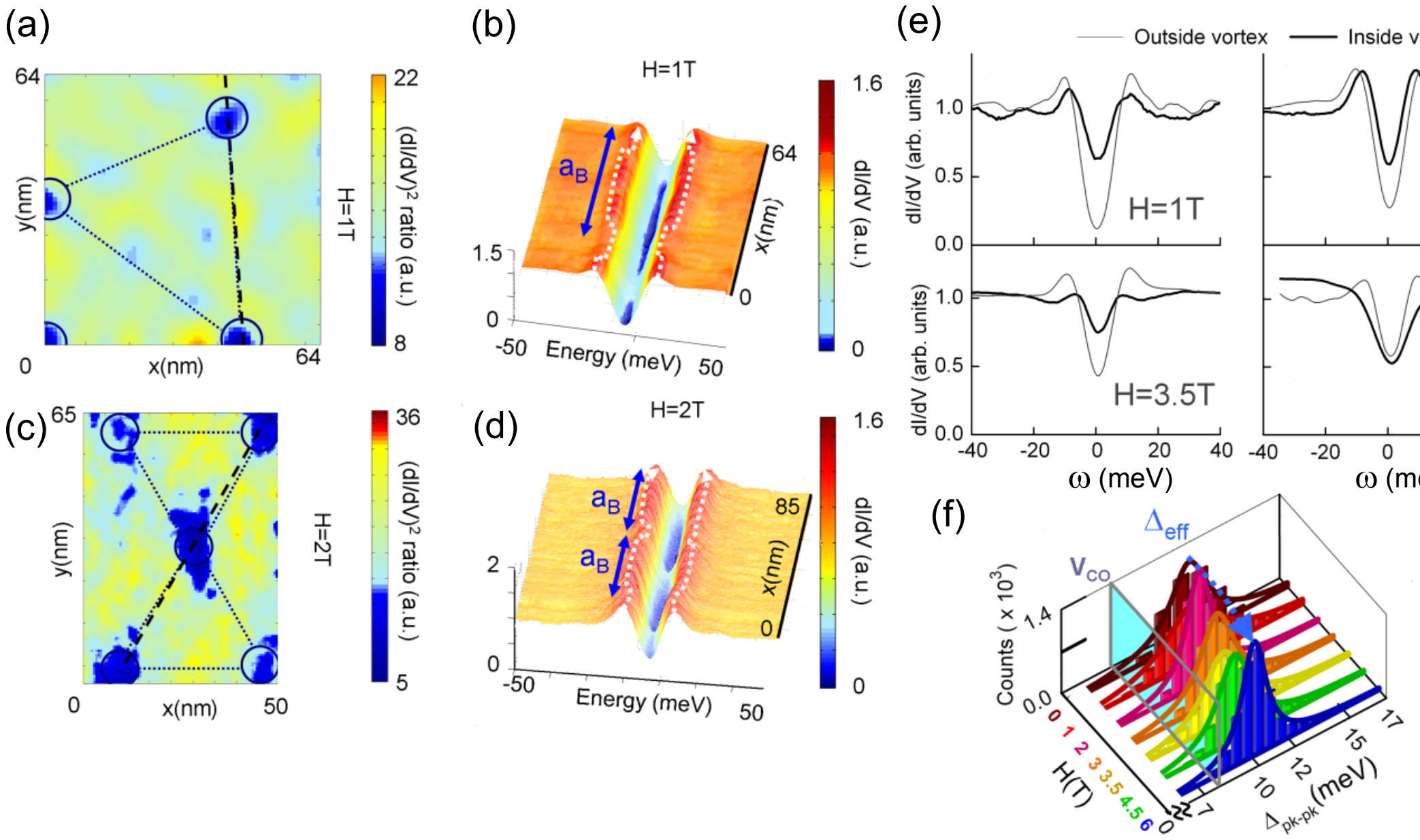

**Fig. 14:** Spatially resolved STS studies of the vortex-state of an optimally doped electron-type cuprate La-112 ($T_c = 43$ K) at $T = 6$ K and for $H \parallel$ c-axis [10,17]: **(a)** A spatial map of the conductance power ratio $r_G$ (in log scale) taken over a $(64{\times}64)$ nm$^2$ area for $H = 1$ T, showing a zoom-in view of vortices separated by an average vortex lattice constant $a_B = 52$nm, which compares favorably with the theoretical value of 49 nm. The average radius of the vortices (indicated by the radius of the circles) is $(4.7{\pm}0.7)$ nm, comparable to the SC coherence length $\xi_{ab} = 4.9$ nm [176]. Here the conductance power ratio is defined as the ratio of $(dI/dV)^2$ at $|\omega| = \Delta_{\text{eff}}$ and that at $\omega = 0$. **(b)** Spatial evolution of the conductance ($dI/dV$) along the black dashed line cutting through two vortices in (a) for $H = 1$T, showing significant modulations in the zero-bias conductance and slight modulations in the peak-to-peak energy gap. **(c)** A spatial map of the conductance power ratio $r_G$ (in log scale) taken over a $(65\times50)$ nm$^2$ area with $H = 2$T, showing a zoom-in view of vortices with an average vortex lattice constant $a_B = 35$ nm, which is consistent with the theoretical value. **(d)** Spatial evolution of the conductance is shown along the black dashed line cutting through three vortices in (b) for $H = 2$T. **(e)** Evolution of the inter- and intra-vortex quasiparticle tunneling spectra with magnetic field in La-112 for $H = 1$, 2, 3.5 and 6 T, where the PG spectra at the center of vortex cores are given by the thick lines and those exterior to vortices are given by the thin lines. **(f)** Energy histograms of La-112 determined from our quasiparticle tunneling spectra of La-112, showing the spectral evolution with $H$. Note that there is no zero-bias conductance peak in the vortex-state, and that a low-energy cutoff at nearly a constant value $V_{\text{CO}} = (8.5{\pm}0.6)$ meV exists for all fields.

In the case of STS studies of the vortex-state of electron-type cuprate superconductors, there are interesting similarities and differences when compared with the vortex-state properties of hole-type cuprates. As shown in Fig. 14 for representative STS studies of the infinite-layer system La-112, long-range disordered vortices similar to those in the Y-123 system are also observed [17]. However, the vortex core radius appears to be comparable to $\xi_{SC}$ [17], which is in contrast to the much larger vortex halo size ($\xi_{halo} \sim 10\ \xi_{SC}$) found in Y-123 [11]. Moreover, spatially resolved tunnelling spectra also exhibit PG features inside the vortex core of La-112, with a PG energy $\Delta_{PG} \sim V_{CO}$ *smaller* than the SC gap, $\Delta_{PG} < \Delta_{SC}$ [17], which in contrast to the hole-type cuprates that reveal $\Delta_{PG} > \Delta_{SC}$ inside the vortex cores [10,11]. Here we note that the gap values $\Delta_{SC}$ and $V_{CO}$ are determined empirically by taking 1/2 of the energy difference between the spectral peak-to-peak, $\Delta_{pk\text{-}pk}$, for the inter- and intra-vortex spectra, respectively. This finding is again consistent with the absence of zero-field PG phenomena above $T_c$ in the electron-type cuprate superconductors. Hence, we conclude that the rich phenomena revealed in the vortex-state STS studies of both hole- and electron-type cuprates can all be consistently understood within the two-gap scenario.

### *2.4.2. Strong quantum, thermal and disorder fluctuations*

As described earlier, cuprate superconductors are doped Mott insulators with strong electronic correlation that can result in a variety of competing orders (COs) in the ground state. Therefore, significant quantum fluctuations and reduced SC stiffness are expected in the cuprates because of the existence of multiple channels of low energy excitations, which are believed to contribute to the extreme type-II nature of the cuprate superconductors [14,15]. Moreover, external variables such as temperature ($T$) and applied magnetic field ($H$) can vary the interplay of SC and CO, such as inducing or enhancing the CO at the price of more rapid suppression of SC [41,52], thereby leading to weakened SC stiffness and strong thermal and field-induced fluctuations [68,77,78,81,82]. These effects are likely the primary cause for the occurrence of a vortex liquid phase below the upper critical field of cuprate superconductors. Moreover, the significantly weakened SC stiffness also implies much stronger susceptibility of vortices to disorder, giving rise to various types of glassy phases at low temperatures, depending on the type and dimensionality of disorder, as exemplified in Fig. 15(a) for the vortex phase diagram of three-dimensional cuprate superconductors with random point disorder and for $H \parallel c$-axis, showing the occurrence of a vortex liquid phase below the upper critical field $H_{c2}(T)$ and additional disordered vortex solid phases (*i.e.*, the "vortex glass" [81,82] and "Bragg glass" [86,87]) below the vortex liquid phase. For comparison, the vortex phase diagram for conventional type-II superconductors is shown in Fig. 15(b), where the ordered vortex solid phase, known as the vortex lattice, extends all the way to $H_{c2}(T)$ without the occurrence of a vortex liquid phase. In the context of strong disorder fluctuations, correlated disorder such as columnar defects and twin boundaries can result in different universality classes of vortex phase transitions [83,84,89-97] relative to the situation of random point defects, as exemplified in Figs. 16(a)-(c) for the vortex glass [81,82], Bose glass [83,84] and splayed glass [96,97] transitions associated with different types of defects. Details for the theory and experimental investigations of disorder-induced novel vortex dynamics in the cuprate superconductors can be found in Refs. [81] – [97].

In addition to the quantum criticality induced by competing orders, the quasi two-dimensional nature of the cuprates may yield a quantum criticality in the limit of decoupling of $CuO_2$ planes [190]. Indeed, recent studies have demonstrated experimental evidence from *macroscopic* magnetization measurements for field-induced quantum fluctuations among a wide variety of cuprate superconductors with different *microscopic* variables such as the doping level ($\delta$) of holes or electrons, the electronic mass anisotropy ($\gamma$), and the number of $CuO_2$ layers per unit cell ($n$) [14,15,77,78]. It is suggested that the manifestation of strong field-induced quantum fluctuations is consistent with a scenario that all cuprates are in close proximity to a quantum critical point (QCP) [77].

To investigate the effect of quantum fluctuations on the vortex dynamics of cuprate superconductors, vortex phase diagrams for different cuprates were studied at $T \to 0$ to minimize the effect of thermal fluctuations, and the magnetic field was applied parallel to the $CuO_2$ planes ($H \parallel ab$) to minimize the effect of random point disorder [77]. The rationale for having $H \parallel ab$ is that the intrinsic pinning effect of layered $CuO_2$ planes generally dominates over the pinning effects of random point disorder [85], so that the commonly observed glassy vortex phases associated with point disorder for $H \parallel c$ (*e.g.*, vortex glass and Bragg glass) [81,82,86,87] can be prevented. In the absence of quantum fluctuations, random point disorder can cooperate with the intrinsic pinning effect to stabilize the low-temperature vortex smectic and vortex solid phases [85], so that the vortex phase diagram for $H \parallel ab$ would resemble that of the vortex-glass and vortex-liquid phases observed for $H \parallel c$ with a glass transition $H_G$ ($T = 0$) approaching $H_{c2}$ ($T = 0$). On the other hand, when field-induced quantum fluctuations are dominant [6,68], the vortex phase diagram for $H \parallel ab$ will deviate substantially from the predictions solely based on thermal fluctuations and intrinsic pinning, so that strong suppression of the magnetic irreversibility field $H_{irr}(T)$ relative to the upper critical field $H_{c2}$ is expected at $T \to 0$ [14,15,77], as schematically shown in Fig. 15(c), because the induced persistent current circulating along both the *c*-axis and the *ab*-plane can no longer be sustained if field-induced quantum fluctuations become too strong to maintain the *c*-axis superconducting phase coherence.

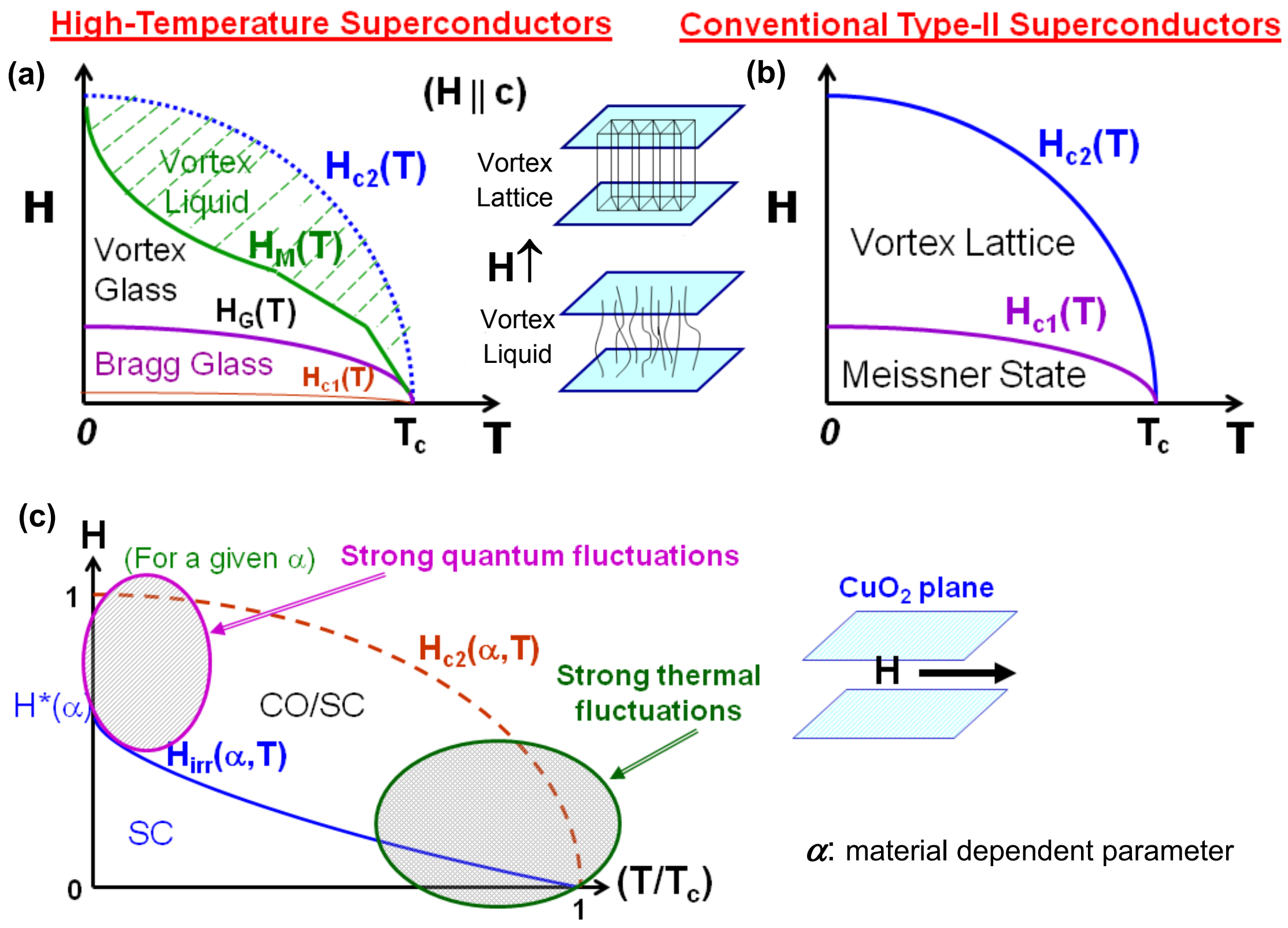

**Fig. 15**: Novel vortex dynamics of high-temperature superconducting cuprates in comparison with conventional type-II superconductors: **(a)** Schematic $H$ vs. $T$ vortex phase diagrams of cuprate superconductors for $H \parallel c$-axis, assuming thermal fluctuations and random point disorder [81,82,86,87]. The vortex phase (neglecting the lower critical field $H_{c1}$ due to the extreme type-II nature) with increasing $H$ and $T$ evolves from the Bragg glass [86,87] to the vortex glass [81,82] through the phase boundary $H_G(T)$ and then to the vortex liquid through the phase boundary $H_M(T)$ before reaching the upper critical field $H_{c2}(T)$. The occurrence of glass and liquid phases below $H_{c2}$ may be attributed to the strong disorder and thermal fluctuations in cuprate superconductors. **(b)** Schematic vortex phase diagram for conventional type-II superconductors is shown for comparison with that of the cuprate superconductors. **(c)** For $H \parallel ab$-plane, assuming dominating quantum fluctuations associated with the proximity to quantum criticality and COs [6,14,15,68,77], it is conjectured and experimentally verified that the application of high in-plane magnetic fields in the $T \to 0$ limit may suppress the phase coherent superconducting (SC) phase at a field characteristic $H^* \equiv H_{irr}(T \to 0)$ much below the upper critical field $H_{c2}$ due to the field-induced currents exceeding the c-axis critical currents and/or field-enhanced competing orders (COs) that result in strong quantum fluctuations [6,14,15,68,77]. The characteristic field is expected to be dependent on the material properties of the cuprates, which may be parameterized by $\alpha$, where $\alpha$ is a function of the doping level ($\delta$), the electronic mass anisotropy ($\gamma$) and the number of $CuO_2$ layers per unit cell ($n$) [14,15, 77].

Indeed, experimental studies on a wide variety of cuprate superconductors revealed consistent findings with the notion that all cuprate superconductors exhibit significant field-induced quantum fluctuations, as manifested by a characteristic field $H_{irr}(T \to 0) \equiv H^* << H_{c2}(T \to 0)$, and exemplified in Fig. 15(c) [77]. The degree of quantum fluctuations for each cuprate may be expressed in terms of a reduced field $h^* \equiv [H^*/H_{c2}(0)]$, with $h^*$ $\to 0$ indicating strong quantum fluctuations and $h^* \to 1$ referring to the mean-field limit. Most importantly, the $h^*$ values of all cuprates appear to follow a trend on a $h^*(\alpha)$-vs.-$\alpha$ plot, where $\alpha$ is a material parameter for a given cuprate that reflects its doping level $\delta$, electronic mass anisotropy $\gamma$, and charge imbalance if the number of $CuO_2$ layers per unit cell $n$ satisfies $n \geq 3$ [57,58]. Specifically, $\alpha$ is defined by the following [77]:

$$\alpha \equiv \gamma^{-1}\delta\left(\delta_o/\delta_i\right)^{-(n-2)}, \qquad (n \geq 3) \tag{12}$$

$$\alpha \equiv \gamma^{-1}\delta . \qquad (n \leq 2) \tag{13}$$

In Eq. (12) the ratio of charge imbalance in multi-layer cuprates with $n \geq 3$ is given by $(\delta_o/\delta_i)$ [57,58] between the doping level of the outer layers ($\delta_o$) and that of the inner layer(s) ($\delta_i$). Finally, in the event that $H_{c2}(0)$ exceeds the paramagnetic field $H_p \equiv \Delta_{SC}(0)/(2^{1/2}\mu_B)$ for highly anisotropic cuprates, where $\Delta_{SC}(0)$ denotes the SC gap at $T = 0$, $h^*$ is defined by $(H^*/H_p)$ because $H_p$ becomes the maximum critical field for superconductivity.

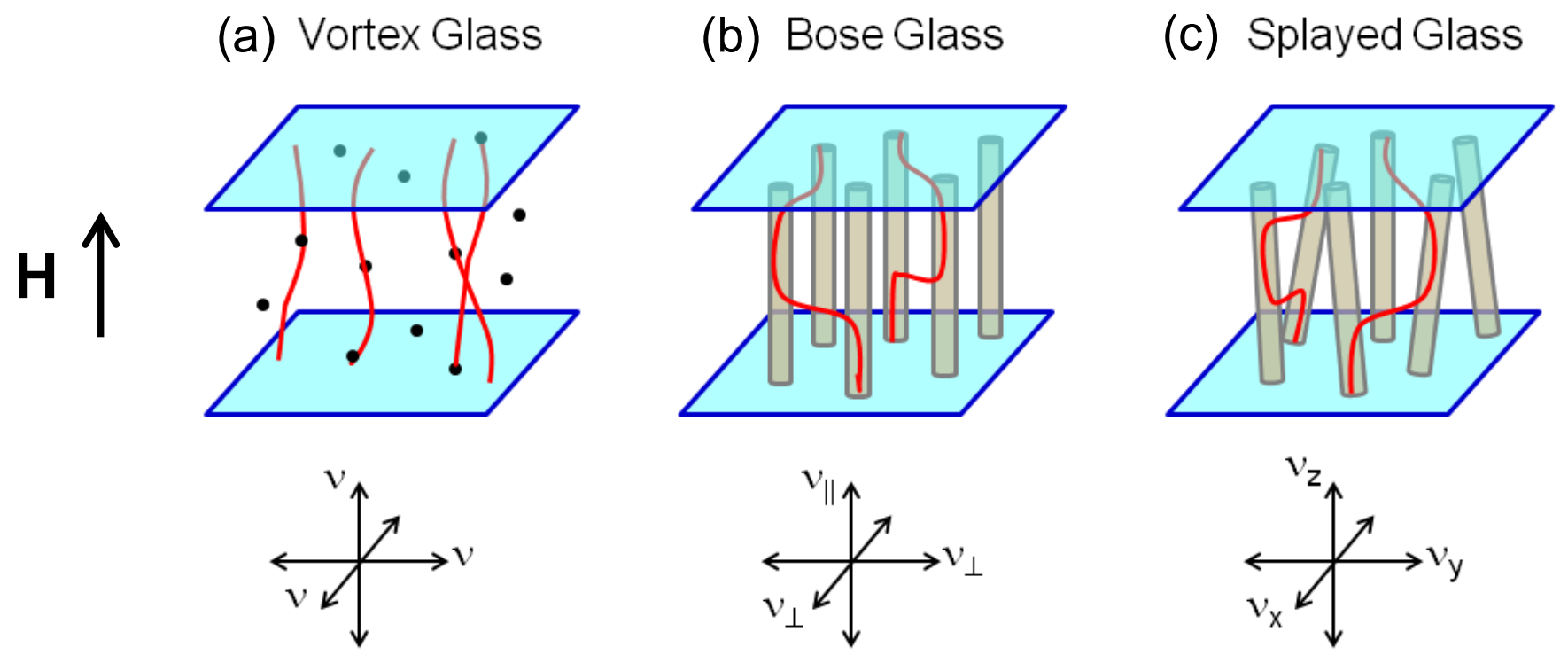

**Fig. 16**: Schematics of different universality classes of vortex phases associated with different types of disorder in cuprate superconductors [96,97]: **(a)** the "vortex glass" due to random point defects and classified by an isotropic static exponent $\nu$ and a dynamic exponent $z$ [81,82,89-94]; **(b)** the "Bose glass" due to correlated parallel columnar defects and classified by two static exponents ($\nu_{\|}, \nu_{\perp}$) and a dynamic exponent $z$ [83,84,94,95]; **(b)** the "splayed glass" due to correlated canted columnar defects and classified by three static exponents ($\nu_x, \nu_y, \nu_z$) and a dynamic exponent $z$ [96,97]. Here the red curves represent vortices and the gray columns represent correlated columnar defects induced by heavy ion irradiation [94,95]. Here the static exponent $\nu$ is defined by the vortex correlation length $\xi$ associated with a second-order vortex phase transition temperature $T_{cr}$ according to the relation $\xi = \xi_0 |1-(T/T_{cr})|^{-\nu}$, where $\xi_0$ is temperature independent; and the dynamic exponent $z$ is defined by the vortex relaxation time $\tau$ near $T_{cr}$ according to the relation $\tau = \tau_0 |1-(T/T_{cr})|^{-z}$, where $\tau_0$ is independent of $T$.

Systematic studies of the in-plane irreversibility fields of various cuprate superconductors (see Fig. 17(a)) revealed a universal trend for $h^*(\alpha)$-vs.-$\alpha$, as shown in Fig. 17(b) [77]. In particular, it is worth noting the $h^*$-vs.-$\alpha$ dependence in the multi-layered hole-type cuprate superconductors $HgBa_2Ca_2Cu_3O_x$ (Hg-1223, $T_c$ = 133 K), $HgBa_2Ca_3Cu_4O_x$ (Hg-1234, $T_c$ = 125 K) and $HgBa_2Ca_4Cu_5O_x$ (Hg-1245, $T_c$ = 108 K): While these cuprate superconductors have the highest $T_c$ and $H_{c2}$ values, they also exhibit the smallest $h^*$ and $\alpha$ values, suggesting maximum quantum fluctuations. These strong quantum fluctuations can be attributed to both their extreme two dimensionality (i.e., $\gamma >> 1$) [198,199] and significant charge imbalance that leads to strong CO in the inner layers [57,58]. This notion is corroborated by the muon spin resonance ($\mu$SR) experiments [56] that revealed increasing AFM ordering in the inner layers of the multi-layer cuprates with $n \geq 3$. Therefore, the investigation of the in-plane magnetic irreversibility in a wide variety of cuprate superconductors reveals strong field-induced quantum fluctuations [77], which is consistent with the notion that cuprate superconductors are in close proximity to quantum criticality as a result of the coexistence of competing orders and superconductivity in the ground state.

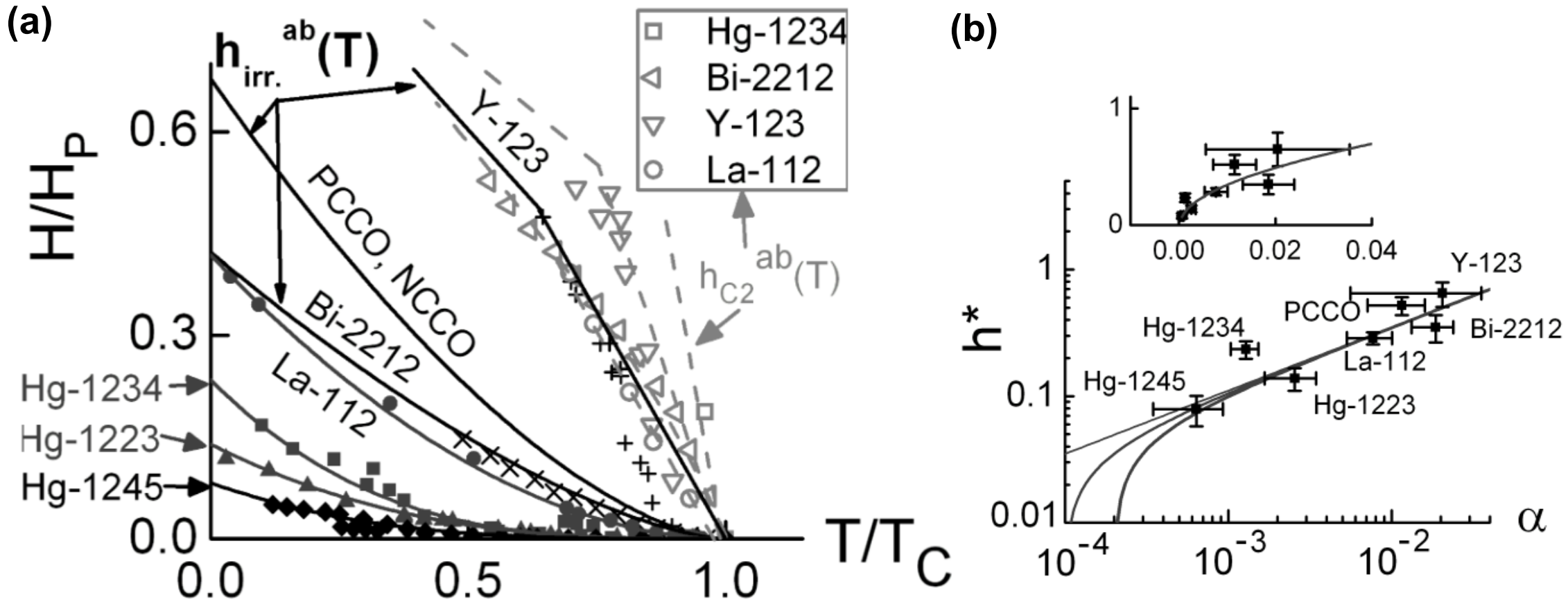

**Fig. 17**: Experimental manifestation of strong field-induced quantum fluctuations in a variety of cuprate superconductors $H \parallel ab$. **(a)** Reduced in-plane fields ($H_{irr}/H_p$) and ($H_{c2}/H_p$) vs. ($T/T_c$) for various cuprates [77]. In the $T \to 0$ limit where $H_{irr} \to H^*$, the reduced fields $h^* \equiv (H^*/H_p) < 1$ are found for all cuprates Y-123, NCCO, Bi-2212, La-112, Hg-1234, Hg-1223, and Hg-1245 (in descending order) [77]. **(b)** $h^*$-vs.-$\alpha$ in logarithmic plot for different cuprates, with decreasing $\alpha$ representing increasing quantum fluctuations. The solid lines are power-law fitting curves given by $5(\alpha - \alpha_c)^{1/2}$, using different $\alpha_c = 0$, $10^{-4}$ and $2\times10^{-4}$ from left to right [77].

### *2.4.3. Anomalous sign-reversal Hall conductivity in the vortex state*

The CO scenario also appears to be relevant to one of the outstanding issues in the cuprate superconductors, namely, the anomalous sign reversal in the vortex-state Hall conductivity ($\sigma_{xy}$) as a function either $T$ or $H$ for both electron- and hole-type cuprate superconductors [106–111]. Although quantitative description for the microscopic theory of the vortex-state Hall conduction remains incomplete, several important facts have been established: First, the sign reversal is associated with the intrinsic physical properties of cuprate superconductors [200–203] and is independent of either the electronic mass anisotropy [110] or the degree of disorder in the superconductors, regardless of random [200] or correlated disorder [110]. Specifically, for a given cuprate superconductor of a mass anisotropy $\gamma \equiv (m_c/m_{ab})$ and for a magnetic field $H$ applied at an angle $\theta$ relative to the crystalline c-axis, the Hall conductivity $\sigma_{xy}(H,T,\theta,\gamma)$ may be scaled into a universal function $\tilde{\sigma}_{xy}\left(\tilde{H},T/T_c\right)$ through the following transformation, independent of the type of disorder or the mass anisotropy [110]:

$$\tilde{\sigma}_{xy}\left(\tilde{H},T/T_c\right) \equiv \sigma_{xy}\left(H,T,\theta,\gamma\right)\sqrt{1+\gamma^{-1}\tan^2\theta}, \qquad \tilde{H} \equiv H\sqrt{\cos^2\theta+\gamma^{-1}\sin^2\theta}\,. \tag{14}$$

Second, the occurrence of sign reversal in $\sigma_{xy}$ is attributed to the non-uniform spatial distribution of carriers within and far outside the vortex core [201–203]. Third, the dc vortex-state Hall conductivity of superconducting cuprates is found to be strongly dependent on the doping level, showing anomalous sign reversal in the underdoped regime and no anomaly in the overdoped regime [204]. This important experimental finding suggests the relevance of the vrtex-core oelectronic structures to the Hall conductivity in the SC state. Given the STS observation of PG phenomena inside the vortex cores of both hole- and electron-type cuprates [10,11,17] that differs fundamentally from the “normal core” approximations in conventional type-II superconductors, we may attribute the occurrence of sign reversal in the vortex-state $\sigma_{xy}$ to the reduced quasiparticle LDOS inside the vortex cores as the result of competing orders (COs). This conjecture is further corroborated by the increasing sign reversal effect with decreasing doping [204] because the effect of COs and therefore the suppression in the vortex-core LDOS becomes more significant in the underdoped limit.

### 2.4.4. Quantum oscillations at low temperatures

The aforementioned ubiquitous presence of strong field-induced quantum fluctuations in a large variety of cuprate superconductors implies that the low-temperature dissipative vortex state is a strongly fluctuating vortex liquid (Fig. 15(c)) [77,78], which may differ from the zero-field normal state above $T_c$. Indeed, recent low temperature high-field quantum oscillations observed in underdoped hole-type cuprates $YB_2Cu_3O_{6+x}$ (Y-123) [98-102] have led to implications of excess Fermi surface structures that differ from those obtained from the zero-field ARPES experiments. Theoretical analysis of the experimental data finds that the assumption that the oscillation period is given by the underlying Fermi-surface area using the Onsager relation becomes invalid [103] in this low-temperature high-field limit. The physical origin for such differences has been attributed by some to reconstructed Fermi surfaces due to the underlying incommensurate SDW [102,104], and by others to the presence of excess electron pockets [105].

Generally speaking, it is not theoretically rigorous to infer the zero-field normal-state properties of cuprate superconductors directly from the studies of quantum oscillations in the low-temperature vortex-liquid state unless the effects of field-induced quantum fluctuations on the Fermi surface of the cuprates and the vortex-state quantum oscillations can be understood. For instance, a SDW state could evolve into a different magnetic order under sufficiently strong magnetic fields at low temperatures, which may give rise to a Fermi surface reconstruction such as the occurrence of four hole pockets created by a $(\pi,\pi)$ folding [102-104]. Additionally, the observation of negative Hall effects [105] in the low-temperature high-field limit is similar to the anomalous sign-reversal Hall conductivity of both hole- and electron-type cuprates in the high-temperature and low-field vortex-liquid state [106-111]. As discussed in Section 2.4.3, the PG phenomena revealed inside the vortex cores of both hole- and electron-type cuprate superconductors [10,11,17] suggest that the vortex cores contributing to the Hall conductivity in the vortex liquid state contain opposite charge carriers to regions outside the vortex cores [7], thereby giving rise to signal reversal Hall conductivity in the vortex liquid state. Given that the PG phenomena inside the vortex core may be attributed to the presence of COs, the attribution of the negative Hall effects [105] together with quantum oscillations to the occurrence of electronic pockets may be naïve. Putting all empirical facts in the vortex state of the cuprates together, it is natural to suggest that both effects of quantum fluctuations and COs must be considered to fully account for the observed mixed-state quantum oscillations.

### 2.4.5. Anomalous Nernst effect under finite fields in the normal state

The anomalous Nernst effect [115,116,205] occurring in the under-doped hole-type cuprates at temperatures will above $T_c$ and the absence of such an effect in all electron-type cuprates remains a mystery. Generally, the Nernst effect refers to a transverse electric field generated by moving vortices in the presence of a thermal gradient. Specifically, for vortices moving with velocity $\boldsymbol{v}$ down a thermal gradient $-\nabla T \parallel \hat{x}$, a Josephson voltage is generated and is observed as a transverse electric field $E_y = Bv_x$, where $B$ is the mean flux density. While the vortex-Nernst effect is well explored and understood in low-$T_c$ superconductors, the observation of the Nernst effect in various hole-type cuprates at temperatures well above $T_c$ has generated much debate over the physical origin for such an effect. For instance, the pre-formed pair model [186-188] suggests that the zero-field superconducting transition temperature $T_c$ is merely the loss of long-range phase rigidity so that pairs may in fact survive up to a much higher temperature. Therefore, vortices could exist above $T_c$ due to local phase coherence, thus giving rise to the observed anomalous Nernst effect. However, given the inconsistency of the one-gap model with most other experimental phenomena, it seems that theoretical investigation for possible contributions from COs to the normal-state Nernst effect is necessary to settle the issue for the physical origin of the anomalous Nernst effect. For instance, the presence of SDW-like CO above $T_c$ may give rise to a non-trivial Berry phase for carriers moving under the influence of a thermal gradient and a finite magnetic field. On the other hand, settling the physical origin for this anomalous Nernst effect may not have a major bearing on our overall understanding of the pairing mechanism of high-temperature superconductivity, because the primary issue appears to be devising means to generate a strong sign-changing pairing potential from repulsive interactions while preventing phase separations and ensuring the itinerant motion of pairs. We shall return to this point in the discussion section.

## 3. Iron-Based Superconductors

The discovery of a new class of iron-based superconductors in 2008 [2] with a maximum transition temperature ($T_c$) of ~ 55 K to date [206-217] has rekindled intense activities in SC research. In particular, there are interesting similarities and contrasts between the cuprates and the ferrous compounds. Parallel studies of the low-energy excitations of both systems have yielded useful insights into the fundamental issue of pair formation in superconductors [218-222]. A list of similarities and differences in some of the important physical properties of these two classes of superconductors are summarized in Table 1.

**Table 1:** Comparison of various important physical properties of cuprate and ferrous superconductors

| | Cuprates (hole-type) | Cuprates (electron-type) | Iron pnictides 1111 & 122 | Iron chalcogenides: 11: $Fe_{1+y}(Te_{1-x}Se_x)$ |
|---|---|---|---|---|
| Record-high $T_c$ | 165 K | 43 K | 55 K | 27 K |
| Parent state (x = 0) | Antiferromagnetic Mott insulator | Antiferromagnetic Mott insulator | Semi-metal with (π,0)-SDW nested to the electron-hole Fermi surfaces | Semi-metal with (π/2,π/2)-SDW (y < 0.11); |
| | | | | Semiconductors w/ incommensurate SDW (y > 0.11) |
| Parent-state electronic configuration | 9 *d*-electrons, single-band approximations | 9 *d*-electrons, single-band approximations | 6 *d*-electrons, five-band approximations | 6 *d*-electrons, five-band approximations |
| Electronic correlation | Strong (U ~ 8 eV) | Strong (U ~ 8 eV) | Weak (U < 2 eV) | Intermediate |
| Pairing symmetry | $d_{x^2-y^2}$ (under- & optimally doped) | $d_{x^2-y^2}$ (all doping levels) | Sign-changing *s*-wave ($s_\pm$) | Sign-changing *s*-wave ($s_\pm$) or nodeless d-wave |
| | $d_{x^2-y^2}+s$ (overdoped) | | | |
| Ground state phases | Coexisting SC & CDW/PDW (under- & optimally doped) | Coexisting SC & SDW (under- & optimally doped) | Coexisting SC & (π,0)-SDW (under-doped 122 systems) | Coexisting SC & (π/2,π/2)-SDW (y < 0.11) |
| | Pure SC (overdoped) | Pure SC (overdoped) | Pure SC (1111 systems) | Coexisting SC & incommensurate SDW (y > 0.11) |
| Energy gaps | SC gap @ $T < T_c$ & pseudogap @ $T^* > T > T_c$ (under- & optimally doped) | SC gap @ $T < T_c$ & pseudogap @ $T < T^* < T_c$ (under- & optimally doped) | Two SC gaps for hole- & electron-pockets @ $T < T_c$ | Two SC gaps for hole- & electron-pockets @ $T < T_c$; or one SC gap for systems with vanishing holes |
| | Pure SC gap @ $T < T_c$ (overdoped) | Pure SC gap @ $T < T_c$ (overdoped) | | |
| Intra-vortex spectra | Pseudogap > SC gap (optimally and under-doped) | Pseudogap < SC gap (optimally doped) | Pseudogap < SC gap (optimally doped) | Pseudogap < SC gap or bound states? Doping dependence? |
| | Pseudogap < SC gap or bound states for overdoped? | | Bound states & no pseudogap for overdoped samples | |

### 3.1. Basic Structural and Magnetic Properties

Similar to the cuprate superconductors, the iron-based superconductors, which include the pnictide [2,206-213] and the iron-chalcogenide [214-217] superconductors, are correlated layered materials with magnetic instabilities. The common chemical building block of these superconductors is FeX, where X = As, P, S, Se, Te. Structurally, FeX forms a tri-layer that consists of a square array of Fe sandwiched between two checkerboard layers of X, as illustrated in Fig. 18(a). These tri-layers are further separated by the "bridging layers" consisting of alkali, alkaline-earth, or rare-earth atoms and oxygen/fluorine. Strong experimental and theoretical evidences have associated the origin of superconductivity in these ferrous superconductors with the *d*-electrons of Fe in the FeX tri-layers, with the X-layer contributing to delocalizing the *d*-electrons [218-220]. Therefore, the FeX tri-layers may be considered as playing the same role in ferrous superconductivity as the $CuO_2$ layers in the cuprates.

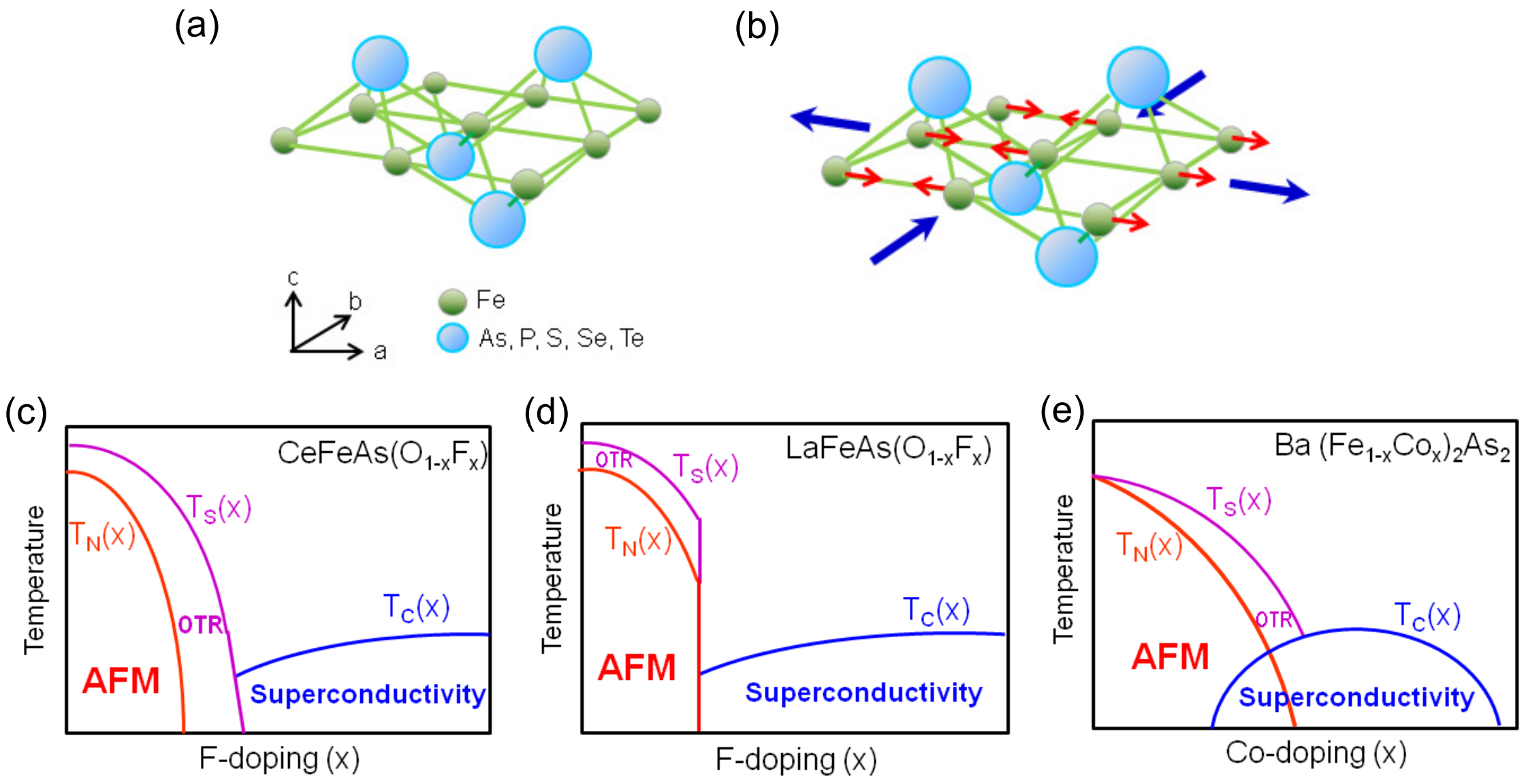

**Fig. 18: (a)** Schematics of the basic building block of the ferrous superconductors: Top view of the FeX trilayer, where X = As, P, S, Se, Te. The triad (a, b, and c) demonstrates the three crystallographic directions. **(b)** The antiferromagnetic order of the stoichiometric iron-based materials. The red arrows represent the magnetic moments, and the blue arrows indicate the directions of structural distortion. **(c)** – **(e)** Schematics of three representative phase diagrams for different types of ferrous superconductors [223-225]. Here $T_S(x)$ denotes the phase boundary for a structural phase transition from a tetragonal phase at $T > T_S$ to an orthorhombic (OTR) crystalline structure at $T < T_S$; $T_N(x)$ is the Néel temperature for the onset of an antiferromagnetic (AFM) phase at $T < T_N$; and $T_c(x)$ represents the doping dependent superconducting transition temperature.

There are three primary structures associated with the layered rare-earth transition-metal oxypnictides. The dominant type ROT*Pn* (R = rare earth elements La, Nd, Sm, Pr, Ce; T = transition metals Fe, Ni, Mn, Co; *Pn* = pnictogen P, As), also denoted as the "1111" system, can be doped with either electrons or holes [2,206-209]. In the case of LaOFeAs, the doping of fluoride ions at the oxygen sites provide electrons from the $La(O_{1-x}F_x)$ layers to the FeAs tri-layers and replacing magnetism with superconductivity (SC) for x = 0.05 to 0.12, leading to a maximum $T_c$ at 26 K [2,206,207]. Replacing La with Pr, Nd, and Sm has been shown to further boost $T_c$ up to 55 K [209]. On the other hand, substituting La with Sr leads to hole-doped compounds $(La_{1-x}Sr_x)OFeAs$ with $T_c$ up to 25 K [208]. The second type of layered compounds known as the 122 system has the formula $(A'_{1-x}A_x)Fe_2As_2$ [210] or $Ba(Fe_{1-x}Co_x)_2As_2$ [211,212], where A′ = Ba or Sr, and A = K or Cs. The $T_c$ of $KFe_2As_2$ and $CsFe_2As_2$ is 3.8 and 2.6 K, respectively, which rises with partial substitution of Sr for K and Cs and peaks at 37 K for 50%-60% Sr substitution [210]. Placing Fe in the 122 system by Co leads to electron-doped 122 with a maximum $T_c$ up to 24 K [211,212]. Moreover, SC and AFM phases are found to coexist for a range of electron doping, similar to the CO phenomena found in the cuprates. The third type of layered compounds MFeAs (or "111") with M = Li or Na are shown to exhibit $T_c$ = 20 K and 18 K [213], respectively, and the 111 system is analogous to the infinite-layer system $SrCuO_2$ in the cuprate superconductors.

Among the iron chalcogenides, the structure is known as the "11" system of $Fe(Se_{1-x}Te_x)$, which is the simplest form among the ferrous superconductors [214-217]. The first discovery of superconductivity in the 11 system was found in $\alpha$-FeSe with $T_c \sim 8$ K [214]. Subsequently, dramatic pressure-enhanced $T_c$ up to ~ 27 K has been reported [215]. Additionally, replacing Se by Te up to 50% can further enhance $T_c$, and the resulting compound exhibits an even stronger pressure effect [216], although FeTe is found to be not superconducting due to structural deformation that simultaneously breaks magnetic symmetry [216]. In a very recent development, a number of intercalated FeSe compounds $A_xFe_{2-y}Se_2$ (where A = K, Cs, Tl) were made, raising $T_c$ from 8 K for FeSe to above 30 K [217]. Overall, the Fermi surface and magnetic properties of the 11 system are very similar to those of the iron pnictides. On the other hand, there are evidences for vanished hole-pockets in the intercalated compounds $A_xFe_{2-y}Se_2$, which would be theoretically favorable for nodeless *d*-wave pairing.

Most of the stoichiometric parent compounds exhibit antiferromagnetism (AFM) at ambient pressure, and the spatial arrangement of the magnetic moments in the FeX tri-layer of the parent compounds (except the 11 system) is schematically shown Fig. 18(b) [223–225]. This magnetic order couples intimately with a tetragonal-to-orthorhombic structural distortion. For the stoichiometric 122 system such as the $BaFe_2As_2$ [224], first-order structural and AFM transitions occur at the same temperature, as illustrated in Fig. 18(e). In the low-temperature phase, the ab plane Fe-Fe distance elongates in the direction parallel to the magnetic moment and contracts in the direction perpendicular to it, as indicated by the blue arrows in Fig. 18(b). On the other hand, the 1111 system such as LaFeAsO [223] and CeFeAsO [225], the structural transition occurs at a slightly higher temperature followed by a magnetic transition, as shown in Fig. 18(c) and (d). Thus, there exists a temperature window in which the stoichiometric 1111 compounds are paramagnetic with fluctuating magnetism, but the four-fold crystalline rotation symmetry is broken by the structural distortion in the orthorhombic (OTR) phase, implying that the electron-lattice coupling will be enhanced in the OTR phase [226]. This coupling could either impede or assist the electron pairing. Moreover, the AFM state is a semi-metal, which is in sharp contrast to the cuprates where the parent AFM compounds are Mott insulators.

### 3.2. Two-Gap Superconductivity, Unconventional Pairing Symmetry and Magnetic Resonances

Calculations based on the density functional theory [212, 227-229] have shown that there are many bands near the Fermi level of these iron-based compounds and that their Fermi surfaces involve multiple disconnected Fermi pockets, as exemplified in Fig. 19. These electronic properties of the ferrous compounds are in contrast to the cuprates, the latter are primarily described by an effective one band model with a large Fermi surface (Fig. 8(a)). The presence of multiple bands and multiple disconnected Fermi surfaces suggests that inter-Fermi surface interactions may be important to the occurrence of ferrous superconductivity. Indeed, calculations of magnetic susceptibility [229,230] have shown that these ferrous compounds have a tendency for AFM order, and the wave vectors associated with the AFM coupling coincide with those connecting the centers of the electron and hole Fermi pockets, as shown in Fig. 19. These theoretical findings have led to the conjecture of two-gap superconductivity mediated by AFM spin fluctuations, with sign-changing *s*-wave ($s_{\pm}$) order parameters for the hole and electron Fermi pockets.

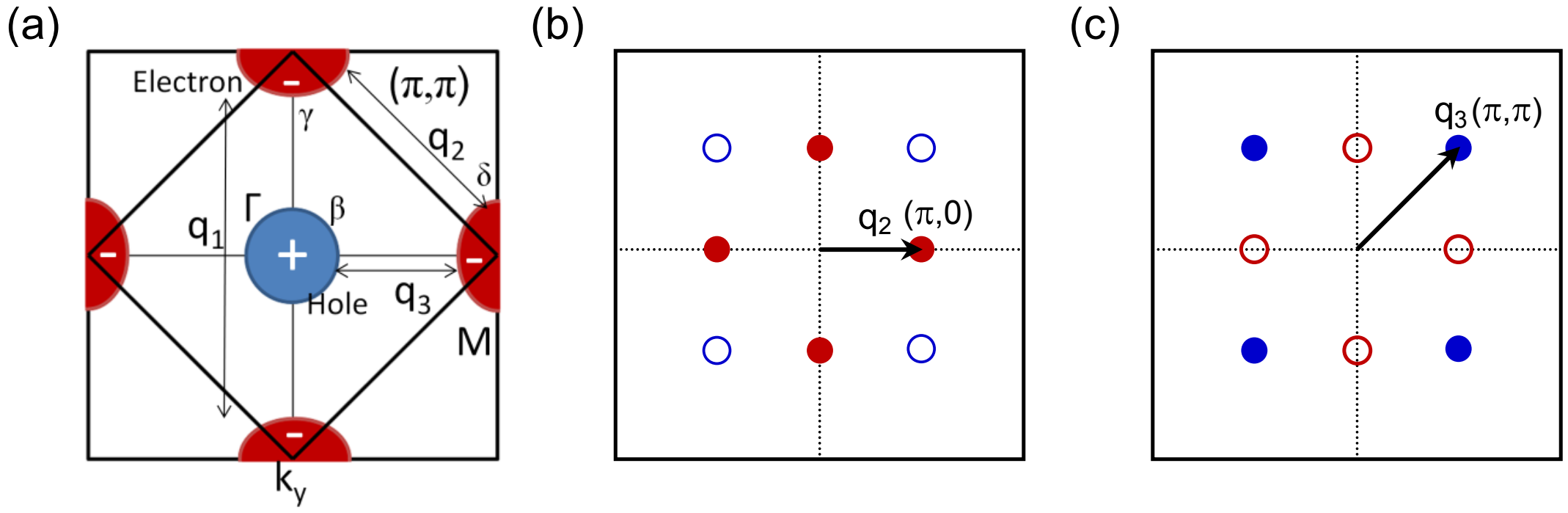

**Fig. 19: (a)** Schematics of the two-dimensional Fermi surfaces (FS) of ferrous superconductors in the one-iron unit cell, showing the presence of *a* and *b* hole pockets at the Γ-point of the Brillouin zone and electron pockets *g* and *d* at the M-points. The SC order parameters are opposite in sign for the hole and electron pockets. Possible quasiparticle interference (QPI) wave-vectors $\mathbf{q}_1$, $\mathbf{q}_2$ and $\mathbf{q}_3$ connecting different parts of the FS are indicated [227,228,231,232]. **(b)** Theoretical prediction for the zero-field QPI intensities as the result of non-magnetic impurity scattering in an $s_{\pm}$-wave superconductor, where $\mathbf{q}_2$ spots should be intense and $\mathbf{q}_3$ should be suppressed if the quasiparticle energy is equal to one of the SC gap values [227,228,231]. Further, $\mathbf{q}_1 = 2\mathbf{q}_2$ may appear due to QPI induced by the charge density wave (CDW) order, where the occurrence of CDW is associated with the AFM order [233]. **(c)** Theoretical prediction for the zero-field QPI intensities due to magnetic impurity scattering in an $s_{\pm}$-wave superconductor, where $\mathbf{q}_3$ spots are intense and $\mathbf{q}_2$ intensities are suppressed if the quasiparticle energy is equal to one of the SC gap values [227,228,231]. Alternatively, for non-magnetic impurities in the presence of magnetic fields, the intensities of $\mathbf{q}_2$ spots would be reduced whereas those of $\mathbf{q}_3$ spots would be enhanced [232].

The manifestation of two-gap superconductivity has been demonstrated by both ARPES [234-236] and STS [237-239] studies. As exemplified in Figs. 20(a)-(b) for under and over-doped $Ba(Fe_{1-x}Co_x)_2As_2$ with $x = 0.06$ and 0.12, two predominant tunneling gap features at $\Delta_\Gamma$ and $\Delta_M$ are apparent for both doping levels. For a given doing level both gaps decrease with increasing temperature and vanish above $T_c$ [237]. Additionally, the tunneling gaps exhibit particle-hole symmetry, confirming that the observed gaps are associated with superconductivity [237].

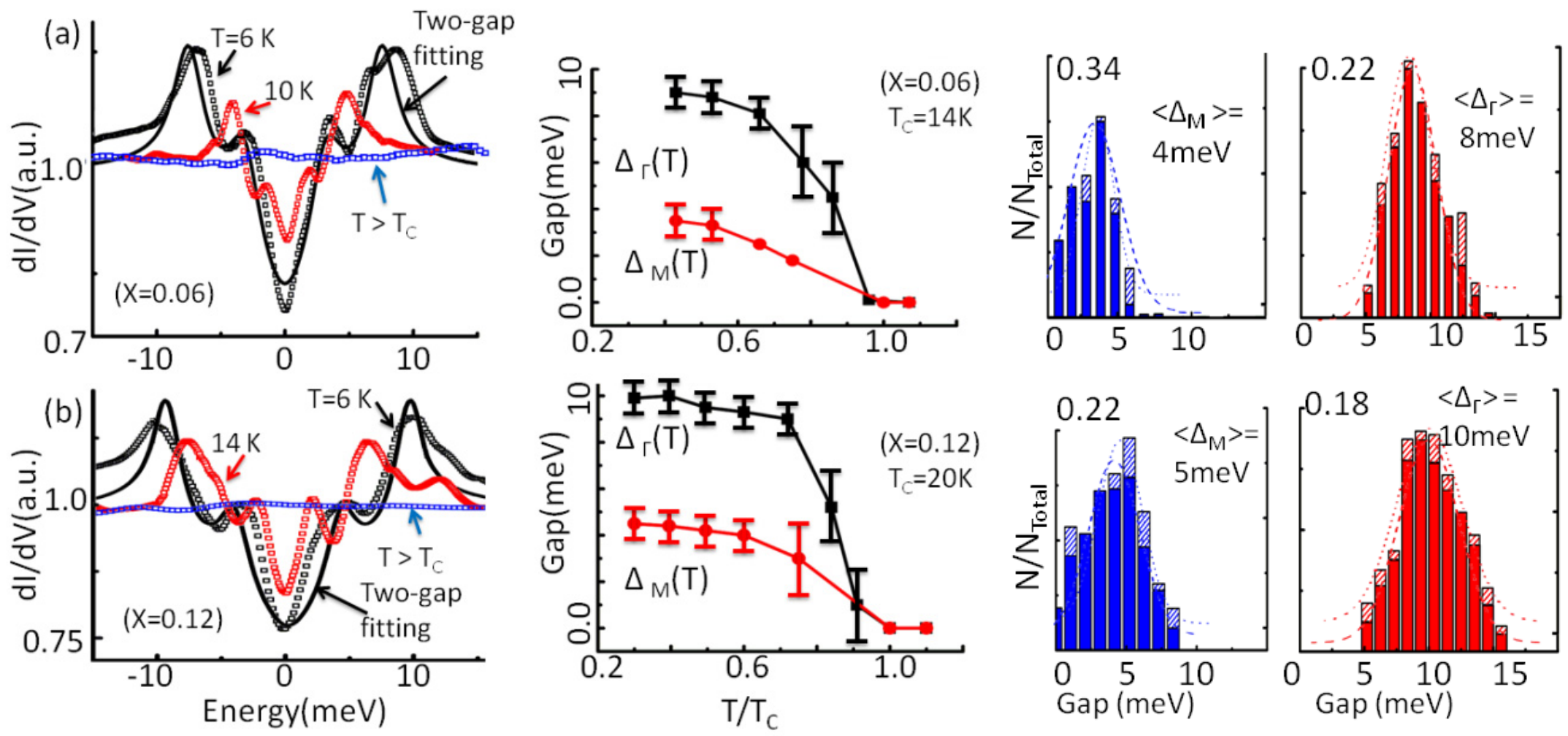

**Fig. 20:** Spectroscopic evidence for two gap superconductivity in $Ba(Fe_{1-x}Co_x)_2As_2$ [237]: **(a)** Left panel, normalized tunneling conductance ($dI/dV$) vs bias voltage ($V$) spectra taken at $T = 6$, 10, and 15 K for the sample with $x = 0.06$ and $T_c = 14$ K. The solid lines represent theoretical fittings to spectra using the Dynes formula [240] modified for two gap BCS superconductors [237]. Two distinct tunneling gaps $\Delta_\Gamma$ and $\Delta_M$ can be identified from the spectrum at $T = 6$ K, and both gaps decrease with increasing temperature and then vanish at $T > T_c$. Central panel, the tunneling gaps $\Delta_\Gamma$ and $\Delta_M$ as a function of the reduced temperature ($T/T_c$) are shown by the symbols and solid lines. Right panels, histograms for the quasiparticle (solid bars) and quasihole (shaded bars) branches, showing particle-hole symmetry and the mean values of $\langle|\Delta_M|\rangle = 4$ meV and $\langle|\Delta_\Gamma|\rangle = 8$ meV. **(b)** Left panel, normalized tunneling conductance ($dI/dV$) vs bias voltage ($V$) spectra taken at $T = 6$, 14, and 21 K for the sample with $x = 0.12$ and $T_c = 20$ K. Central panel, the tunneling gaps $\Delta_\Gamma$ and $\Delta_M$ as a function of the reduced temperature ($T/T_c$). Right panels, histograms for the quasiparticle (solid bars) and quasihole (shaded bars) branches, showing particle-hole symmetry and the mean values of $\langle|\Delta_M|\rangle = 5$ meV and $\langle|\Delta_\Gamma|\rangle = 10$ meV.

The theoretical prediction for $s_\pm$-wave pairing symmetry in the ferrous superconductors has been confirmed by the inelastic neutron scattering (INS) spectroscopy [241-244], STS [237,245] and a phase sensitive experiment [246]. In the case of INS experiments, a neutron resonance at the AFM ordering wave vector was theoretically expected below $T_c$ for $s_\pm$-wave pairing [241,242] and experimentally verified [243,244]. Specifically, the magnetic susceptibility in the SC state of a multi-band superconductor is governed by the sign change of the SC gaps at the ''hot spots'' of the Fermi surface and the following energy conservation formula for inelastic scattering of the Bogoliubov quasiparticles on the Fermi surface [247,248]:

$$\Omega^{\nu\nu'}(\mathbf{k}_F, \mathbf{q}) = |\Delta_\nu(\mathbf{k}_F)| + |\Delta_{\nu'}(\mathbf{k}_F + \mathbf{q})|, \tag{15}$$

where $\nu$ and $\nu'$ represent different energy bands, and the wave-vectors $\mathbf{q}$ are between various Fermi surface pieces with opposite signs in the SC pairing potential $\Delta_\nu(\mathbf{k}_F + \mathbf{q})$.

For the STS studies, the elastic scattering of quasiparticles by impurities will be dependent on whether the SC order parameter has opposite signs on electron and hole Fermi pockets [247,248] so that

$$\Omega^{\nu\nu'}(\mathbf{k}_F, \mathbf{q}) = |\Delta_\nu(\mathbf{k}_F)| = |\Delta_{\nu'}(\mathbf{k}_F + \mathbf{q})|. \tag{16}$$

Specifically, non-magnetic impurities will result in strong scattering of quasiparticles between the Fermi pockets of different signs in the pairing potential while suppressing the scattering between pockets of the same sign in the pairing potential [227,228,231], thus giving rise to the quasiparticle interference (QPI) patterns shown in Fig. 19(b). On the other hand, the presence of magnetic impurities or magnetic field would yield the QPI patterns shown in Fig. 19(c) [227,228,231,232]. This behavior has been confirmed in $Fe_{1+x}(Se,Te)$ compounds [244] and in $Ba(Fe_{1-x}Co_x)_2As_2$ [237].

As exemplified in Figs. 21(a) and 21(b), the FT-LDOS of underdoped $Ba(Fe_{1-x}Co_x)_2As_2$ with $x = 0.06$ in the reciprocal space for the one-iron unit cell is shown for $\omega = \Delta_{\alpha,\gamma/\delta}$ and $\omega = \Delta_\beta$, respectively. Two QPI wave-vectors $\mathbf{q}_1$ and $\mathbf{q}_2$ are identified [237], and the $\omega$-dependence of $F(\mathbf{q}_2,\omega)$ is shown in Fig. 21(c) [238]. The pronounced peaks of $F(\mathbf{q}_2,\omega)$ at $\omega = \Delta_\beta$, $\Delta_{\alpha,\gamma/\delta}$ and an energy associated with the magnetic resonance $\Omega_{r1} \sim (\Delta_\beta + \Delta_{\gamma/\delta})$ [238] following Eqs. (15) and (16) support the notion that $\mathbf{q}_2$ is associated with the QPI wave-vector between the electron- and hole-pockets rather than due to Bragg diffraction because the latter would have been $\omega$-independent. The absence of $\mathbf{q}_3$ in Figs. 21(a)-(b) further corroborates the $s_\pm$-wave pairing [237]. Moreover, the strong $\mathbf{q}_1$ intensity is consistent with the presence of charge density waves (CDW) [233].

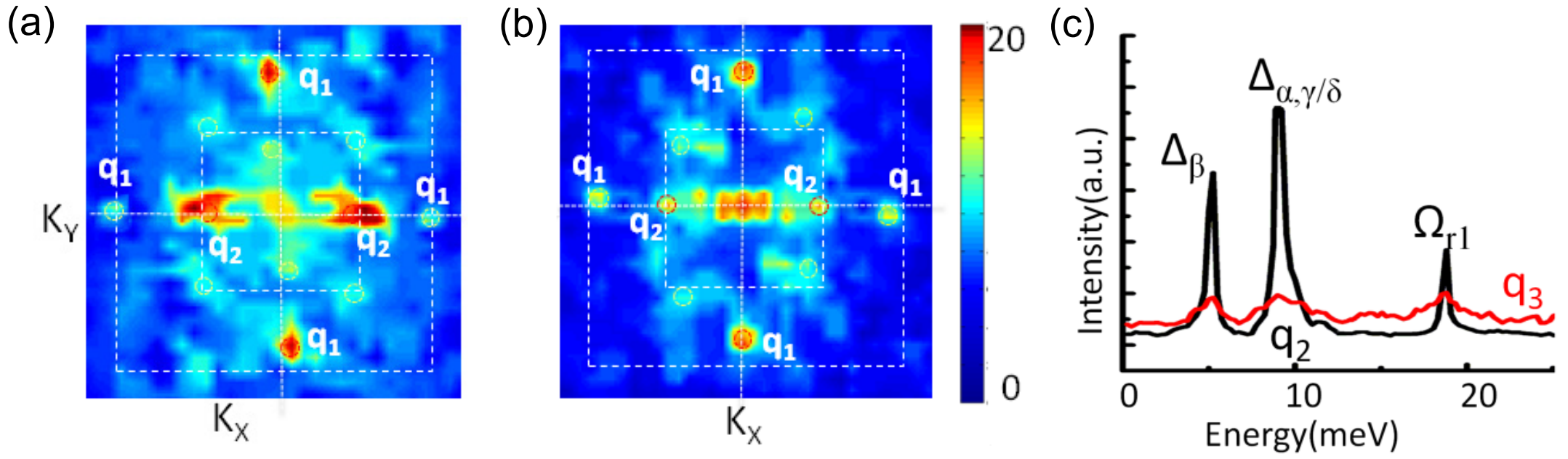

**Fig. 21:** Spectroscopic evidence for $s_\pm$-wave pairing symmetry in $Ba(Fe_{1-x}Co_x)_2As_2$ [237,238]: **(a)** FT-LDOS of an underdoped sample $Ba(Fe_{0.94}Co_{0.06})_2As_2$ for $H = 0$ at $\omega = \Delta_{\alpha,\gamma/\delta} \sim 8$ meV is shown in the 2D reciprocal space. Strong intensities at $\mathbf{q} = \mathbf{q}_2$ and $\mathbf{q}_1$ together with an additional nematic order are found [237]. The strong intensities at $\mathbf{q} = \mathbf{q}_2$ is consistent with the theoretical predictions for QPI patterns associated with $s_\pm$-wave pairing potential, whereas the strong intensities found at $\mathbf{q}_1 \sim 2\mathbf{q}_2$ are in agreement with the presence of CDW [233]. **(b)** $F(\mathbf{q},\omega)$-*vs.*-$\omega$ of underdoped $Ba(Fe_{0.94}Co_{0.06})_2As_2$ ($x = 0.06$) for $H = 0$ at $\mathbf{q} = \mathbf{q}_2$, showing sharp peaks only at $\omega = \Delta_\beta$, $\Delta_{\alpha,\gamma/\delta}$ and $\Omega_{r1}$, where $F(\mathbf{q},\omega)$ denotes the intensity of the FT-LDOS. The sharp QPI intensities occurring only at the SC gaps and magnetic resonance exclude the possibility of attributing these wave-vectors to Bragg diffractions of the reciprocal lattice vectors.

In addition to the information obtained from QPI due to elastic impurity scattering, the quasiparticle tunneling spectra also contain inelastic scattering information at higher quasiparticle energies. Indeed, STS studies of $Ba(Fe_{1-x}Co_x)_2As_2$ [237,238] single crystals have revealed spectral features that are consistent with the magnetic resonances. The characteristic energies identified from the quasiparticle tunneling spectra of $Ba(Fe_{1-x}Co_x)_2As_2$ and attributed to the magnetic resonances in these samples are found to satisfy the relationship $\Omega_{r1} \sim (\Delta_\beta + \Delta_{\gamma/\delta}) \sim 1.5\Delta_{\alpha,\gamma/\delta} \sim 3\Delta_\beta$ and $\Omega_{r2} \sim (\Delta_\alpha + \Delta_{\gamma/\delta}) \sim 2\Delta_{\alpha,\gamma/\delta}$ [238]. Therefore, only one magnetic resonance $\Omega_{r1}$ is observed for optimally and overdoped $Ba(Fe_{1-x}Co_x)_2As_2$ due to vanished $\alpha$-pocket, whereas $\Omega_{r2} \sim (16\pm1)$ meV for underdoped $Ba(Fe_{1-x}Co_x)_2As_2$. The correlation between the SC gaps and magnetic resonances is manifested in Figs. 22(a)-(c) for three different doping levels. These findings from the STS studies for the magnetic resonances are consistent with the observation in INS experiments [244,247], which again confirm the presence of sign changes in the pairing potential associated with different disconnected Fermi surfaces [248].

In general, the magnetic resonance behavior directly probed by the INS spectroscopy provides valuable information about the pairing mechanism of unconventional superconductors. In the cuprate superconductors, INS exhibits a clear signature of a resonance mode that becomes strongly enhanced below $T_c$ in addition to its characteristic dispersion known as the ‘‘hourglass’’ behavior, and the resonant energy scales universally with the SC gap amplitude [249]. The generalization of the observed magnetic resonances from the single-band cuprate superconductors to the multi-band multi-gap ferrous superconductors [238,248] points to the importance of spin fluctuations in the SC state of these unconventional superconductors.

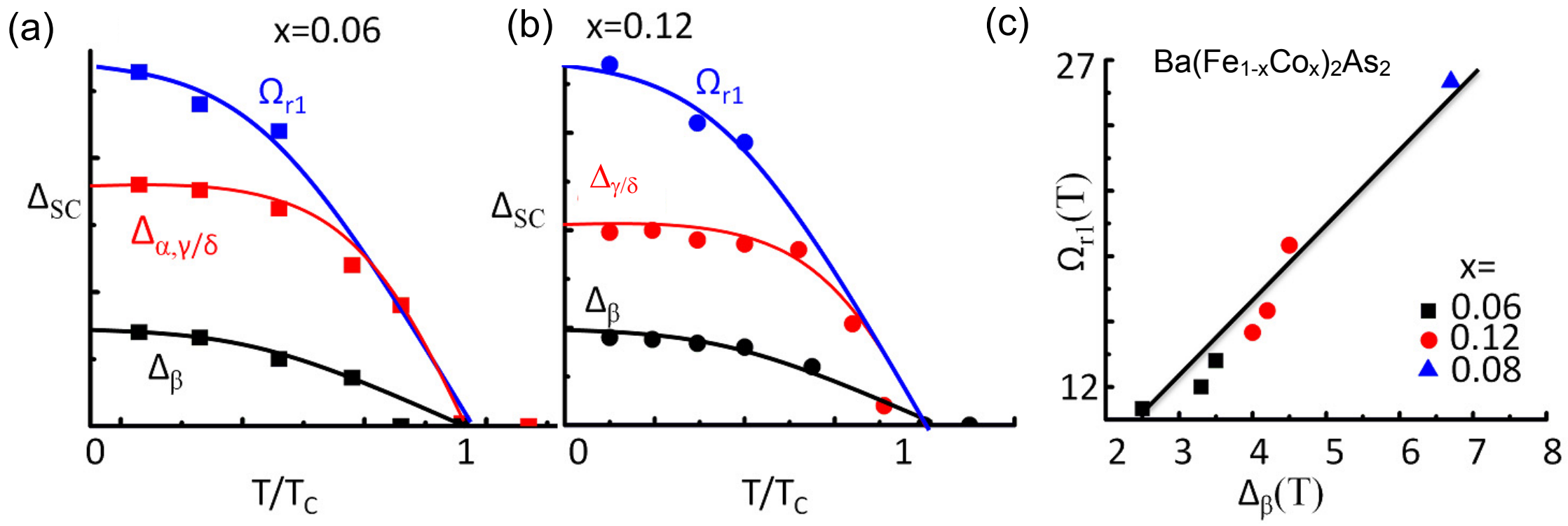

**Fig 22:** Correlation between the SC gaps $\Delta_{\alpha,\gamma/\delta}$, $\Delta_\beta$ and the magnetic resonant mode $\Omega_{r1}$ of the $Ba(Fe_{1-x}Co_x)_2As_2$ superconductors [238]: (a) $T$-dependence of $\Delta_{\alpha,\gamma/\delta}$, $\Delta_\beta$ and $\Omega_{r1}$ for the underdoped sample ($x$ = 0.06). (b) $T$-dependence of $\Delta_{\gamma/\delta}$, $\Delta_\beta$ and $\Omega_{r1}$ for the overdoped sample ($x$ = 0.12). (c) Correlation of the magnetic resonant mode $\Omega_{r1}(T)$ with the SC gap $\Delta_\beta(T)$ for Co-122 samples with three different doping levels of $x$ = 0.06, 0.08 and 0.12. The slope agrees with the relation $\Omega_{r1} \sim 3\Delta_\beta \sim 1.5\Delta_{\gamma/\delta}$. This universal relationship among samples of $Ba(Fe_{1-x}Co_x)_2As_2$ with doping levels and at different temperatures below $T_c$ suggests the relevance of AFM spin fluctuations to the Cooper pairing in these superconductors [238,248].

### 3.3. Vortex-State Characteristics

As mentioned in Section 3.1 and exemplified in Figs. 18(c)-(e), there are a number of non-universal properties among different ferrous superconductors. For instance, the SC state appears to be exclusive of the AFM phase in the 1111 system, whereas the electron-type 122 system appears to have coexisting SC and AFM phases for a finite range of doping levels, similar to the finding of competing orders in cuprate superconductors. In this context, spatially resolved vortex-state STS studies of different ferrous superconductors should exhibit different types of intra-vortex quasiparticle spectra, depending on whether AFM coexists with SC in the ground state. Such results can also provide useful comparison of the iron-based superconductors with the cuprates.

To date there have only been two published reports with varying findings from the vortex-state STS studies of the ferrous superconductors [239,250]. In one report, asymmetric vortex bound states appearing as subgap peaks inside the vortex cores were observed in a hole-type 122 system $(Ba_{0.6}K_{0.4})Fe_2As_2$ [239], which implies pure SC both CO in this compound. In contrast, STS studies of an electron-type 122 system $Ba(Fe_{0.9}Co_{0.1})_2As_2$ found complete suppression of SC coherence peaks but no apparent subgap peaks inside the vortex cores [250]. However, in the latter case of investigation, the zero-field tunneling spectra of the specific sample only revealed one spatially varying SC gap for all scanned areas [250], which differed from the expected two-gap SC characteristics [237-239] that have been exemplified in Fig. 20. The missing two-gap phenomena may be the result of surface reconstructions due to the reactive nature of the sample surface. Therefore, it is not conclusive whether vortex bound states or pseudogap features exist inside the vortex core of the electron-type $Ba(Fe_{1-x}Co_x)_2As_2$ system. Moreover, whether the vortex core states of both the electron-type and hole-type 122 systems may exhibit doping dependence remains an open issue because no doping dependence of the vortex core states has been investigated. A comprehensive STS experimental survey of the doping dependent vortex-state spectral characteristics in various families of iron-based superconductors can provide useful information about possible correlations between the vortex-core states and the existence of AFM spin fluctuations. These studies are nonetheless quite challenging, because STS experiments are extremely time consuming, whereas the surface layers of most ferrous superconductors are very reactive and are prone to surface reconstructions as well as surface degradation with time.

## 4. Implications for the High-Temperature Superconducting Mechanism

In this section we discuss the physical implications on the pairing mechanism of high-temperature superconductivity based on the aforementioned comparative studies of the cuprate and ferrous superconductors.

The quest for the pairing mechanism of high-temperature superconductivity [2-4,251,252] and the debate over the role of electron-phonon coupling [196,253] has never ceased since the discovery of cuprate superconductors. Based on comprehensive experimental surveys in both the cuprate and ferrous superconductors, there are strong evidences that the relevant pairing interaction must be repulsive in these two classes of high-temperature superconductors. In fact, the same situation applies to the heavy-fermion superconductors that also exhibit unconventional pairing symmetry, spin resonances, layered structures, competing orders, and quantum criticality [254-263]. Hence, the pairing mechanism for all these unconventional superconductors must involve sign-changing pairing potentials in different parts of the Fermi surfaces [123,124,237,238,245]. The apparent link between the Fermi surface topology and the sign-reversal SC pairing potential [222,248] as well as the proportionality between the SC transition temperature $T_c$ and the spin fluctuation temperature $T_0$ [249,264] for all these unconventional superconductors strongly suggests that the presence of a magnetic mode mediates the Cooper pairing via a repulsive interaction. In this context, the attractive and small-magnitude pairing energy mediated by the electron-phonon coupling is unlikely the pairing mechanism for high-temperature superconductivity. Rather, the pairing mechanism would favor repulsive electronic interactions with Coulomb like correlations, and it is tempting to suggest that stronger correlated cuprates with an insulating parent state would yield larger pairing potentials and therefore higher $T_c$ values than the iron-based compounds that have a semi-metallic parent state.

However, the formation of Cooper pairing in unconventional superconductors must involve a subtle balance between the large repulsive interaction and the tendency to localize charge carriers [220]. In particular, the repulsive interaction must be gingerly arranged among pairs of carriers in different parts of the Fermi surfaces with opposite signs in the pairing potential, which may be achieved by either pairing the carriers in the same band with an orbital angular momentum larger than 0 [118], or pairing the carriers with zero orbital angular momentum in different bands of opposite signs in potential [229]. The latter pairing arrangement is consistent with most iron-based superconductors other than a special class of iron chalcogenides $A_xFe_{2-y}Se_2$, whereas the former pairing scenario is consistent with the situation in cuprate superconductors and also in $A_xFe_{2-y}Se_2$ where the hole pockets are found to completely vanish and so nodeless *d*-wave pairing has been predicted theoretically [265,266]. In addition to the orbital degree of freedom for pairing, AFM coupling is most favorable for the spin degree of freedom to achieve singlet Cooper pairs. Thus, a parent system with AFM order appears to be an important common feature among the cuprate, ferrous and heavy-fermion superconductors. While long-range AFM must be suppressed before SC can appear, dynamic AFM fluctuations [252] may be favorable for high-temperature superconductivity. Moreover, the cuprate, ferrous and heavy-fermion superconductors all exhibit very similar doping dependent phase diagrams as illustrated in Fig. 1(a) and Figs. 18(c)-(e). The proximity of SC to AFM instabilities implies the relevance of quantum criticality and quantum fluctuations to the occurrence of SC. Finally, all these three types of unconventional superconductors exhibit quasi-two dimensionalities, with effective layered structures playing the key role in superconductivity.

On the other hand, the aforementioned clever arrangements for pairing the carriers in different parts of the Fermi surfaces under a repulsive interaction are not sufficient to yield superconductivity, because the complexity of many-body interactions in a correlated electronic system could favor different instabilities from superconductivity upon lowering the temperature. It is therefore not difficult to understand the complications of competing orders in the underdoped cuprate superconductors where the electronic correlations are most significant. Similarly, in the case of heavy-fermion superconductors where the $T_c$ values are much lower than those of the cuprate and ferrous superconductors, long-range magnetic orders are generally prevailing at higher temperatures and superconductivity cannot appear until the temperature lowers below the Kondo temperature so that localized magnetic moments become completely shielded by conduction electrons. Thus, the occurrence of competing orders in these unconventional superconductors seems to be a natural consequence of strong electronic correlations rather than a necessary condition for high temperature superconductivity.

Although the electron-phonon interaction is unlikely the “glue” for Cooper pairs in the cuprate and ferrous superconductors, the presence of electron-phonon coupling under special circumstances may assist stronger charge transfer and thus enhance the electronic density of states at the Fermi level and result in an increase of $T_c$ under favorable conditions [196,253]. For instance, in the underdoped hole-type cuprate superconductors, isotope effects on $T_c$ are found to decrease with increasing doping and then completely vanish at the optimal doping [267-271]. In contrast, no isotope effects have been reported in any of the electron-type cuprate superconductors. Such findings may be attributed to the slower carrier mobility in underdoped cuprate superconductors so that coupling between typically faster moving carriers in the dynamic mixture of $3d^9 2p^5$ and $3d^9 2p^6$ electronic configurations to the slower-moving longitudinal optical phonon modes along the Cu-O bond becomes possible. This stronger electron-phonon coupling could assist better charge transfer along the anti-nodal direction of the pairing potential [196]. In contrast, the electronic configurations of electron-doped cuprate superconductors consist of a dynamic mixture of $3d^{10} 2p^6$ and $3d^9 2p^6$, which is less favorable for optical phonon-assisted charge transfer along the Cu-O bond [7] and therefore is consistent with the empirical finding of insignificant isotope effect. Overall, it appears that electron-phonon coupling is unlikely to be solely responsible for the occurrence of high-temperature superconductivity, at least not in the case of the cuprate, ferrous and heavy-fermion superconductors, even though in some special cases the coupling may help enhance the superconducting transition.

The realization of repulsive pairing potentials and competing orders appears to have settled many puzzling issues and debates in the cuprate and ferrous superconductivity. However, a conclusive experiment to unambiguously point to the AFM spin fluctuations as the mediator for Cooper pairing in these high-temperature superconductors is yet to be devised. Moreover, in search for a good "recipe" for high-temperature superconductivity, it is important to ask whether other pairing mechanisms different from the AFM spin fluctuations may be viable candidates. These profound and yet unsettled issues will certainly keep the research field of superconductivity intellectually challenging and exciting.

## 5. Conclusion

We have reviewed the experimental findings and the corresponding theoretical understandings of various unconventional low-energy excitation phenomena in two types of high-temperature superconductors, the cuprate and iron-based superconductors. In the cuprate superconductors that are known as doped Mott insulators with strong electronic correlations and are in close proximity to an antiferromagnetic (AFM) instability, sign-changing unconventional $d_{x^2-y^2}$-wave and $(d_{x^2-y^2}+s)$-wave pairing symmetries are established among different cuprates with varying doping levels, and the $s$-component appears to increase with increasing doping. The unconventional $d_{x^2-y^2}$-wave and $(d_{x^2-y^2}+s)$-wave pairing symmetries minimize the on-site Coulomb repulsion, and have significant consequences on the low-energy excitations and impurity-induced quasiparticle scattering in the cuprates. The strong correlation in the cuprates and their proximity to an AFM instability result in coexisting superconductivity (SC) with various competing orders (COs) in the ground state, yielding non-universal phenomena (such as the pseudogap and Fermi arc phenomena) among different cuprates as well as the occurrence of quantum criticality, strong quantum fluctuations, and weakened superconducting stiffness. A phenomenology based on coexisting COs and SC in the cuprates appears to provide consistent account for a wide range of experimental findings, and is also compatible with the possibility of pre-formed Cooper pairs and significant phase fluctuations in cuprate superconductors.

In the case of iron-based superconductors whose parent states are AFM semi-metals, studies of the low-energy quasiparticle and spin excitations reveal unconventional sign-changing $s_{\pm}$-wave or nodeless $d$-wave pairing symmetries with two SC gaps and two magnetic resonant modes that scale with the SC gaps. Our comparative studies therefore suggest that the commonalities among the cuprate and the ferrous superconductors include the proximity to AFM instabilities, the existence of AFM spin fluctuations and magnetic resonances in the SC state, the unconventional pairing symmetries with sign-changing order parameters on different parts of the Fermi surfaces, the layered structures, and the appearance of multi-channel low-energy excitations in the SC state either due to COs as in the cuprates or due to multi-band pairing as in the iron-based superconductors. These common features imply that the pairing potential in these high-temperature superconductors is repulsive and therefore is predominantly electronic in nature. Moreover, the apparent link between the Fermi surface topology and the sign-reversal SC pairing potential as well as the proportionality between the SC transition temperature $T_c$ and the spin fluctuation temperature $T_0$ strongly suggests that the Cooper pairing in these high-temperature superconductors is mediated by a magnetic mode through repulsive interactions. Although under special circumstances the electron-phonon interaction may help enhance the $T_c$ value, it seems that the attractive and relatively small electron-phonon interaction is unlikely the sole pairing mechanism for high-$T_c$ superconductivity.

In the context of singlet pairing through repulsive electronic interaction, it is tempting to suggest that the strongly correlated cuprates with an insulating parent state are likely to acquire larger pairing potentials and higher SC transition temperatures, whereas the semi-metallic parent state of the iron-based superconductors may result in overall lower $T_c$ values relative to the cuprates. On the other hand, strong electronic correlations have the tendency to localize carriers and/or to induce other instabilities than superconductivity. Hence, proper balance between the electronic correlation and the itinerancy of pairs is essential to the occurrence of high-temperature superconductivity. Finally, whether pairing mechanisms other than spin fluctuations may be feasible for high-temperature superconductivity remains an open issue for exploration.

## Acknowledgement

Much of the research results presented here are in collaboration with Marcus L. Teague, Dr. Andrew D. Beyer, Dr. C.-T. Chen, M. S. Grinolds, Professor Setsuko Tajima, Professor Hai-Hu Wen and Professor Jochen Mannhart. The author also acknowledges research support from the National Science Foundation and the Kavli Foundation through the facilities at the Kavli Nanoscience Institute at Caltech.

## References

[1] J. G. Bednorz and K. A. Muller, Z. Phys. B **64**, 189 (1986).
[2] Y. Kamihara, T. Watanabe, M. Hirano and H. Hosono, J. Am. Chem. Soc. **130**, 3296 (2008).
[3] P. A. Lee, N. Nagaosa and X.-G. Wen, Rev. Mod. Phys. **78**, 17 (2006); and references therein.
[4] S. A. Kivelson, I. P. Bindloss, E. Fradkin, V. Oganesyan, J. M. Tranquada, A. Kapitulnik, and C. Howald, Rev. Mod. Phys. **75**, 1201 (2003).
[5] S. Sachdev, Rev. Mod. Phys. **75**, 913 (2003).
[6] E. Demler, W. Hanke and S.-C. Zhang, Rev. Mod. Phys. **76**, 909 (2004).
[7] N.-C. Yeh and A. D. Beyer, Int. J. Mod. Phys. B **23**, 4543 (2009).
[8] M. A. Kastner, R. J. Birgeneau, G. Shirane and Y. Endoh, Rev. Mod. Phys. **70**, 897 (1998).
[9] M. B. Maple, MRS Bulletin **15**(6), 60 (1990).
[10] N.-C. Yeh, A. D. Beyer, M. L. Teague, S.-P. Lee, S. Tajima and S. I. Lee, J. Supercond. Nov. Magn. **23**, 757 (2010).
[11] A. D. Beyer, M. S. Grinolds, M. L. Teague, S. Tajima and N.-C. Yeh, Europhys. Lett. **87**, 37005 (2009).
[12] C.-T. Chen, A. D. Beyer and N.-C. Yeh, Solid State Commun. **143**, 447 (2007).
[13] A. D. Beyer, C.-T. Chen and N.-C. Yeh, Physica C **468**, 471 (2008).
[14] N.-C. Yeh, A. D. Beyer, M. L. Teague, S.-P. Lee, S. Tajima, and S. I. Lee, Int. J. Mod. Phys. B **19**, 285 (2005).
[15] N.-C. Yeh, C. T. Chen, A. D. Beyer and S. I. Lee, Chinese J. Phys. **45**, 263 (2007).
[16] B.-L. Yu, J. Wang, A. D. Beyer, M. L. Teague, J. S. A. Horng, S.-P. Lee and N.-C. Yeh, Solid State Commun. **149**, 261 (2009).
[17] M. L. Teague, A. D. Beyer, M. S. Grinolds, S. I. Lee and N.-C. Yeh, Europhys. Lett. **85**, 17004 (2009).
[18] T. Timusk and B. Statt, Rep. Prog. Phys. **62**, 61 (1999); and references therein.
[19] A. Damascelli, Z. Hussain and Z.-X. Shen, Rev. Mod. Phys. **75**, 473 (2003).
[20] Ø. Fischer, Ø. Fischer, M. Kugler, I. Maggio-Aprile, C. Berthod and Ch. Renner, Rev. Mod. Phys. **79**, 353 (2007); and references therein.
[21] W. S. Lee, I. M. Vishik, K. Tanaka, D. H. Lu, T. Sasagawa, N. Nagaosa, T. P. Devereaux, Z. Hussain and Z.-X. Shen, Nature **450**, 81 (2007).
[22] Ch. Renner, B. Revaz, J.-Y. Genoud, K. Kadowaki, and Ø. Fischer, Phys. Rev. Lett. **80**, 149 (1998).
[23] M. Vershinin, S. Misra, S. Ono, Y. Abe, Y. Ando and A. Yazdani, Science **303**, 1995 (2004).
[24] M. C. Boyer, W. D. Wise, K. Chatterjee, M. Yi, T. Kondo, T. Takeuchi, H. Ikuta, Y. Wang, and E. W. Hudson, Nature Phys. **3**, 802 (2007).
[25] K. K. Gomes, A. N. Pasupathy, A. Pushp, S. Ono, Y. Ando and A. Yazdani, Nature **447**, 569 (2007).
[26] R. E. Walstedt, R. F. Bell and D. B. Mitzi, Phys. Rev. B **44**, 7760 (1991).
[27] R. Stern, M. Mali, I. Mangelschots, J. Roos, D. Brinkmann, Phys. Rev. B **50**, 426 (1994).
[28] R. Stern, M. Mali, J. Roos, D. Brinkmann, J. Y. Genoud, T. Graf, and J. Muller, Phys. Rev. B **52**, R15734 (1995).
[29] R. L. Corey, N. J. Curro, K. OHara, T. Imai, C. P. Slichter, K. Yoshimura, M. Katoh and K. Kosuge, Phys. Rev. B **53**, 5907 (1996).
[30] M.-H. Julien, P. Carretta, M. Horvatic, C. Berthier, Y. Berthier, P. Segransan, A. Carrington and D. Colson, Phys. Rev. Lett. **76**, 4238 (1996).

[31] Y.-W. Hsueh, B. W. Statt, M. Reedyk, J. S. Xue and J. E. Greedan, Phys. Rev. B **56**, R8511 (1997).
[32] A. V. Puchkov, P. Fournier, D. N. Basov, T. Timusk, A. Kapitulnik and N. N. Kolesnikov, Phys. Rev. Lett. **77**, 3212 (1996).
[33] M. Opel, R. Nemetschek, C. Hoffmann, R. Philipp, P. F. Müller, R. Hackl, I. Tüttő, A. Erb, B. Revaz, and E. Walker, H. Berger and L. Forró, Phys. Rev. B **61**, 9752 (2000).
[34] M. LeTacon, A. Georges, G. Kotliar, Y. Gallais, D. Colson, and A. Forget, Nature Phys. **2**, 537 (2006).
[35] C.-T. Chen, P. Seneor, N.-C. Yeh, R. P. Vasquez, L. D. Bell, C. U. Jung, J. Y. Kim, Min-Seok Park, Heon-Jung Kim, and Sung-Ik Lee, Phys. Rev. Lett. **88**, 227002 (2002).
[36] H. Matsui, K. Terashima, T. Sato, T. Takahashi, M. Fujita, and K. Yamada, Phys. Rev. Lett. **95**, 017003 (2005).
[37] S. Kleefisch, B. Welter, A. Marx, L. Alff, R. Gross and M. Naito, Phys. Rev. B **63**, 100507 (2001).
[38] L. Alff, Y. Krockenberger, B. Welter, M. Schonecke, R. Gross, D. Manske, and M. Naito, Nature **422**, 698 (2003).
[39] G. Blumberg, A. Koitzsch, A. Gozar, B. S. Dennis, C. A. Kendziora, P. Fournier and R. L. Greene, Phys. Rev. Lett. **88**, 107002 (2002).
[40] K. Yamada, S. Wakimoto, G. Shirane, C.-H. Lee, M. A. Kastner, S. Hosoya, M. Greven, Y. Endoh and R. J. Birgeneau, Phys. Rev. Lett. **75**, 1626 (1995).
[41] B. Lake, H. M. Ronnow, N. B. Christensen, G. Aeppli, K. Lefmann, D. F. McMorrow, P. Vorderwisch, P. Smeibidl, N. Mangkorntong, T. Sasagawa, M. Nohara, H. Takagi and T. E. Mason, Nature **415**, 299 (2002).
[42] C.-H. Lee, K. Yamada, Y. Endoh, G. Shirane, R. J. Birgeneau, M. A. Kastner, M. Greven and Y. J. Kim, J. Phys. Soc. Japan **69**, 1170 (2000).
[43] K. Yamada, K. Kurahashi, T. Uefuji, M. Fujita, S. Park, S.-H. Lee, and Y. Endoh, Phys. Rev. Lett. **90**, 137004 (2003).
[44] M. Fujita, M. Matsuda, S. Katano and K. Yamada, Phys. Rev. Lett. **93**, 147003 (2004).
[45] H. J. Kang et al., Phys. Rev. B **71**, 214512 (2005).
[46] Y. Onos, Y. Taguchi, K. Ishizaka and Y. Tokura, Phys. Rev. Lett. **87**, 217001 (2001).
[47] Y. Gallais, A. Sacuto, T. P. Devereaux and D. Colson, Phys. Rev. B **71**, 012506 (2005).
[48] K. Yamada, C. H. Lee, K. Kurahashi, J. Wada, S. Wakimoto, S. Ueki, H. Kimura, Y. Endoh, S. Hosoya, G. Shirane, R. J. Birgeneau, M. Greven, M. A. Kastner and Y. J. Kim, Phys. Rev. B **57**, 6165 (1998).
[49] J. M. Tranquada, B. J. Sternlieb, J. D. Axe, Y. Nakamura, and S. Uchida, Nature **375**, 561 (1995).
[50] J. M. Tranquada, J. D. Axe, N. Ichikawa, A. R. Moodenbaugh, Y. Nakamura and S. Uchida, Phys. Rev. Lett. **78**, 338 (1997).
[51] B. O. Wells, Y. S. Lee, M. A. Kastner, R. J. Christianson, R. J. Birgeneau, K. Yamada, Y. Endoh, and G. Shirane, Science **277**, 1067 (1997).
[52] B. Lake, G. Aeppli, K. N. Clausen, D. F. McMorrow, K. Lefmann, N. E. Hussey, N. Mangkorntong, M. Nohara, H. Takagi, T. E. Mason, and A. Schroder, Science **291**, 1759 (2001).
[53] H. A. Mook, P. C. Dai, and F. Dogan, Phys. Rev. Lett. **88**, 097004 (2002).
[54] M. Fujita, H. Goka, K. Yamada and M. Matsuda, Phys. Rev. Lett. **88**, 167008 (2002).
[55] M. Matsuda, S. Katano, T. Uefuji, M. Fujita, and K. Yamada, Phys. Rev. B **66**, 172509 (2002).
[56] K. Tokiwa, H. Okumoto, T. Imamura, S. Mikusu, K. Yuasa, W. Higemoto, K. Nishiyama, A. Iyo, Y. Tanaka, and T. Watanbe, Int. J. Mod. Phys. B **17**, 3540 (2003).
[57] H. Kotegawa, Y. Tokunaga, K. Ishida, G.-q. Zheng, Y. Kitaoka, K. Asayama, H. Kito, A. Iyo, H. Ihara, K. Tanaka, K. Tokiwa, T. Watanabe, J. Phys. Chem. Solids **62**, 171 (2001).
[58] H. Kotegawa Y. Tokunaga, K. Ishida, G.-q. Zheng, Y. Kitaoka, H. Kito, A. Iyo, K. Tokiwa, T. Watanabe, Phys. Rev. B **64**, 064515 (2001).
[59] T. Kondo, T. Takeuchi, A. Kaminski, S. Tsuda, and S. Shin, Phys. Rev. Lett. **98**, 267004 (2007).
[60] M. Hashimoto, R.-H. He, K. Tanaka, J.-P. Testaud, W. Meevasana, R. G. Moore, D. Lu, Ho. Yao, Y. Yoshida, H. Eisaki, T. P. Devereaux, Z. Hussain and Z.-X. Shen, Nat. Phys. **6**, 414 (2010).
[61] J. E. Hoffman, E. W. Hudson, K. M. Lang, V. Madhavan, H. Eisaki, S. Uchida, and J. C. Davis, Science **295**, 466 (2002).
[62] C. Howald, H. Eisaki, N. Kaneko, M. Greven and A. Kapitulnik, Phys. Rev. B **67**, 014533 (2003).
[63] T. Hanaguri, C. Lupien, Y. Kohsaka, D.-H. Lee, M. Azuma, M. Takano, H. Takagi, and J. C. Davis, Nature **430**, 1001 (2004).
[64] W. D. Wise, M. C. Boyer, K. Chatterjee, T. Kondo, T. Takeuchi, H. Ikuta, Y. Wang, and E. W. Hudson, Nature Phys. **4**, 696 (2008).
[65] S.-C. Zhang, Science **275**, 1089 (1997).
[66] M. Vojta, Y. Zhang and S. Sachdev, Phys. Rev. B **62**, 6721 (2000).
[67] C. M. Varma, Phys. Rev. B **55**, 14554 (1997).
[68] E. Demler, S. Sachdev and Y. Zhang, Phys. Rev. Lett. **87**, 067202 (2001).
[69] A. Polkovnikov, M. Vojta and S. Sachdev, Phys. Rev. B **65**, 220509(R) (2002).

[70] Y. Chen, H. Y. Chen and C. S. Ting, Phys. Rev. B **66**, 104501 (2002).
[71] H. D. Chen, J.-P. Hu, S. Capponi, E. Arrigoni, and S.-C. Zhang, Phys. Rev. Lett. **89**, 137004 (2002).
[72] H.-D. Chen et al., Phys. Rev. Lett. **93**, 187002 (2004).
[73] S. Chakravarty, R. B. Laughlin, D. K. Morr and C. Nayak, Phys. Rev. B **63**, 094503 (2001).
[74] U. Schollwöck, S. Chakravarty, J. O. Fjærestad, J. B. Marston, and M. Troyer, Phys. Rev. Lett. **90**, 186401 (2003).
[75] J. X. Li, C. Q.Wu and D.-H. Lee, Phys. Rev. B **74**, 184515 (2006).
[76] S. Chakravarty, H.-Y. Kee, and K. Volker, Nature (London) **428**, 53 (2004).
[77] A. D. Beyer, V. S. Zapf, H. Yang, F. Fabris, M. S. Park, K. H. Kim, S.-I. Lee, and N.-C. Yeh, Phys. Rev. B **76**, 140506(R) (2007).
[78] V. S. Zapf, N.-C. Yeh, A. D. Beyer, C. R. Hughes, C. Mielke, N. Harrison, M.-S. Park, K.-H. Kim, and S.-I. Lee, Phys. Rev. B **71**, 134526 (2005).
[79] V. J. Emery and S. A. Kivelson, Nature **374**, 434 (1995).
[80] J. Corson, R. Mallozzi, J. Orenstein and J. N. Eckstein, Nature **398**, 221 (1999).
[81] G. Blatter, M. V. Feigel'man, V. B. Geshkenbein, A. I. Larkin, and V. M. Vinokur, Rev. Mod. Phys. **66**, 1125 (1994).
[82] D. S. Fisher, M. P. A. Fisher and D. Huse, Phys. Rev. B **47**, 130 (1991).
[83] D. R. Nelson and V. M. Vinokur, Phys. Rev. Lett. **68**, 2398 (1992).
[84] D. R. Nelson and V. M. Vinokur, Phys. Rev. B **48**, 13060 (1993).
[85] L. Balents and D. R. Nelson, Phys. Rev. Lett. **73**, 2618 (1994).
[86] T. Giamarchi and P. Le Doussal, Phys. Rev. Lett. **72**, 1530 (1994).
[87] T. Giamarchi and P. Le Doussal, Phys. Rev. B **52**, 1242 (1995).
[88] N.-C. Yeh, D. S. Reed, W. Jiang, U. Kriplani, C. C. Tsuei, C. C. Chi, and F. Holtzberg, Phys. Rev. Lett. **71**, 4043 (1993).
[89] N.-C. Yeh, D. S. Reed, W. Jiang, U. Kriplani, M. Konczykowski, F. Holtzberg, C. C. Tsuei and C. C. Chi, Physica A **200**, 374 (1993).
[90] N.-C. Yeh, D. S. Reed, W. Jiang, U. Kriplani, and F. Holtzberg, Physica C **235--240**, 2659 (1994).
[91] D. S. Reed, N.-C. Yeh, W. Jiang, U. Kriplani, and F. Holtzberg, Phys. Rev. B **47**, 6150 (1993).
[92] W. Jiang, N.-C. Yeh, D. S. Reed, U. Kriplani, T. A. Tombrello, A. P. Rice and F. Holtzberg, Phys. Rev. B **47**, 8308 (1993).
[93] D. S. Reed, N.-C. Yeh, W. Jiang, U. Kriplani, D. A. Beam and F. Holtzberg, Phys. Rev. B **49**, 4384 (1994).
[94] W. Jiang, N.-C. Yeh, D. S. Reed, D. A. Beam, U. Kriplani, M. Konczykowski, and F. Holtzberg, Phys. Rev. Lett. **72**, 550 (1994).
[95] D. S. Reed, N.-C. Yeh, M. Konczykowski, A. V. Samoilov, and F. Holtzberg, Phys. Rev. B **51**, 16448 (1995).
[96] N.-C. Yeh, D. S. Reed, W. Jiang, U. Kriplani, D. A. Beam, M. Konczykowski, F. Holtzberg, and C. C. Tsuei, in "Advances in Superconductivity – VII", Vol. **1**, pg. 455--461, Spinger-Verlag, Tokyo (1995).
[97] N.-C. Yeh, W. Jiang, D. S. Reed, U. Kriplani, M. Konczykowski, F. Holtzberg, and C. C. Tsuei, Ferroelectrics **177**, pg. 143--159 (1996).
[98] N. Doiron-Leyraud, D. LaBoeuf, J. Levallois, J.-B. Bonnemaison, R. Liang, D. A. Bonn, W. N. Hardy, C. Proust, and L. Taillefer, Nature (London) **447**, 565 (2007).
[99] A. Bangura, J. D. Fletcher, A. Carrington, J. Levallois, M. Nardone, B. Vignolle, D. J. Heard, N. Doiron-Leyraud, D. LaBoeuf, L. Taillefer, S. Adachi, C. Proust, and N. E. Hussey, Phys. Rev. Lett. **100**, 047004 (2008).
[100] E. A. Yelland, J. Singleton, C. H. Mielke, N. Harrison, F. F. Balakirev, B. Dabrowski, and J. R. Cooper, Phys. Rev. Lett. **100**, 047003 (2008).
[101] C. Jaudet, D. Vignolles, A. Audouard, J. Levallois, D. LaBoeuf, N. Doiron-Leyraud, B. Vignolle, M. Nardone, A. Zitouni, R. Liang, D. A. Bonn, W. N. Hardy, L. Taillefer, and C. Proust, Phys. Rev. Lett. **100**, 187005 (2008).
[102] S. E. Sebastian, N. Harrison, E. Palm, T. P. Murphy, C. H. Mielke, R. Liang, D. A. Bonn, W. N. Hardy, and G. G. Lonzarich, Nature (London) **454**, 200 (2008).
[103] K. T. Chen and P. A. Lee, Phys. Rev. B **79**, 180510 (2009).
[104] A. J. Millis and M. R. Norman, Phys. Rev. B **76**, 220503(R) (2007).
[105] D. LeBoeuf, N. Doiron-Leyraud, J. Levallois, R. Daou, J.-B. Bonnemaison, N. E. Hussey, L. Balicas, B. J. Ramshaw, R. Liang, D. A. Bonn, W. N. Hardy, S. Adachi, C. Proust, and L. Taillefer, Nature (London) **450**, 533 (2007).
[106] Y. Iye, S. Nakamura and T. Tamegai, Physica C **159**, 616 (1989).
[107] S. J. Hagen, A. W. Smith, M. Rajeswari, J. L. Peng, Z. Y. Li, R. L. Greene, S. N. Mao, X. X. Xi, S. Bhattacharya, Q. Li, and C. J. Lobb, Phys. Rev. B **47**, 1064 (1993).
[108] J. M. Harris, N. P. Ong and Y. F. Yan, Phys. Rev. Lett. **71**, 1455 (1993).

[109] T. W. Clinton, A. W. Smith, Q. Li, J. L. Peng, R. L. Greene, C. J. Lobb, M. Eddy, and C. C. Tsuei, Phys. Rev. B **52**, R7046 (1995).
[110] D. A. Beam, N.-C. Yeh and F. Holtzberg, J. Phys.: Condens. Matter **10**, 5955 (1998).
[111] D. A. Beam, N.-C. Yeh and R. P. Vasquez, Phys. Rev. B **60**, 601 (1999).
[112] K. McElroy, D.-H. Lee, J. E. Hoffman, K. M. Lang, J. Lee, E. W. Hudson, H. Eisaki, S. Uchida and J. C. Davis, Phys. Rev. Lett. **94**, 197005 (2005).
[113] X. J. Zhou, T. Yoshida, D.-H. Lee, W. L. Yang, V. Brouet, F. Zhou, W. X. Ti, J. W. Xiong, Z. X. Zhou, T. Sasagawa, T. Kakeshita, H. Eisaki, S. Uchida, A. Fujimori, H. Hussain, and Z.-X. Shen, Phys. Rev. Lett. **92**, 187001 (2004).
[114] S. H. Pan, E. W. Hudson, A. K. Gupta, K.-W. Ng, H. Eisaki, S. Uchida, and J. C. Davis, Phys. Rev. Lett. **85**, 1536 (2000).
[115] Y. Wang, N. P. Ong, Z. A. Xu, T. Kakeshita, S. Uchida, D. A. Bonn, R. Liang, and W. N. Hardy, Phys. Rev. Lett. **88**, 257003 (2002).
[116] Y. Wang, L. Li and N. P. Ong, Phys. Rev. B **73**, 024510 (2006).
[117] M. R. Norman, A. Kanigel, M. Randeria, U. Chatterjee and J. C. Campuzano, Phys. Rev. B **76**, 174501 (2007).
[118] F.-C. Zhang and T. M. Rice, Phys. Rev. B **37**, 3759 (1988).
[119] D. J. van Harlingen, Rev. Mod. Phys. **67**, 515 (1995); and references therein.
[120] C. C. Tsuei and J. R. Kirtley, Rev. Mod. Phys. **72**, 969 (2000), and references therein.
[121] C. C. Tsuei and J. R. Kirtley, Phys. Rev. Lett. **85**, 182 (2000).
[122] J. R. Kirtley, C. C. Tsuei, and K. A. Moler, Science **285**, 1373 (1999).
[123] J. Y. T. Wei, N.-C. Yeh, D. F. Garrigus, and M. Strasik, Phys. Rev. Lett. **81**, 2542 (1998).
[124] (80) N.-C. Yeh, C.-T. Chen, G. Hammerl, J. Mannhart, A. Schmehl, C. W. Schneider, R. R. Schulz, S. Tajima, K. Yoshida, D. Garrigus, and M. Strasik,, Phys. Rev. Lett. **87**, 087003 (2001).
[125] N.-C. Yeh, C.-T. Chen, G. Hammerl, J. Mannhart, S. Tajima, K. Yoshida, A. Schmehl, C. W. Schneider and R. R. Schulz, Physica C **364-365**, 450 (2001).
[126] J. Y. T. Wei, N.-C. Yeh, W. D. Si, and X. X. Xi, Physica B **284**, 973 (2000).
[127] N.-C. Yeh, C.-T. Chen, R. P. Vasquez, C. U. Jung, S. I. Lee, K. Yoshida, and S. Tajima, J. Low Temp. Phys. **131**, 435 (2003).
[128] C. R. Hu, Phys. Rev. Lett. **72**, 1526 (1994).
[129] Y. Tanaka and S. Kashiwaya, Phys. Rev. Lett. **74**, 3451 (1995).
[130] S. Kashiwaya and Y. Tanaka, Phys. Rev. B **53**, 2667 (1996).
[131] G. E. Blonder, M. Tinkham and T. M. Klapwijk, Phys. Rev. B **25**, 4515 (1982).
[132] N.-C. Yeh, Bulletin of Assoc. Asia Pacific Phys. Soc. **12**, 2 (2002).
[133] A. G. Sun, D. A. Gajewski, M. B. Maple and R. C. Dynes, Phys. Rev. Lett. **72**, 2267 (1994).
[134] A. G. Sun, A. Truscott, A. S. Katz, R. C. Dynes, B. W. Veal and C. Gu, Phys. Rev. B **54**, 6734 (1996).
[135] Q. Li, Y. N. Tsay, M. Suenaga, R. A. Klemm, G. D. Gu and N. Koshizuka, Phys. Rev. Lett. **83**, 4160 (1999).
[136] R. A. Klemm, Philos. Mag. **85**, 801 (2005).
[137] R. Kleiner, A. S. Katz, A. G. Sun, R. Summer, D. A. Gajewski, S. H. Han, S. I. Woods, E. Dantsker, B. Chen, K. Char, M. B. Maple, R. C. Dynes and J. Clarke, Phys. Rev. Lett. **76**, 2161 (1996).
[138] T. Masui, M. Limonov, H. Uchiyama, S. Lee, and S. Tajima, and A. Yamanaka, Phys. Rev. B **68**, 060506(R) (2003).
[139] R. Khasanov, A. Shengelaya, A. Maisuradze, F. La Mattina, A. Bussmann-Holder, H. Keller and K. A. Mueller, Phys. Rev. Lett. **98,** 057007 (2007).
[140] R. Khasanov, S. Straessle, D. Di Castro, T. Masui, S. Miyasaka, S. Tajima, A. Bussmann-Holder and H. Keller, Phys. Rev. Lett. **99**, 237601 (2007).
[141] A. A. Abrikosov and L. P. Gor'kov, Soviet Phys. JETP **12**, 1243 (1961).
[142] H. Shiba, Prog. Theor. Phys. **40**, 435 (1968).
[143] P. Fulde and R. A. Ferrell, Phys. Rev. **135**, A550 (1964).
[144] M. A. Wolf and F. Reif, Phys. Rev. **137**, A557 (1965).
[145] P. Schlottmann, Phys. Rev. B **13**, 1 (1976).
[146] A. Yazdani, B. A. Jones, C. P. Lutz, M. F. Crommie, and D. A. Eigler, Science **275**, 1767 (1997).
[147] P. W. Anderson, J. Phys. Chem. Solids **11**, 26 (1959).
[148] G.-q. Zheng, T. Odaguchi, T. Mito, Y. Kitaoka, K. Asayama, and Y. Kodama, *J. Phys. Soc. Japan* **62**, 2591 (1989).
[149] H. Alloul, P. Mendels, H. Casalta, J. F. Marucco, and J. Arabski, Phys. Rev. Lett. **67**, 3140 (1991).
[150] T. Miyatake, K. Yamaguchi, T. Takata, N. Koshizuka, S. Tanaka, Phys. Rev. B **44**, 10139 (1991).
[151] G.-q. Zheng, T. Odaguchi, Y. Kitaoka, K. Asayama, Y. Kodama, K. Mizuhashi, and S. Uchida, Physica C **263**, 367 (1996).

[152] N. L. Wang, S. Tajima, A. I. Rykov, and K. Tomimoto, Phys. Rev. B **57**, R11081 (1999).
[153] K. Tomimoto, I. Terasaki, A. I. Rykov, T. Mimura, and S. Tajima, Phys. Rev. B **60**, (1999).
[154] Y. Sidis et al., Phys. Rev. Lett. **84**, 5900 (2000).
[155] H. F. Fong et al. Phys. Rev. Lett. **82**, 1939 (1999).
[156] J. Figueras, T. Puig, A. E. Carrillo, and X. Obradors, Supercond. Sci. Technol. **13**, 1067 (2000).
[157] K. Ishida, Y. Kitaoka, K. Yamazoe, K. Asayama, and Y. Yamada, Phys. Rev. Lett. **76**, 531 (1996).
[158] S. H. Pan, E. W. Hudson, K. M. Lang, H. Eisaki, S. Uchida, and J. C. Davis, Nature **403**, 746 (2000).
[159] E. W. Hudson, K. M. Lang, V. Madhavan, S. H. Pan, H. Eisaki, S. Uchida, and J. C. Davis, Nature **411**, 920 (2001).
[160] A. V. Balatsky, M. I. Salkola, and A. Rosengren, Phys. Rev. B **51**, 15547 (1995).
[161] M. I. Salkola, A. V. Balatsky, and D. J. Scalapino, Phys. Rev. Lett. **77**, 1841(1996).
[162] M. E. Flatte and J. M. Byers, Phys. Rev. B **56**, 11213 (1997); M. E. Flatte and J. M. Byers, Phys. Rev. Lett. **80**, 4546 (1998).
[163] M. I. Salkola, A. V. Balatsky, and J. R. Schrieffer, Phys. Rev. B **55**, 12648(1997).
[164] M. E. Flatte, Phys. Rev. B **61**, 14920 (2000).
[165] J. Bobroff, W. A. MacFarlane, H. Alloul, P. Mendels, N. Blanchard, G. Collin, and J. F. Marucco, Phys. Rev. Lett. **83**, 4381 (1999).
[166] A. Polkovnikov, S. Sachdev and M. Vojta Phys. Rev. Lett. **86**, 296 (2001).
[167] M. Vojta and R. Bulla, Phys. Rev. B **65**, 014511 (2001).
[168] R. Kilian, S. Krivenko, G. Khaliullin and P. Fulde, Phys. Rev. B **59**, 14432 (1999).
[169] A. V. Chubukov and N. Gemelke, Phys. Rev. B **61**, R6467 (2000).
[170] C. L. Wu, C. Y. Mou and D. Chang, Phys. Rev. B **63**, 172503 (2001).
[171] A. Lanzara, P. V. Bogdanov, X. J. Zhou, S. A. Kellar, D. L. Feng, E. D. Lu, T. Yoshida, H. Eisaki, A. Fujimori, K. Kishio, J.-I. Shimoyama, T. Noda, S. Uchida, Z. Hussain, Z.-X. Shen, Nature **412**, 510 (2001).
[172] J. Bobroff, H. Alloul, S. Ouazi, P. Mendels, A. Mahajan, N. Blanchard, G. Collin, V. Guillen, and J. F. Marucco, Phys. Rev. Lett. **89**, 157002 (2002).
[173] K. M. Lang, V. Madhavan, J. E. Hoffman, E. W. Hudson, H. Eisaki, S. Uchida, and J. C. Davis, Nature **415**, 412 (2002).
[174] S. H. Pan, J. P. O'Neal, R. L. Badzey, C. Chamon, H. Ding, J. R. Engelbrecht, Z. Wang, H. Eisaki, S. Uchida, A. K. Gupta, K.-W. Ng, E. W. Hudson, K. M. Lang, and J. C. Davis, Nature **413**, 282 (2001).
[175] C. U. Jung, J. Y. Kim, M. S. Park, M. S. Kim, H. J. Kim, S. Y. Lee and S. I. Lee, Phys. Rev. B **65,** 172501 (2002).
[176] C. U. Jung, J. Y. Kim, Mun-Seog Kim, Min-Seok Park, Heon-Jung Kim, Yushu Yao, S. Y. Lee, and S.-I. Lee, Physica C **366**, 299 (2002).
[177] G. Blumberg, A. Koitzsch, A. Gozar, B. S. Dennis, C. A. Kendziora, P. Fournier and R. L. Greene, Phys. Rev. Lett. **88**, 107002 (2002).
[178] J. R. Schrieffer, X.-G. Wen and S.-C. Zhang, Phys. Rev. B **39**, 11663 (1989).
[179] A. Polkovnikov, S. Sachdev and M. Vojta, Physica C **388**, 19 (2003).
[180] P. Dai, H. A. Mook, R. D. Hunt and F. Dogan, Phys. Rev. B **63**, 054525 (2001).
[181] E. M. Motoyama, P. K. Mang, D. Petitgrand, G. Yu, O. P. Vajk, I. M. Vishik and M. Greven, Phys. Rev. Lett. **96**, 137002 (2006).
[182] T. Kondo, R. Khasanov, T. Takeuchi, J. Schmalian and A. Kaminski, Nature **457**, 296 (2009).
[183] C.-T. Chen and N.-C. Yeh, Phys. Rev. B **68**, 220505(R) (2003).
[184] A. Kanigel, M. R. Norman, M. Randeria, U. Chatterjee, S. Souma, A. Kaminski, H. M. Fretwell, S. Rosenkranz, M. Shi, T. Sato, T. Takahashi, Z. Z. Li, H. Raffy, K. Kadowaki, D. Hinks, L. Ozyuzer, and J. C. Campuzano, Nature Phys. **2**, 447 (2006).
[185] P. W. Anderson, Science 235, 1196 (1987).
[186] M. R. Norman, A. Kanigel, M. Randeria, U. Chatterjee and J. C. Campuzano, Phys. Rev. B **76**, 174501 (2007).
[187] C. Gros, B. Edegger, V. N. Muthukumar and P. W. Anderson, Proc. Natl. Acad. Sci. **103**, 14298 (2006).
[188] A. V. Chubukov, M. R. Norman, A. J. Millis and E. Abrahams, Phys. Rev. B **76**, 180506(R) (2007).
[189] F. Onufrieva and P. Pfeuty, Phys. Rev. Lett. **92**, 247003 (2004).
[190] G. Kotliar and C. M. Varma, Phys. Rev. Lett. **77**, 2296 (1996).
[191] A. A. Abrikosov, Sov. Phys. JETP **5**, 1174 (1957).
[192] C. Caroli, P. G. deGennes and J. Matricon, J. Phys. Lett. **9**, 307 (1964).
[193] F. Gygi and M. Schluter, Phys. Rev. B **43**, 7609 (1991).
[194] H. Hess, R. B. Rubinson and J. V. Waszczak, Phys. Rev. Lett. **64**, 2711 (1990).
[195] S. H. Pan, E. W. Hudson, A. K. Gupta, K.-W. Ng, H. Eisaki, S. Uchida, and J. C. Davis, Phys. Rev. Lett. **85**, 1536 (2000).

[196] M. Tachiki, M. Machida and T. Egami, Phys. Rev. B **67**, 174506 (2003).
[197] N.-C. Yeh, M. L. Teague, A. D. Beyer, B. Shen and H.-H. Wen, conference proceedings for the 26th International Low-Temperature Physics Conference (LT26), accepted for publication in J. Phys. Conf. Series (2011); [arXiv:1107.0697].
[198] M.-S. Kim, S.-I. Lee, S. C. Yu, I. Kuzemskaya, E. S. Itskevich, and K. A. Lokshin, Phys. Rev. B **57**, 6121 (1998).
[199] M.-S. Kim, C. U. Jung, S.-I. Lee and A. Iyo, Phys. Rev. B **63**, 134513 (2001).
[200] V. M. Vinokur, V. B. Geshkenbein, M. V. Feigelman and G. Blatter, Phys. Rev. Lett. **71**, 1242 (1993).
[201] M. V. Feigelman, V. B. Geshkenbein, A. I. Larkin and V. M. Vinokur, Pisma Zh. Eksp. Teor. Fiz. **62**, 811 (1995).
[202] M. V. Feigelman, V. B. Geshkenbein, A. I. Larkin and V. M. Vinokur, JETP Lett. **62**, 834 (1995).
[203] A. van Otterlo, M. V. Feigelman, V. B. Geshkenbein and G. Blatter, Phys. Rev. Lett. **75**, 3736 (1995).
[204] T. Nagaoka et al., Phys. Rev. Lett. **80**, 3594 (1998).
[205] L. Li, Y. Wang, M. J. Naughton, S. Komiya, S. Ono, Y. Ando, and N. P. Ong, J. Magn. Magn. Mater. **310**, 460 (2007).
[206] H. Takahashi, K. Igawa, K. Arii, Y. Kamihara, M. Hirano and H. Hosono, Nature **453**, 376 (2008).
[207] X. H. Chen, T. Wu, G. Wu, R. H. Liu, H. Chen and D. F. Fang, Nature (London) **453**, 761 (2008).
[208] H. H. Wen, C. Mu, L. Fang, H. Yang, and X. Zhu, Europhys. Lett. **82**, 17009 (2008).
[209] Z. A. Ren, J. Yang, W. Lu, X.-L. Shen, Z.-C. Li, G.-C. Che, X.-L. Dong, L.-L. Sun, F. Zhou and Z.-X. Zhao, Europhys. Lett. **82**, 57002 (2008).
[210] K. Sasmal, K. Sasmal, B. Lv, B. Lorenz, A. M. Guloy, F. Chen, Y. Y. Xue and C. W. Chu, Phys. Rev. Lett. **101**, 107007 (2008).
[211] M. Rotter, M. Tegel and D. Johrendt, Phys. Rev. Lett. **101**, 107006 (2008).
[212] A. S. Sefat, Rongying Jin, Michael A. McGuire, Brian C. Sales, David J. Singh, and David Mandrus, Phys. Rev. Lett. **101**, 117004 (2008).
[213] J. H. Tapp, Z. Tang, B. Lv, K. Sasmal, B. Lorenz, C. W. Chu, and A. M. Guloy, Phys. Rev. B **78**, 060505(R) (2008).
[214] K.-W. Yeh, T.-W. Huang, Y.-L. Huang, T.-K. Chen, F.-C. Hsu, P. M. Wu, Y.-C. Lee, Y.-Y. Chu, C.-L. Chen, J.-Y. Luo, D.-C. Yan and M.-K. Wu, Europhys. Lett. **84**, 37002 (2008).
[215] F.-C. Hsu, J.-Y. Luo, K.-W. Yeh, T.-K. Chen, T.-W. Huang, P. M. Wu, Y.-C. Lee, Y.-L. Huang, Y.-Y. Chu, D.-C. Yan and M.-K. Wu, Proc. Nat. Acad. Sci. **105**, 14262 (2008).
[216] Y. Mizuguchi, F. Tomioka, S. Tsuda, T. Yamaguchi, Y. Takano, Appl. Phys. Lett. **93**, 152505 (2008).
[217] J. Guo, S. Jin, G. Wang, S. Wang, K. Zhu, T. Zhou, M. He, and X. Chen, Phys. Rev. B **82**, 180520(R) (2010).
[218] V. Cvetkovic and Z. Tesanovic, Europhys. Lett. **85**, 37002 (2009).
[219] M. R. Norman, Physics **1**, 21 (2008).
[220] Z. Tesanovic, Physics **2**, 60 (2009).
[221] A. V. Balatsky and D. Parker, Physics **2**, 59 (2009).
[222] F. Wang and D.-H. Lee, Science **332**, 200 (2011).
[223] C. de la Cruz, Q. Huang, J. Li, W. Ratcliff, II, J. L. Zarestky, H. A. Mook, G. F. Chen, J. L. Luo, N. L. Wang and P. C. Dai, Nature **453**, 899 (2008).
[224] Q. Huang, Y. Qiu, W. Bao, M. A. Green, J. W. Lynn, Y. C. Gasparovic, T. Wu, G. Wu and X. H. Chen, Phys. Rev. Lett. **101**, 257003 (2008).
[225] J. Zhao, Q. Huang, C. de la Cruz, S. L. Li, J. W. Lynn, Y. Chen, M. A. Green, G. F. Chen, G. Li, Z. Li, J. L. Luo, N. L. Wang and P. C. Dai, Nat. Mater. **7**, 953 (2008).
[226] T. Yildirim, Physica C **469**, 425 (2009).
[227] F. Wang, H. Zhai and D.-H. Lee, Europhys. Lett. **85**, 37005 (2009).
[228] F. Wang, H. Zhai, Y. Ran, A. Vishwanath and D.-H. Lee, Phys. Rev. Lett. **102**, 047005 (2009).
[229] I. I. Mazin, D. J. Singh, M. D. Johannes, and M. H. Du, Phys. Rev. Lett. **101**, 057003 (2008).
[230] K. Kuroki et al., Phys. Rev. Lett. **101**, 087004 (2008).
[231] Y.-Y. Zhang et al., Phys. Rev. B **80**, 094528 (2009).
[232] E. Plamadeala, T. Pereg-Barnea, and G. Refael, Phys. Rev. B **81**, 134513 (2010).
[233] A. V. Balatsky, D. N. Basov and J.-X. Zhu, Phys. Rev. B **82**, 144522 (2010).
[234] D. H. Lu, M. Yi, S.-K. Mo, A. S. Erickson, J. Analytis, J.-H. Chu, D. J. Singh, Z. Hussain, T. H. Geballe, I. R. Fisher and Z.-X. Shen, Nature (London) **455**, 81 (2008).
[235] H. Ding, P. Richard, K. Nakayama, K. Sugawara, T. Arakane, Y. Sekiba, A. Takayama, S. Souma, T. Sato, T. Takahashi, Z. Wang, X. Dai, Z. Fang, G. F. Chen, J. L. Luo and N. L. Wang, Europhys. Lett. **83**, 47001 (2008).

[236] K. Terashima, Y. Sekibab, J. H. Bowenc, K. Nakayamab, T. Kawaharab, T. Satob,d, P. Richarde, Y.-M. Xuf, L. J. Lig, G. H. Caog, Z.-A. Xug, H. Dingc, and T. TakahashiProc. Natl. Acad. Sci. U.S.A. 106, 7330 (2009).
[237] M. L. Teague, G. K. Drayna, G. P. Lockhart, P. Cheng, B. Shen, H.-H. Wen and N.-C. Yeh, Phys. Rev. Lett. **106**, 087004 (2011).
[238] N.-C. Yeh, M. L. Teague, A. D. Beyer, B. Shen and H.-H. Wen, Conference Proceedings for the 26$^{th}$ International Low-Temperature Physics Conference (LT26), accepted for publication in J. Phys. Conf. Series (2012); [arXiv:1107.0697].
[239] L. Shan, Y.-L.Wang, B. Shen, B. Zeng, Y. Huang, A. Li, D. Wang, H. Yang, C. Ren, Q.-H. Wang, S. H. Pan and H.-H. Wen, Nat. Phys. **7**, 325 (2011).
[240] R. C. Dynes, V. Narayanamurti, and J. P. Garno, Phys. Rev. Lett. **41**, 1509 (1978).
[241] M. M. Korshunov, I. Eremin, Phys. Rev. B **78**, 140509(R) (2008).
[242] T. A. Maier, D. J. Scalapino, Phys. Rev. B **78**, 020514(R) (2008).
[243] M. D. Lumsden et al., Phys. Rev. Lett. **102**, 107005 (2009).
[244] A. D. Christianson, E. A. Goremychkin, R. Osborn, S. Rosenkranz, M. D. Lumsden, C. D. Malliakas, I. S. Todorov, H. Claus, D. Y. Chung, M. G. Kanatzidis, R. I. Bewley and T. Guidi, Nature **456**, 930 (2008).
[245] T. Hanaguri, S. Niitaka, K. Kuroki, H. Takagi, Science **328**, 474 (2010).
[246] C.-T. Chen, C. C. Tsuei1, M. B. Ketchen, Z.-A. Ren and Z. X. Zhao, Nature Phys. **6**, 260 (2010).
[247] I. Eremin et al., Phys. Rev. Lett. 94, 147001 (2005).
[248] T. Das and A. V. Balatsky, Phys. Rev. Lett. **106**, 157004 (2011).
[249] G. Yu, Y. Li, E. M. Motoyama and M. Greven, Nature Phys. **5**, 873 (2009).
[250] Y. Yin, M. Zech, T. L. Williams, X. F. Wang, G. Wu, X. H. Chen, and J. E. Hoffman, Phys. Rev. Lett. **102**, 097002 (2009).
[251] P. W. Anderson, "The Theory of Superconductivity in the High-$T_c$ Cuprates", Princeton University Press, Princeton (1997).
[252] D. J. Scalapino, Phys. Rep. **250**, 329 (1995); and many references therein.
[253] X.-J. Chen, V. V. Struzhkin, Z. Wu, H.-Q. Lin, R. J. Hemley and H.-k. Mao, PNAS **104**, 3732 (2007).
[254] J. L. Sarrao, L. A. Morales, J. D. Thompson, B. L. Scott, G. R. Stewart, F. Wastin, J. Rebizant, P. Boulet, E. Colineau and G. H. Lander, Nature **420**, 297 (2002).
[255] E. D. Bauer, J. D. Thompson, J. L. Sarrao, L. A. Morales, F. Wastin, J. Rebizant, J. C. Griveau, P. Javorsky, P. Boulet, E. Colineau, G. H. Lander and G. R. Stewart, Phys. Rev. Lett. **93**, 147005 (2004).
[256] J. Custers, P. Gegenwart, H. Wilhelm, K. Neumaier, Y. Tokiwa, O. Trovarelli, C. Geibel, F. Steglich, C. Pepin and P. Coleman, Nature **424**, 524 (2003).
[257] J. R. Jeffries, N. A. Frederick, E. D. Bauer, H. Kimura, V. S. Zapf, K. D. Hof, T. A. Sayles and M. B. Maple, Phys. Rev. B **72**, 024551 (2005).
[258] P. Gegenwart, Q. Si, and F. Steglich, Nature Physics **4**, 186 (2008).
[259] T. Park, F. Ronning, H. Q. Yuan, M. B. Salamon, R. Movshovich, J. L. Sarrao and J. D. Thompson, Nature **440**, 65 (2006).
[260] N. K. Sato et al., Nature **410**, 340 (2001).
[261] C. Stock, C. Broholm, J. Hudis, H. J. Kang, and C. Petrovic, Phys. Rev. Lett. **100**, 087001 (2008).
[262] F. Steglich, J. Arndt, S. Friedemann, C. Krellner, Y. Tokiwa, T. Westerkamp, M. Brando, P. Gegenwart, C. Geibel, S. Wirth and O. Stockert, J. Phys.: Condens. Matter **22**, 164202 (2010).
[263] E. D. Bauer, R. P. Dickey, V. S. Zapf and M. B. Maple, J. Phys.: Cond. Matter **13**, L759 (2001).
[264] N. J. Curro, T. Caldwell, E. D. Bauer, L. A. Morales, M. J. Graf, Y. Bang, A. V. Balatsky, J. D. Thompson, J. L. Sarrao, Nature **434**, 622 (2005).
[265] T. Das and A. V. Balatsky, Phys. Rev. B **84**, 014521 (2011).
[266] T. Das and A. V. Balatsky, Phys. Rev. B **84**, 115117 (2011).
[267] T. Egami, J.-H. Chung, R. J. McQueeney, M. Yethiraj, H. A. Mook, C. Frost, Y. Petrov, F. Dogan, Y. Inamura, M. Arai, S. Tajima, and Y. Endoh, Physica B **316-317**, 62 (2002).
[268] D. Zech, K. Conder, H. Keller, E. Kaldis, K. A. Muller, Physica B **219&220**, 136 (1996).
[269] M. K. Crawford, M. N. Kunchur, W. E. Farneth, E. M. McCarron III, S. J. Poon, Phys. Rev. B **41**, 282 (1990).
[270] D. J. Pringle, G. V. M. Williams and J. L. Tallon, Phys. Rev. B **62**, 12527 (2000).
[271] G. M. Zhao, V. Kirtikar and D. E. Morris, Phys. Rev. B **63**, 220506 (2001).